\documentclass[12pt,preprint]{aastex}

\begin{document}

\shortauthors{Luhman et al.}
\shorttitle{}

\title{A Census of Young Stars and Brown Dwarfs in IC~348 and
NGC~1333\altaffilmark{1}}

\author{
K. L. Luhman\altaffilmark{2,3},
T. L. Esplin\altaffilmark{2},
and N. P. Loutrel\altaffilmark{4}
}

\altaffiltext{1}
{Based on data from the NASA Infrared Telescope Facility, Gemini Observatory, 
Canada-France-Hawaii Telescope, Keck Observatory, Subaru Telescope,
the Digitized Sky Survey, and the Two Micron All-Sky Survey.}

\altaffiltext{2}{Department of Astronomy and Astrophysics, The Pennsylvania
State University, University Park, PA 16802; kluhman@astro.psu.edu.}

\altaffiltext{3}{Center for Exoplanets and Habitable Worlds, The
Pennsylvania State University, University Park, PA 16802, USA}

\altaffiltext{4}{Department of Physics, Montana State University, 
Bozeman, MT 59715, USA}

\begin{abstract}

We have obtained optical and near-infrared spectra of candidate members
of the star-forming clusters IC~348 and NGC~1333.
We classify 100 and 42 candidates as new members of the clusters,
respectively, which brings the total numbers of known members to 478 and 203.
We also have performed spectroscopy on a large majority of the previously
known members of NGC~1333 in order to provide spectral classifications
that are measured with the same scheme that has been applied to IC~348
in previous studies.
The new census of members is nearly complete for $K_s<16.8$ at $A_J<1.5$ in
IC~348 and for $K_s<16.2$ at $A_J<3$ in NGC~1333, which correspond to masses of
$\gtrsim$0.01~$M_\odot$ for ages of 3~Myr according to theoretical evolutionary
models. The faintest known members extend below these completeness limits
and appear to have masses of $\sim$0.005~$M_\odot$.  
In extinction-limited samples of cluster members, NGC~1333 exhibits a higher
abundance of objects at lower masses than IC~348.
It would be surprising if the initial mass functions of these clusters differ
significantly given their similar stellar densities and formation environments.
Instead, it is possible that average extinctions are lower for less massive
members of star-forming clusters, 
in which case extinction-limited samples could be biased in favor of low-mass
objects in the more heavily embedded clusters like NGC~1333.
In the H-R diagram,
the median sequences of IC~348 and NGC~1333 coincide with each other for
the adopted distances of 300 and 235~pc, which would suggest that they 
have similar ages. However, NGC~1333 is widely believed to be younger
than IC~348 based on its higher abundance of disks and protostars and its
greater obscuration. 
Errors in the adopted distances may be responsible for this discrepancy.

\end{abstract}


\section{Introduction}
\label{sec:intro}

A thorough census of young stars and brown dwarfs in nearby star-forming
regions is important for measuring the global properties of young stellar
populations (e.g., initial mass functions, disk fractions) and for providing
well-defined samples of targets for a variety of studies of star and planet
formation.
The Perseus molecular cloud is one of the nearest and richest of these regions
\citep[$\sim300$~pc,][]{sch14}.
It contains several hundred young stars, most of which reside in two clusters,
IC~348 and NGC~1333 \citep{her08,wal08}.
Age estimates for IC~348 based on evolutionary models range from
2 to 6~Myr \citep{luh03,bel13}.
NGC~1333 appears to be younger than IC~348 based on its greater obscuration
and higher abundance of circumstellar disks and protostars \citep{mue07,gut08}.

Candidate members of IC~348 have been identified via
proper motions \citep{fre56,sch99},
H$\alpha$ emission \citep{her54,her98},
X-ray emission \citep{pre96,pre01,pre02,pre04,for11,ste12},
outflow signatures \citep{wal06,hat09},
optical and near-infrared (IR) photometry
\citep{str74b,lad95,luh98,luh03,luh99,naj00,mai03,bur09,mue03,alv13},
mid-IR excess emission \citep{lad06,jor06,jor07,cie07,mue07,reb07,cur09,eva09,you15},
variability \citep{fla12,fla13,cod14},
and kinematics \citep{cot15}.
Optical and near-IR spectroscopy has been used to measure spectral types
and confirm membership for many of those candidates
\citep{str74b,har54,her98,luh98,luh99,luh03,luh05flam,mue07,alv13}.
Similar diagnostics of youth and membership have been applied
to NGC~1333, including
X-ray emission \citep{pre97,pre03,get02,win10,for11}, outflows \citep{hat09},
optical and near-IR photometry
\citep{str76,asp94,lad96,asp03,wil04,gre07,ots08,sch09},
mid-IR excesses \citep{jor06,jor07,cie07,reb07,gut08,eva09,you15},
kinematics \citep{fos15}, and variability \citep{reb15b}.
As in IC~348, spectroscopic classification has been performed on many
of the resulting candidates
\citep{asp03,wil04,gre07,win09,win10,sch09,sch12a,sch12b}.

The current census of IC~348 is incomplete in the outer portions of the
cluster and among the least massive brown dwarfs.
The census of NGC~1333 has significant incompleteness as well,
particularly at substellar masses.
In addition, the methods of spectral classification that have been applied
to NGC~1333 are less uniform than in IC~348. 
To address these issues, we have performed a survey for new
members down to $\sim0.005$~$M_\odot$ across the full extent of each cluster,
and we have measured spectral types for a large fraction of the known members
of NGC~1333 with the classification scheme that we have previously applied
to IC~348 and other nearby star-forming regions. 
In our presentation of this work, we begin by compiling lists of all known
members of these clusters from previoius studies (Section~\ref{sec:previous}).
We then select candidate cluster members based on X-ray emission,
mid-IR excess emission, optical and near-IR color-magnitude diagrams, and
proper motions (Section~\ref{sec:select}) and use optical and near-IR
spectra to measure their spectral types and determine whether they are members
(Section~\ref{sec:spectra}). We also measure new spectral types for a
large number of the known members of NGC~1333. We conclude by analyzing
several aspects of the new samples of members of IC~348 and NGC~1333, which
include their completeness, ages, mass functions, disk fractions, and spatial
distributions (Section~\ref{sec:analysis}).

\section{Census from Previous Studies}
\label{sec:previous}

We have searched previous studies of IC~348 and NGC~1333 for objects
that exhibit evidence of membership. For IC~348, we began with the
census of 288 members compiled by \citet{luh03}. In that tabulation, pairs
of objects with separations of $<1\arcsec$ appeared as single entries.
We adopt the same approach in this work.
\citet{luh03} noted that LRL~1434\footnote{When referring to objects
in IC~348, we use the number identifications from our previous studies and
from this work, which are found in the second column of Table~\ref{tab:mem348}.}
was unusually faint for a cluster member
near its spectral type, which would indicate that it is either a field star
or a member that is detected in scattered light, as in the case of an edge-on
disk. Because its H$\alpha$ emission and Na~I absorption in a low
signal-to-noise (S/N) spectrum seemed to indicate that it was young,
\citet{luh03} adopted it as a member. However, it was not detected in
subsequent mid-IR images of the cluster \citep{lad06,mue07}, indicating that
it does not have a disk, and hence is unlikely to be seen in scattered light. 
Therefore, we omit LRL~1434 from our sample of members.
In a low S/N spectrum from \citet{luh03}, LRL~624 appeared
to have the weak Na~I absorption that is expected for a young, low-gravity
cluster member. However, we now treat it as a field star based on its proper
motion, which is inconsistent with membership in IC~348 (Section~\ref{sec:pm}).
\citet{luh05wfpc} measured spectral types for a likely companion in the census
from \citet{luh03}, LRL~78~B, and for the candidate companion LRL~166~B.
Because LRL~166~B and its primary are separated by less than $1\arcsec$,
they appear as a single entry in our tabulation of members.
We have added to our sample the 16, 45, and 16 members identified by
\citet{luh05flam}, \citet{mue07}, and \citet{alv13}, respectively.
The latter study presented spectra for several additional objects whose
membership was uncertain. Based on our inspection of those spectra, we
have classified five of those candidates as likely members 
(Section~\ref{sec:class}), consisting of sources 3, 5, 14, 20, and 31
from \citet{alv13} (LRL~670, LRL~5209, LRL~10378, LRL~22443, LRL~54229).
We include in our census the protostar HH~211-IR, the candidate protostellar
brown dwarf IC~348-SMM2E \citep{pal14}, and LRL 1898, LRL 54361, LRL 54362,
LRL 54419, LRL 54459, LRL 54460, LRL 55400, and LRL 57025 \citep{mue07}.
The latter eight objects are probable protostars based on their spectral
energy distributions and their proximity to millimeter cores and 
known protostars. Through the above steps, we arrived at a sample
of 378 known members of IC~348 based on previous studies.
Later in this work, we describe the identification of 100 new members,
resulting in a total of 478 known members.
We present the full sample of members in Table~\ref{tab:mem348}.
The new members can be identified by the presence of a spectral type from
this study alone with the exception of LRL~60~B and LRL~187~B.
Although the latter stars lacked classifications prior
to our spectroscopy, they are counted as previously known members rather
than new ones since they appeared within the sample of members from
\citet{luh03}.

To construct a census of known members of NGC~1333, we assessed the evidence
of membership for all objects proposed to be members in previous surveys
of the cluster. The evidence consisted of signatures of youth in the form
of strong emission lines, Li absorption, X-ray emission,
mid-IR excess emission, and the shape of the gravity-sensitive steam bands.
We also examined whether the proposed members exhibited radial velocities
\citep{fos15} and proper motions (Section~\ref{sec:pm}) that are
consistent with those of the larger population of objects that show evidence
of youth.
The radial velocity of 2MASS~03290289+3116010 and the proper motion of source
38 from \citet{sch12a} indicate that they are unlikely to be cluster members.
The previously reported members that we find have sufficient evidence of
membership in NGC~1333 are listed in Table~\ref{tab:mem1333}.
We also include in that tabulation the new members found in our study.
This census contains a total of 203 members. For 42 sources, our
new spectral classifications are the only ones available, some of which
were identified as candidate members in previous studies.

In our tabulations of members of IC~348 and NGC~1333, we have included
all known young stars and brown dwarfs within the fields encompassed by the
maps in Figure~\ref{fig:map}. However, one could argue that the
objects at the largest distances from the centers of the clusters
should instead be assigned to the distributed population that
is present across the Perseus cloud \citep{you15}.

\section{Identification of Candidate Members}
\label{sec:select}

To improve the completeness of the census of known members of IC~348
and NGC~1333, we have obtained spectra of candidate members that
have been identified through several signatures of cluster membership.
In this section, we describe the selection of these candidates.

\subsection{Survey Fields}
\label{sec:fields}

\begin{figure}
\epsscale{1}
\plotone{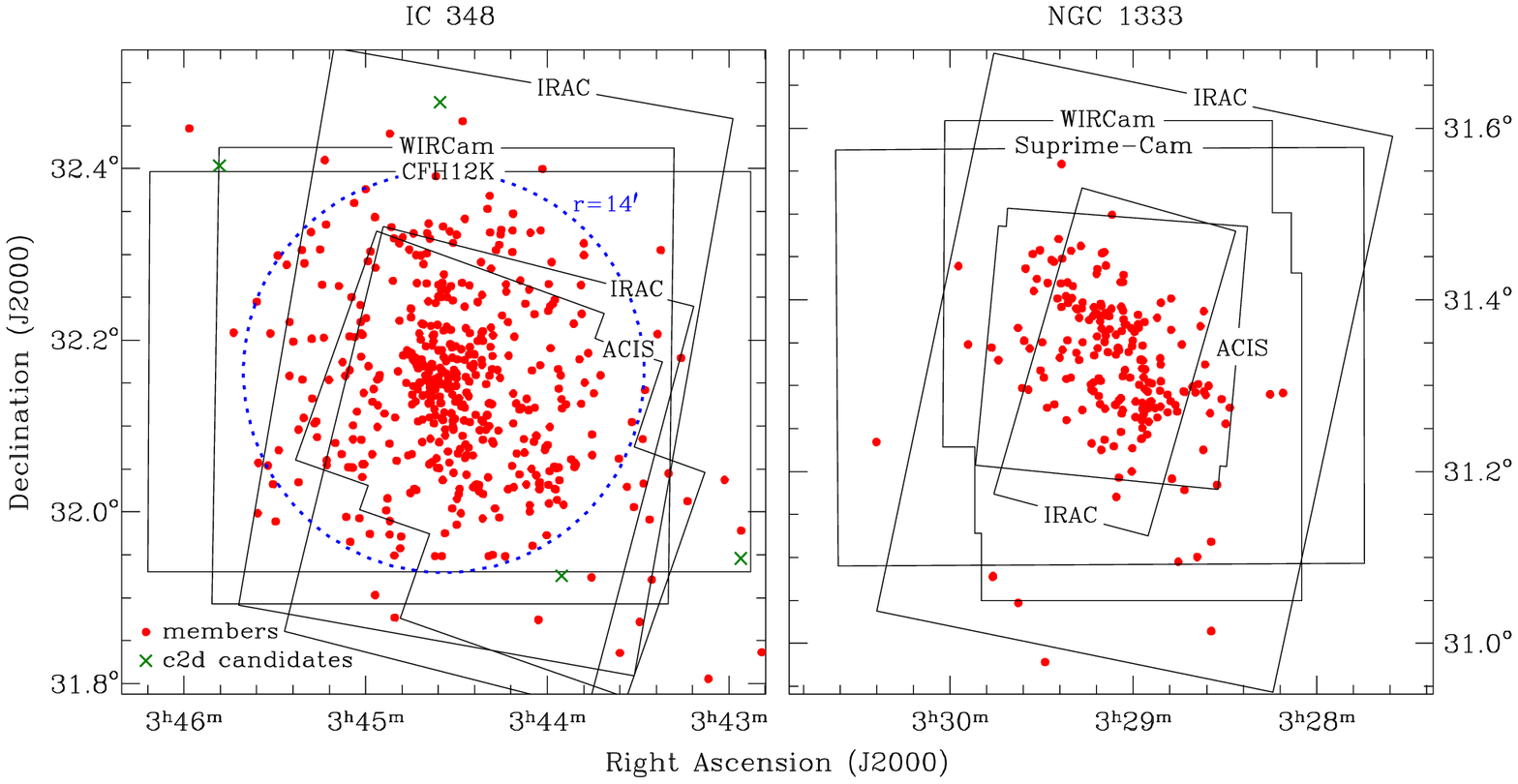}
\caption{
The positions of the known members of the IC~348 and NGC~1333 clusters
(filled circles, Tables~~\ref{tab:mem348} and \ref{tab:mem1333}) and
additional candidate young stars in the surrounding molecular cloud from
the c2d survey \citep[crosses,][]{eva09,you15}.
We have marked the boundaries of fields that have been observed by ACIS-I on
{\it Chandra} \citep{pre01,pre02,get02,win10,for11,ste12},
IRAC on {\it Spitzer} \citep{lad06,mue07,gut08,fla13,reb15b},
Suprime-Cam on Subaru \citep{sch09},
and CFH12K and WIRCam on the CFHT \citep[][this work]{luh03,alv13}.
MegaCam on the CFHT and UKIDSS imaged the entire field encompassed by the map
of IC~348 \citep{alv13}.
We have searched for new members of these clusters primarily within
a radius of $14\arcmin$ from the B5 star BD+$31\arcdeg$643 in IC~348
(dotted circle) and within the ACIS-I field in NGC~1333.
}
\label{fig:map}
\end{figure}

We have made use of several imaging surveys of IC~348 and NGC~1333
for identifying candidate members. In Figure~\ref{fig:map}, we show maps
of the positions of the known members of the clusters and the boundaries
of the fields in those surveys.
IC~348 was observed with the imaging array of the Advanced CCD Imaging
Spectrometer (ACIS-I) on the {\it Chandra} X-ray Observatory
\citep{pre01,pre02,for11,ste12}, CFH12K on the 
Canada-France-Hawaii Telescope (CFHT) \citep[$IZ$,][]{luh03},
and WIRCam and MegaCam on the CFHT
\citep[$z\arcmin$$JHK_s$,][]{alv13}.
The MegaCam data encompass the entire field surrounding IC~348 in
Figure~\ref{fig:map}.
Data in $ZYJHK$ are also available for all of IC~348 from Data Release 10
of the United Kingdom Infrared Telescope Infrared Deep Sky Survey
\citep[UKIDSS,][]{law07}.
NGC~1333 was observed with ACIS-I \citep{get02,win10,for11} and
Suprime-Cam on Subaru Telescope \citep[$i\arcmin$$z\arcmin$,][]{sch09}.
Unpublished images from WIRCam are also available for NGC~1333.
We reduced all of the WIRCam images in $JHK_s$ that are available
for IC~348 and NGC~1333 from the CFHT archive, which were obtained through
programs 07BH20 (K. Allers), 07BH12 (B. Biller), 09BD95 (L. Albert),
O6BF23, 08BF98, and 09BF50 (J. Bouvier). The data from O6BF23 were those
in IC~348 analyzed by \citet{alv13}. We also reduced the Suprime-Cam images
from \citet{sch09}, which were retrieved from the Subaru Telescope data archive.

Both clusters have been observed on many occasions at mid-IR wavelengths
with the Infrared Array Camera \citep[IRAC;][]{faz04} on the
{\it Spitzer Space Telescope} \citep{wer04}.
Through the c2d {\it Spitzer} Legacy project \citep{eva03}, shallow IRAC images
were obtained for much of the Perseus cloud, including all of 
IC~348 and NGC~1333 \citep{jor06,eva09}.
Deeper images were taken for the larger IRAC fields indicated in
Figure~\ref{fig:map} \citep{lad06,mue07,gut08}. The smaller IRAC
fields in Figure~\ref{fig:map} were monitored for approximately one
month \citep{fla13,reb15b}. In the latest IRAC observations,
most of each cluster was imaged at an additional epoch through program
90071 (A. Kraus) to facilitate the identification of candidate members
via proper motions. The IRAC observations prior to May 2009 were performed with
bands at 3.6, 4.5, 5.8, and 8.0~\micron, denoted as [3.6], [4.5], [5.8],
and [8.0], respectively. The later images were collected only in the 
[3.6] and [4.5] bands because of the depletion of the liquid helium coolant.
The Multiband Imaging Photometer for {\it Spitzer} \citep[MIPS;][]{rie04}
also has been used to fully map each cluster at 24~\micron\ 
\citep{lad06,reb07,gut08,cur09}.

In addition to the known members and the survey fields, we also show in
Figure~\ref{fig:map} the positions of candidate disk-bearing stars that were
identified by the c2d survey and that have not been observed with spectroscopy
to confirm membership. Only a few of these candidates are present within
the map of IC~348 and none are found in the map of NGC~1333,
indicating that the distributions of known members trace the full extent of
the clusters. The distribution for IC~348 has a radius of $\sim14\arcmin$,
which is consistent with previous estimates of 10--$15\arcmin$
for the cluster radius \citep{sch99,mue03}.
We wish to achieve a census of the members of these clusters
within fields that have well-defined boundaries, are large enough to
encompass most of the members, and are small enough to be covered by as
many imaging surveys as possible. Given these considerations, we have
searched for new members primarily within a radius of $14\arcmin$ from the B5
star BD+$31\arcdeg$643 in IC~348 and within the $18\arcmin\times18\arcmin$
field in NGC~1333 that was observed by ACIS-I.
We will assess the completeness of our new census within these fields in
Section~\ref{sec:completeness}.

\subsection{X-ray Emission}

Because young stars are bright in X-rays, one can search for
members of star-forming regions via their X-ray emission \citep{fei87,wal88}.
The X-ray studies of IC~348 and NGC~1333 (see Section~\ref{sec:intro})
have done so by checking for X-ray sources that have optical and near-IR
data that are consistent with those expected for cluster members.
Most of the resulting candidate members have been observed with the
spectroscopy that is needed to confirm membership.
We have pursued spectroscopy of the remaining candidates that lack spectra.
For this sample, we selected X-ray sources identified in {\it Chandra}
images of IC~348 and NGC~1333 by \citet{ste12} and K. Getman (in preparation),
respectively, that are not rejected as non-members by our color-magnitude
diagrams (Section~\ref{sec:cmd}). Those two studies have generated
catalogs of sources found in all available ACIS-I images of the two clusters.
In Table~\ref{tab:mem348}, the members of IC~348 that have X-ray detections
can be identified by the presence of source names from \citet{ste12}.
For the members of NGC~1333, we indicate in Table~\ref{tab:mem1333}
the X-ray detections from the new catalog of K. Getman under the column
for membership evidence.

\subsection{Mid-IR Excess Emission}
\label{sec:midir}

When a star is born, it is surrounded by an accretion disk and an infalling
envelope.
The stars in a young cluster that still retain these structures can be
identified via mid-IR emission in excess above that expected from a stellar
photosphere.
The {\it Spitzer} images described in Section~\ref{sec:fields} 
have been previously used to search for new members of IC~348 and NGC~1333 in
that manner \citep{mue07,gut08,eva09,you15}. 
As with the X-ray candidates, we have sought spectroscopy for the small
fraction of mid-IR candidates that lack previous spectral classifications.
When assembling this sample of candidates, we rejected those
that appear to be knots of extended emission rather than stars based
on visual inspection of the IRAC and MIPS images, which consist of
sources 171, 216, 222, and 227 from \citet{eva09} and \citet{you15}.
If a young star is seen primarily in scattered light, as in
the case of an edge-on disk, it is likely to appear unusually faint for
its color compared to other cluster members. As a result, it would be prone
to rejection as a field star in optical and near-IR color-magnitude diagrams.
Therefore, we have retained mid-IR candidates in our spectroscopic sample
regardless of their locations in the color-magnitude diagrams in
Section~\ref{sec:cmd}.
Active galactic nuclei and stars on the asymptotic giant branch are
common types of contaminants in a sample of this kind that is selected via
red mid-IR colors.
The absence or presence of mid-IR excess emission is indicated for
each known member of IC~348 and NGC~1333 in Tables~\ref{tab:mem348}
and \ref{tab:mem1333}, except for a few of the faintest members that
lack sufficiently accurate mid-IR photometry.

\subsection{Optical and Near-IR Color-Magnitude Diagrams}
\label{sec:cmd}

In the Hertzsprung-Russell (H-R) diagram, the members of a young, coeval
stellar population appear along the main sequence at higher masses and diverge
above the main sequence at lower masses, which is manifested in color-magnitude
diagrams as a band that becomes redder at fainter magnitudes.
For a nearby young cluster, that stellar sequence is brighter than most
foreground and background stars. As a result, color-magnitude diagrams can
be used to select a sample of candidate cluster members that has relatively
little contamination from field stars.

\begin{figure}
\epsscale{1.1}
\plotone{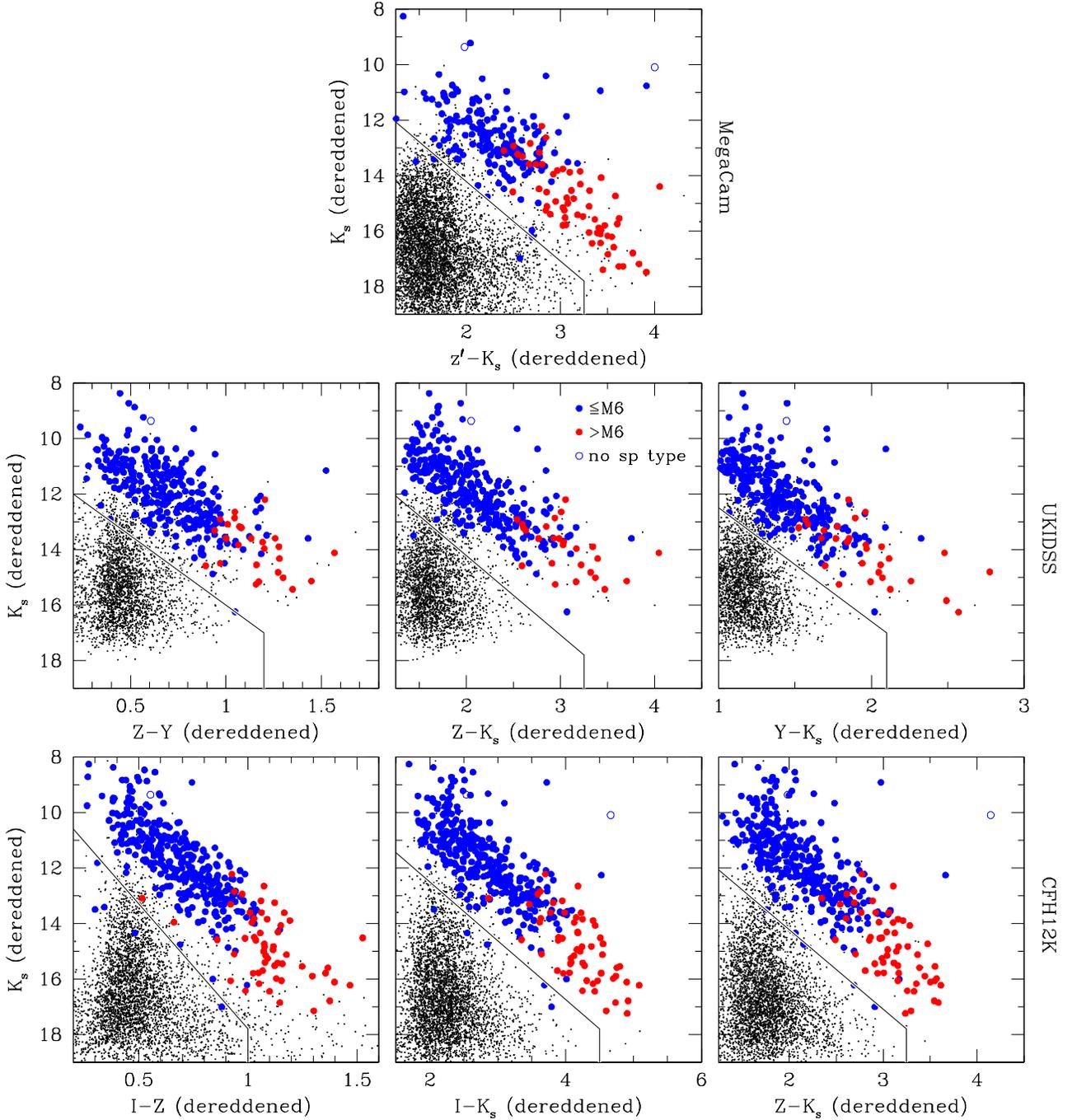}
\caption{
Extinction-corrected color-magnitude diagrams for the known members of
IC~348 (large filled and open circles) and other sources (small points)
within the area encompassed by the cluster map in Figure~\ref{fig:map}
(top and middle) and within the CFH12K field (bottom).
These data are from MegaCam, UKIDSS, CFH12K, 2MASS, WIRCam, and \citet{mue03}.
Candidate members
have been selected based on positions above the solid boundaries.
}
\label{fig:cmd348}
\end{figure}

To identify candidate members of IC~348, we have used the diagrams
of $I$ versus $R-I$ and $m_{791}$ versus $m_{791}-m_{850}$
from \citet{luh99} and \citet{luh05wfpc}, respectively.
We also have constructed the following extinction-corrected diagrams:
$K_s$ versus $I-Z$, $I-K_s$, and $Z-K_s$ based on the $I$ and $Z$ data from
\citet{luh03}, $K_s$ versus $Z-Y$, $Z-K_s$, and $Y-K_s$ based on the $Z$ and $Y$
data from UKIDSS, and $K_s$ versus $Z-K_s$ based on the $Z$ data that we have
measured from MegaCam images. The $K_s$ (or $K$) measurements are from the
Two Micron Point Source Catalog \citep[2MASS,][]{skr06}, \citet{mue03}, 
UKIDSS, and the WIRCam images described in Section~\ref{sec:fields}.
The extinctions of stars in these diagrams were estimated in the manner
described by \citet{luh03tau} using the extinction law from \citet{car89}.
In Figure~\ref{fig:cmd348}, we show the extinction-corrected diagrams
for the known members of IC~348 (including the new ones from this work)
and all other sources with the exception of those that have been
spectroscopically classified as field stars in this work and in previous
studies. In each diagram, we have marked a boundary along the lower envelope
of the locus of known members, which we have used for selecting candidate
members for spectroscopy.
The small number of known members that appear below
the cluster sequence in some of the color-magnitude diagrams are likely seen
in scattered light. They consist of LRL 435, LRL 725, LRL 904, LRL 1287, and
LRL 4011. \citet{luh03} previously noted that LRL 425 and LRL 725 were
unusually faint for their spectral types and colors.

\begin{figure}
\epsscale{1.1}
\plotone{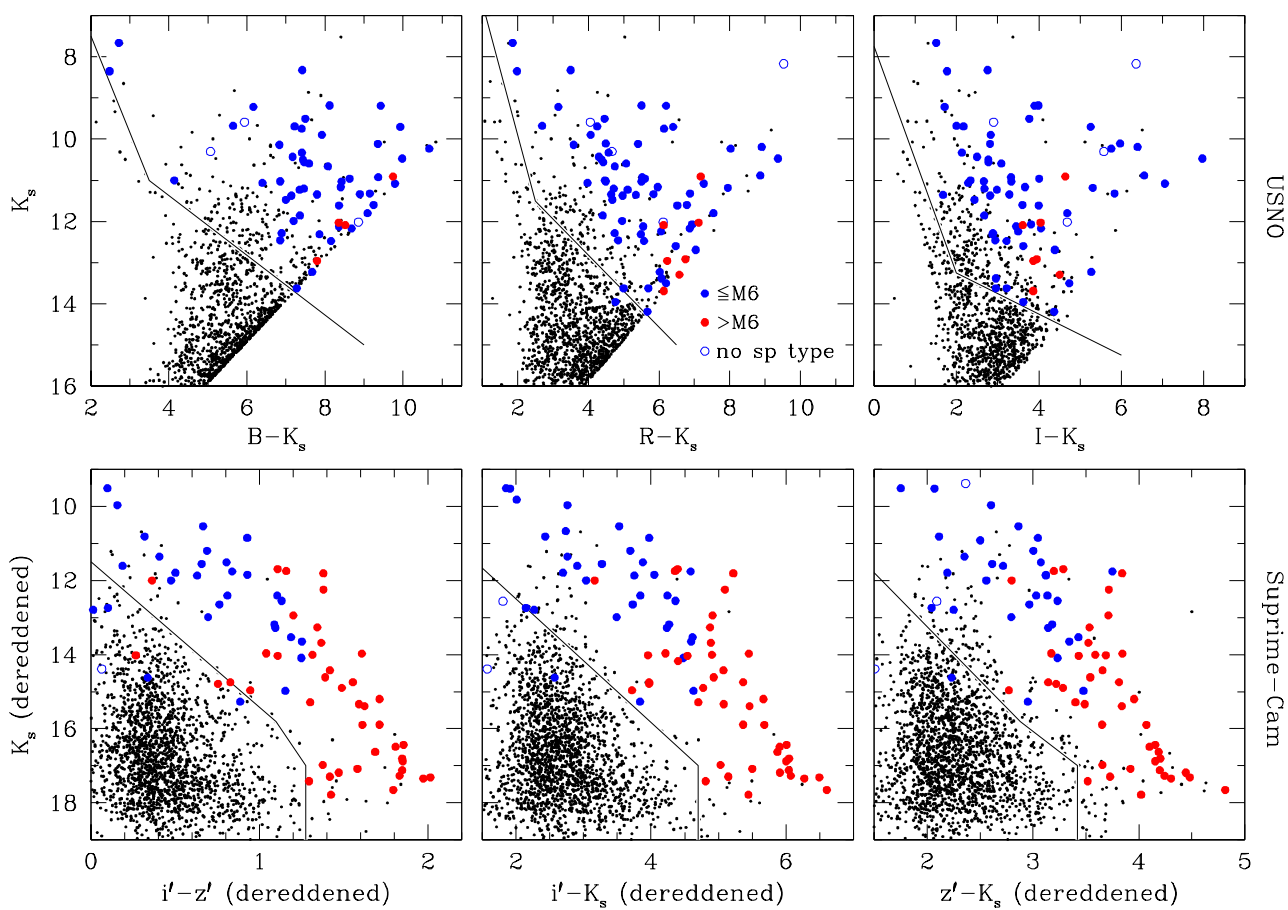}
\caption{
Color-magnitude diagrams for the known members of NGC~1333 (large filled and
open circles) and other sources (small points) within the area encompassed by
the cluster map in Figure~\ref{fig:map} (top) and within the Suprime-Cam field
(bottom). 
These data are from USNO-B1.0, Suprime-Cam, 2MASS, and WIRCam.
Candidate members have been selected based on positions above the solid
boundaries.
}
\label{fig:cmd1333}
\end{figure}

For NGC~1333, we have constructed diagrams of $K_s$ versus $B-K_s$, $R-K_s$, and
$I-K_s$ where $B$, $R$, and $I$ are photographic data from 
the USNO-B1.0 Catalog \citep{mon03} and extinction-corrected diagrams
of $K_s$ versus $i\arcmin-z\arcmin$, $i\arcmin-K_s$, and $z\arcmin-K_s$ based
on the $i\arcmin$ and $z\arcmin$ data from Suprime-Cam.
The $K_s$ data are from 2MASS, UKIDSS, and WIRCam.
These diagrams are shown in Figure~\ref{fig:cmd1333} for the known members
of NGC~1333 and all other detected sources, excluding known field stars.
As with IC~348, we have plotted a boundary that follows the lower envelope
of the cluster sequence in each diagram for use in selecting candidate members.
The known members that are below those boundaries consist of
2MASS J03291228+3123065 and sources 39, 66, 92, 102, 110, 113, and 122 from
\citet{gut08}. All of these objects exhibit mid-IR excess emission, so it is
plausible that they are seen in scattered light because of occulting disks.

For each cluster, a source is considered a candidate member
if it is above a boundary in any diagram and is not below a boundary
in any diagram. As an exception to the latter criterion, we do not
reject candidates identified based on mid-IR excesses, as mentioned
in the previous section.

\subsection{Proper Motions}
\label{sec:pm}

Nearby young clusters (150--300~pc) have proper motions of
10--20~mas~yr$^{-1}$, and the members of a given cluster exhibit dispersions
of $\sim1$~mas~yr$^{-1}$ ($\sim1$~km~s$^{-1}$). In comparison, foreground
and background stars typically have much larger and smaller motions,
respectively. 
As a result, precise measurements of proper motions can be used to identify
possible cluster members. To measure proper motions in IC~348 and NGC~1333, we
have made use of the multiple epochs of IRAC observations that are available
for the clusters, which were performed through programs 6 (G. Fazio), 36
(G. Fazio), 178 (N. Evans), 30516 (L. Looney), 50596 (G. Rieke), 60160
(J. Muzerolle), 61026 (J. Stauffer), 80174 (K. Flaherty), and 90071 (A. Kraus).
We measured proper motions for all sources in these images by applying the
astrometric techniques and distortion corrections from \cite{esp16}.
Because the different epochs of astrometry have been registered using
the stars that are in common among them (most of which are
background stars), our analysis has produced relative proper motions.
We have ignored proper motion measurements for sources with median values
of S/N in the final epoch of exposures at 4.5~\micron\ that are below
4 and 7.5 and that have errors larger than 6 and 8~mas~yr$^{-1}$
for IC~348 and NGC~1333, respectively. We were able to adopt a lower threshold
of S/N for IC~348 because more epochs of IRAC images are available for it,
which provides lower proper motion errors at a given S/N.
We also inspected the IRAC images of the members that exhibited discrepant
motions compared to the bulk of the population to identify and exclude
measurements that were erroneous due to extended emission or blending of stars.
Two previously identified members, LRL 624 in IC~348
and source 38 from \citet{sch12a} in NGC~1333, appear to have reliable
proper motion measurements that are inconsistent with membership, and hence
have been rejected from the sample of members for each cluster, as mentioned
in Section~\ref{sec:previous}.

\begin{figure}
\epsscale{1.1}
\plotone{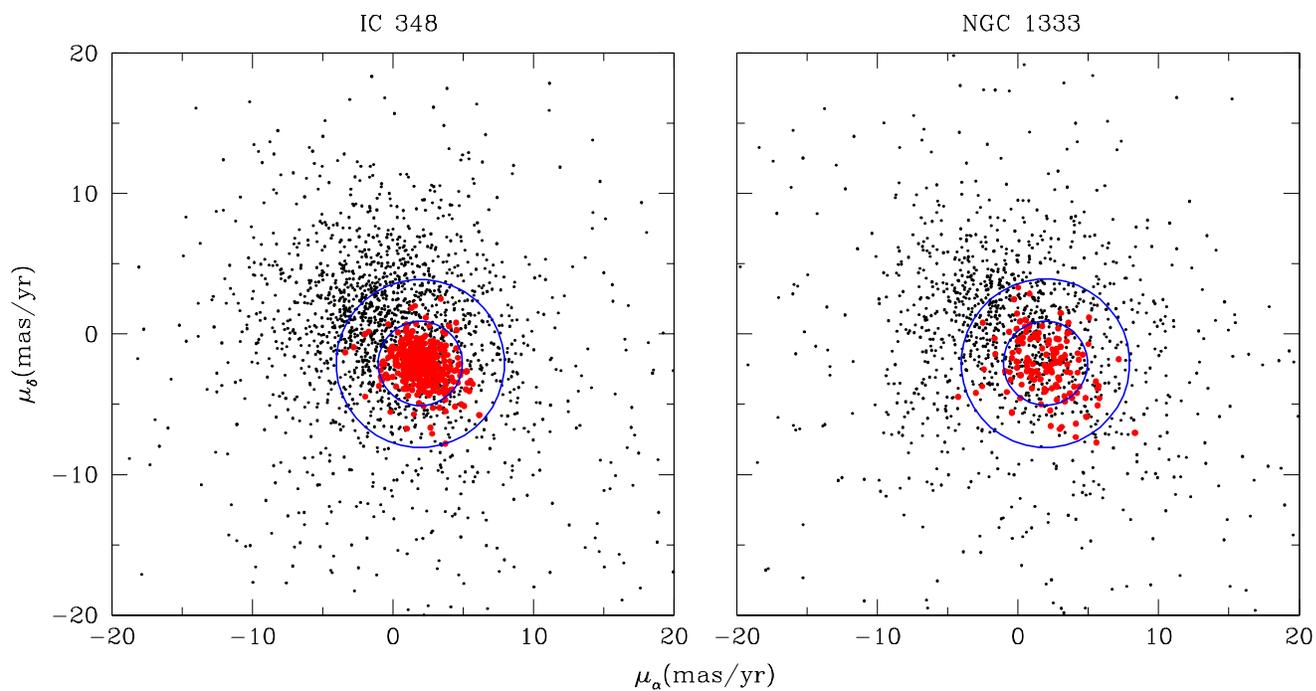}
\caption{
Relative proper motions for known members of IC~348 and NGC~1333
(large filled circles) and other sources detected in multi-epoch
IRAC images (small points). When identifying candidate members of each
cluster with photometry, we rejected objects with motions that differed
by more than 6~mas~yr$^{-1}$ from the cluster medians (beyond outer circles).
Photometric candidates that are within 3 and 3--6~mas~yr$^{-1}$ of the median
motions (inner circle and outer annulus) are labeled as ``pm" and ``pm?"
in Tables~\ref{tab:cand348} and \ref{tab:cand1333}.
}
\label{fig:pm}
\end{figure}

Among the 478 and 203 adopted members of IC~348 and NGC~1333, 405 and 141
sources (85 and 69\%) have useful proper motions, respectively.
A larger fraction of the stellar population in IC~348 has measured motions
because of its greater number of IRAC epochs. The proper motions
measured for known members are included in Tables~\ref{tab:mem348} and
\ref{tab:mem1333}. Those data have median values of
($\mu_{\alpha}, \mu_{\delta}=1.9, -2.1$~mas~yr$^{-1}$) for IC~348
and ($\mu_{\alpha}, \mu_{\delta}=2.3, -3.0$~mas~yr$^{-1}$) for NGC~1333.
The motions for the known members and all other sources with measured
motions are plotted in Figure~\ref{fig:pm}.
For each cluster, most of the members are within 3~mas~yr$^{-1}$ of the
median value, and nearly all members are within 6~mas~yr$^{-1}$ of it.
Therefore, we consider objects within 6~mas~yr$^{-1}$ of the median motions
to be candidates, and those within 3~mas~yr$^{-1}$ are the most promising ones.
Sources with motions that differ by $>6$~mas~yr$^{-1}$ are rejected
as non-members (although the previously known members that are only slightly
beyond that threshold are retained as members).
It is evident from Figure~\ref{fig:pm} that the known members overlap with
a large number of field stars in their proper motion measurements.
As a result, the sample of proper motion candidates has significant
contamination from field stars, and it would be inefficient to pursue
spectroscopy of candidates identified based on these measurements alone.
However, these proper motions are valuable for refining the samples of
candidates selected with the other methods that we have already described.
If a candidate found with other diagnostics appeared to be a non-member
based on its motion, we inspected its IRAC images prior to rejection
to check for blends and extended emission that might lead to an erroneous
proper motion measurement.
Many of our spectra of candidates were obtained before we measured the
IRAC proper motions. As a result, some of those objects in our spectroscopic
sample would have been rejected by proper motions if they had been available,
as indicated in Table~\ref{tab:non}.

\section{Spectroscopy}
\label{sec:spectra}

\subsection{Observations}
\label{sec:obs}

We have obtained spectra of 152 candidate members of IC~348 from the analysis
in the previous section and 16 previously known members.
We observed 130 sources with the near-IR spectrograph SpeX
\citep{ray03} at the NASA Infrared Telescope Facility (IRTF).
They consisted of 122 candidates (80 new members, 42 non-members) and eight
known members. Three of the latter objects (LRL~464, LRL~659, LRL~10111)
were candidates at the time of our spectroscopy and were independently 
identified as members by \citet{alv13}. One known member in
the SpeX sample (LRL~233) was placed in the slit during the observation
of a candidate at a separation of $3\arcsec$ (LRL~3171).
The spectrum of the candidate was not useful because of low S/N, but
we later successfully observed it with Gemini North. Another known member
observed with SpeX (LRL~62) has a discrepant radial velocity relative to other
cluster members \citep{cot15}, so we sought to verify the evidence of youth
previously found through optical spectroscopy.
The final three members observed with SpeX consist of the companions LRL~60~B
and LRL~187~B, which lacked previous spectral classifications, and LRL~60~A.
The SpeX data were collected in the prism mode with the $0\farcs8$ slit
(0.8--2.5~\micron, $R=150$).
We performed optical spectroscopy on 15 candidates (8 new members,
7 non-members) and six known members (LRL~141, LRL~174, LRL~294, LRL~334,
LRL~366, LRL~10094) with the Inamori Magellan Areal Camera and Spectrograph
(IMACS) on the Magellan I telescope at Las Campanas Observatory. Those data
were taken during the multi-slit observations described by \citet{mue07} for
their sample of IR-selected candidates. The IMACS spectra spanned from
6300--8900~\AA\ and exhibited a resolution of 3~\AA. One of the known members
observed with IMACS (LRL~10094) was originally selected as a candidate member
and was subsequently classified as a member by \citet{alv13}.
We used the near-IR camera on the Keck~I telescope \citep[NIRC,][]{mat94}
to obtain low-resolution ($\sim$100) spectra of one candidate (LRL~6005) and
two known members from \citet{luh05flam} that had uncertain spectral types
(LRL~1050, LRL~2103). They were observed with the gr120 and gr150 grisms, which
together provided coverage from 1--2.5~\micron.
We observed two and 12 candidates with the
Gemini Near-Infrared Imager \citep[NIRI,][]{hod03} 
and the Gemini Near-Infrared Spectrograph \citep[GNIRS,][]{eli06},
respectively, at the Gemini North telescope. NIRI was operated with
the $H$-band grism and the $0\farcs75$ slit ($R=500$). 
The GNIRS data were collected in the cross-dispersed mode with the
31.7~l~mm$^{-1}$ grating and the $0\farcs67$ slit (1--2.5~\micron, $R=800$).
The 152 candidates observed by these spectrographs consist of 100 new members
and 52 non-members based on the classifications from the next section.

In NGC~1333, we have performed spectroscopy on 55 candidate members and 124
known members with SpeX and 14 candidates with GNIRS. The instrument
configurations were the same as those employed for the targets in IC~348
except for BD+$30\arcdeg$547, for which we used the SXD mode of SpeX
with the $0\farcs8$ slit (0.7--2.5~\micron, $R=750$) because its earlier
type required higher resolution for classification.
We included a large number of known members in our spectroscopic sample
because we wish to maximize the number of members that are classified in
the same manner as the members of IC~348.
In the next section, we find that 42 and 26 candidates are members
and non-members, respectively, and that one of the candidates
(J03284883+3117537) has an uncertain classification because of low S/N.

The SpeX data were reduced with the Spextool package \citep{cus04} and
corrected for telluric absorption in the manner described by \citet{vac03}.
For the IMACS and NIRC spectra, we used routines within IRAF
to apply bias subtraction and flat fielding to the two-dimensional images,
extract spectra of the targets from those images, and perform
wavelength calibration on the extracted data with spectra of arc lamps.
Each NIRC spectrum was also corrected for telluric absorption using the
spectrum of an A star that was observed at a similar airmass.

\begin{figure}
\epsscale{1}
\plotone{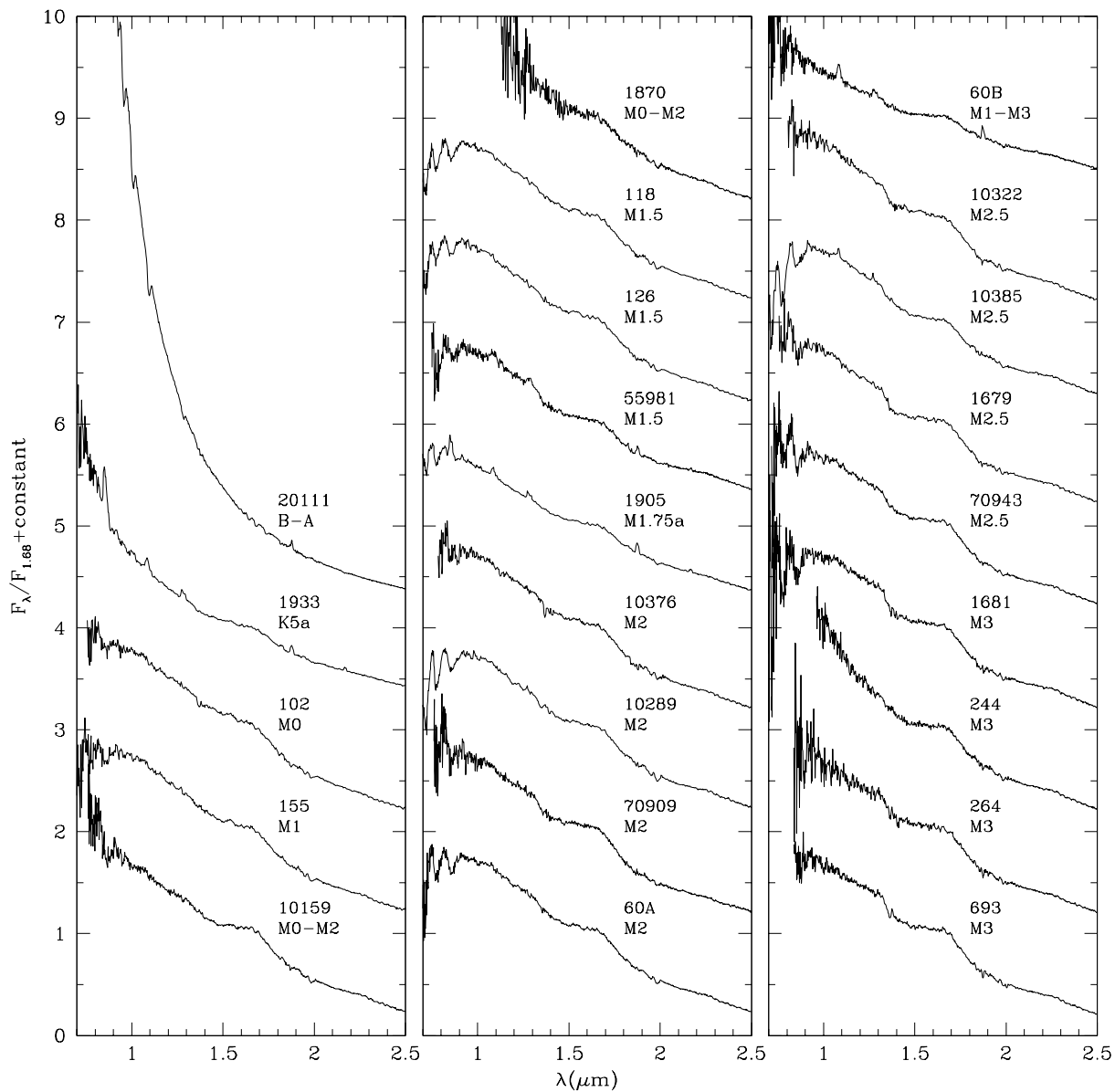}
\caption{
Near-IR spectra of members of IC~348 from this work and \citet{mue07}.
The spectral types denoted with ``a" have been adopted from optical spectra
because accurate types could not be measured from these IR data (LRL~1933,
LRL~1905) or the objects serve as standards for classifying our IR spectra
(LRL~147, LRL~201, LRL~405). The remaining types have been measured from these
spectra. The spectra have been dereddened to match the slopes of 
standards near 1~\micron. These data have a resolution of $R=150$.
The data behind this figure have been provided as FITS files.
}
\label{fig:sp348a}
\end{figure}

\begin{figure}
\epsscale{1}
\plotone{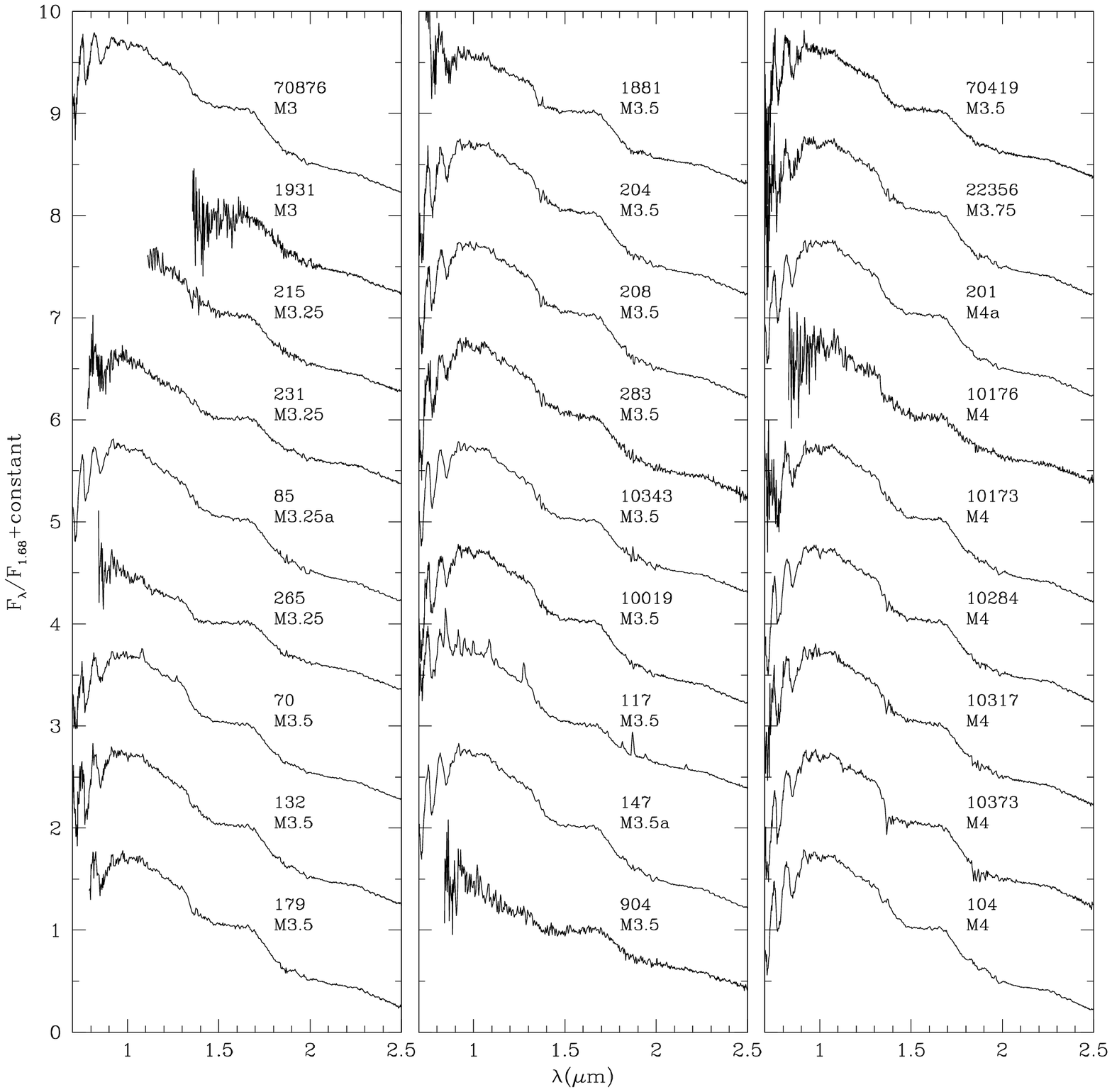}
\caption{
More near-IR spectra of members of IC~348 (see Figure~\ref{fig:sp348a}).
The data behind this figure have been provided as FITS files.
}
\label{fig:sp348b}
\end{figure}

\begin{figure}
\epsscale{1}
\plotone{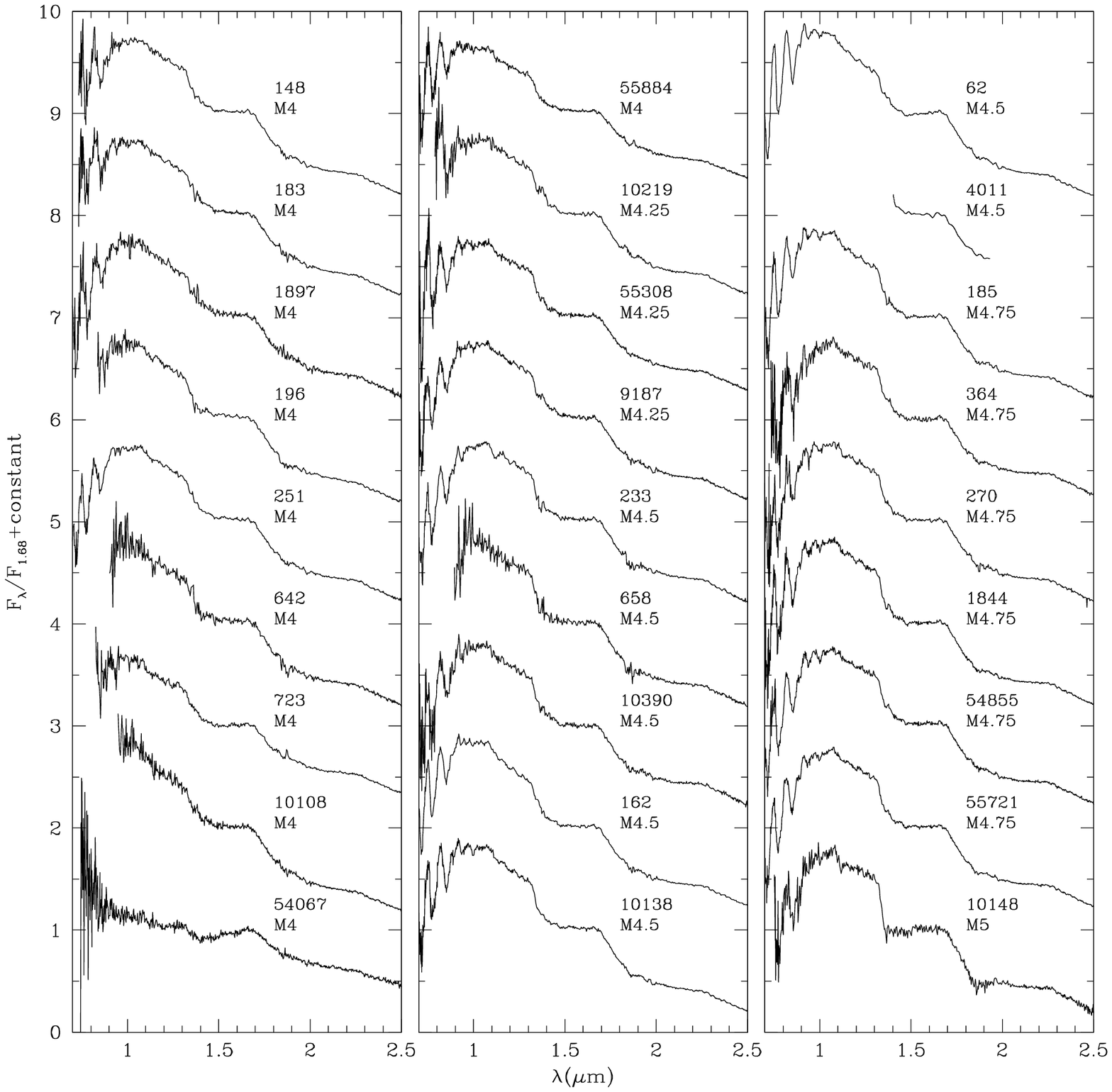}
\caption{
More near-IR spectra of members of IC~348 (see Figure~\ref{fig:sp348a}).
The data behind this figure have been provided as FITS files.
}
\label{fig:sp348c}
\end{figure}

\begin{figure}
\epsscale{1}
\plotone{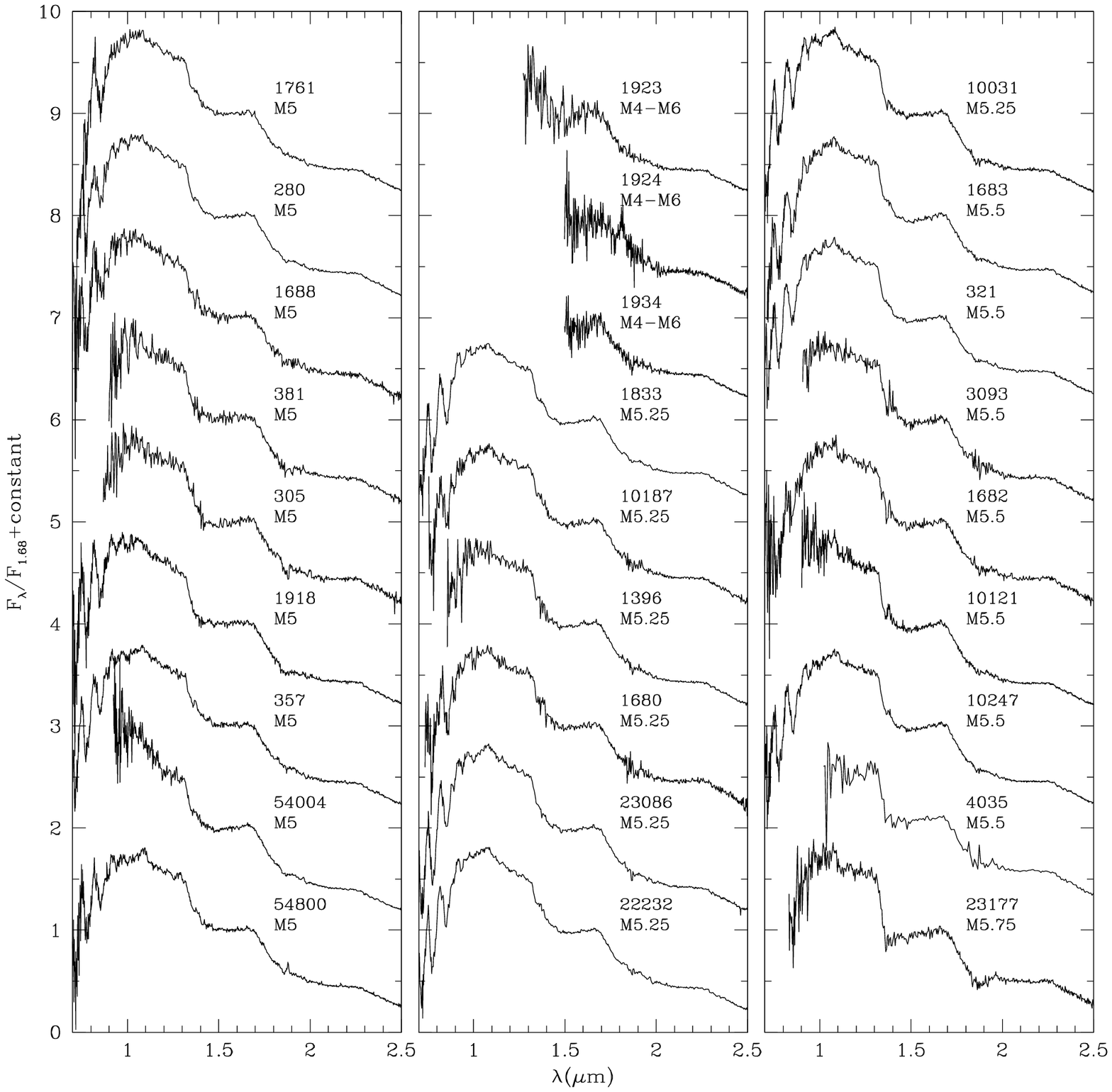}
\caption{
More near-IR spectra of members of IC~348 (see Figure~\ref{fig:sp348a}).
The data behind this figure have been provided as FITS files.
}
\label{fig:sp348d}
\end{figure}

\begin{figure}
\epsscale{1}
\plotone{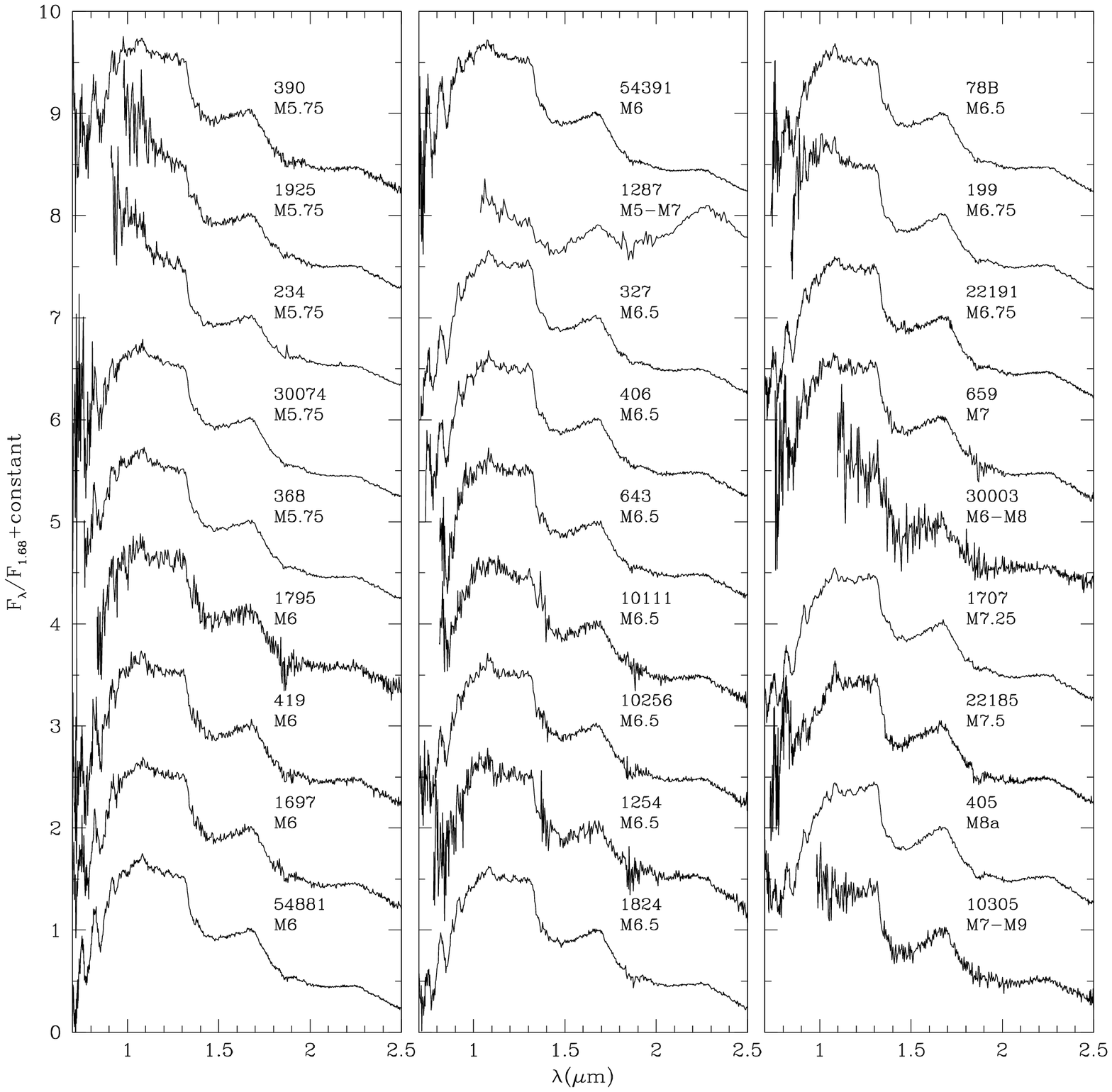}
\caption{
More near-IR spectra of members of IC~348 (see Figure~\ref{fig:sp348a}).
The data behind this figure have been provided as FITS files.
}
\label{fig:sp348e}
\end{figure}

\begin{figure}
\epsscale{1}
\plotone{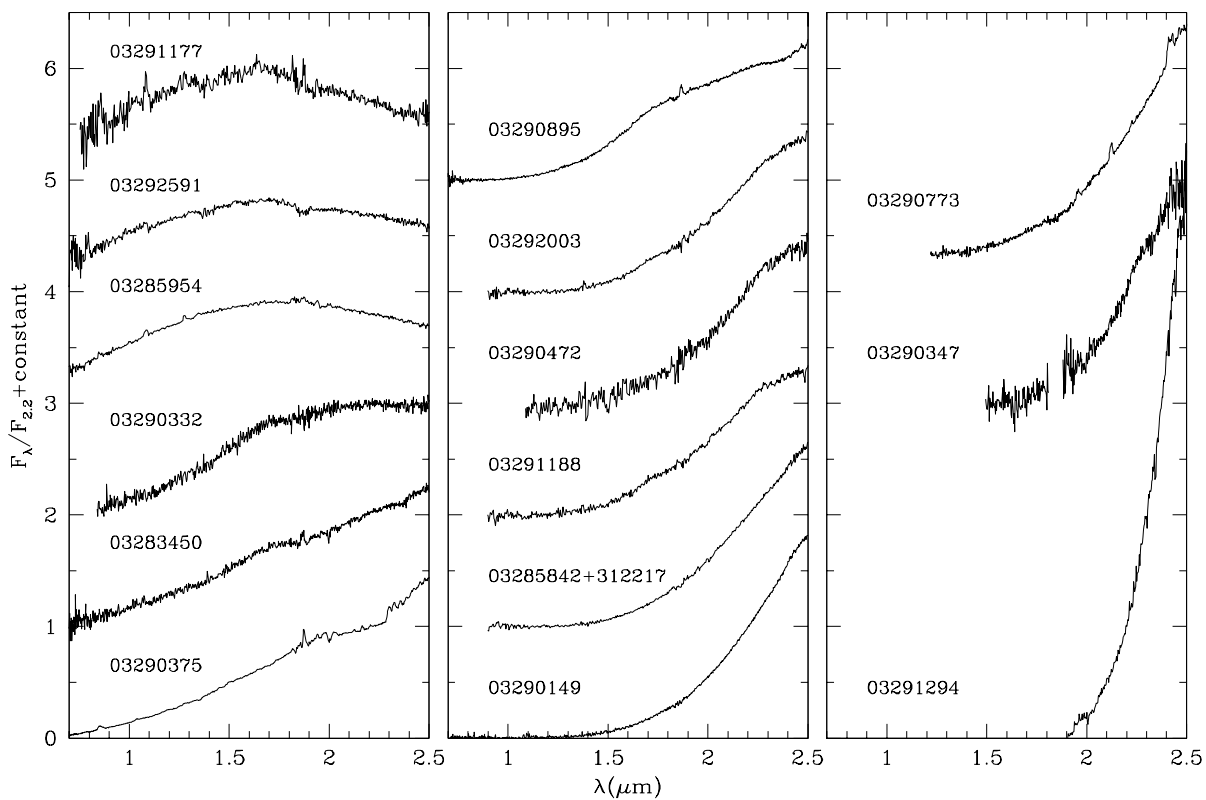}
\caption{
Near-IR spectra of members of NGC~1333 that lack measured spectral types
because photospheric absorption features were not detected (i.e., protostars).
These data have a resolution of $R=150$. Each spectrum is labeled with
the portion of the source name from Table~\ref{tab:mem1333} that
corresponds to right ascension. For one object, additional digits
of declination are listed to uniquely identify it among the members.
The data behind this figure have been provided as FITS files.
}
\label{fig:sp1333a}
\end{figure}

\begin{figure}
\epsscale{1}
\plotone{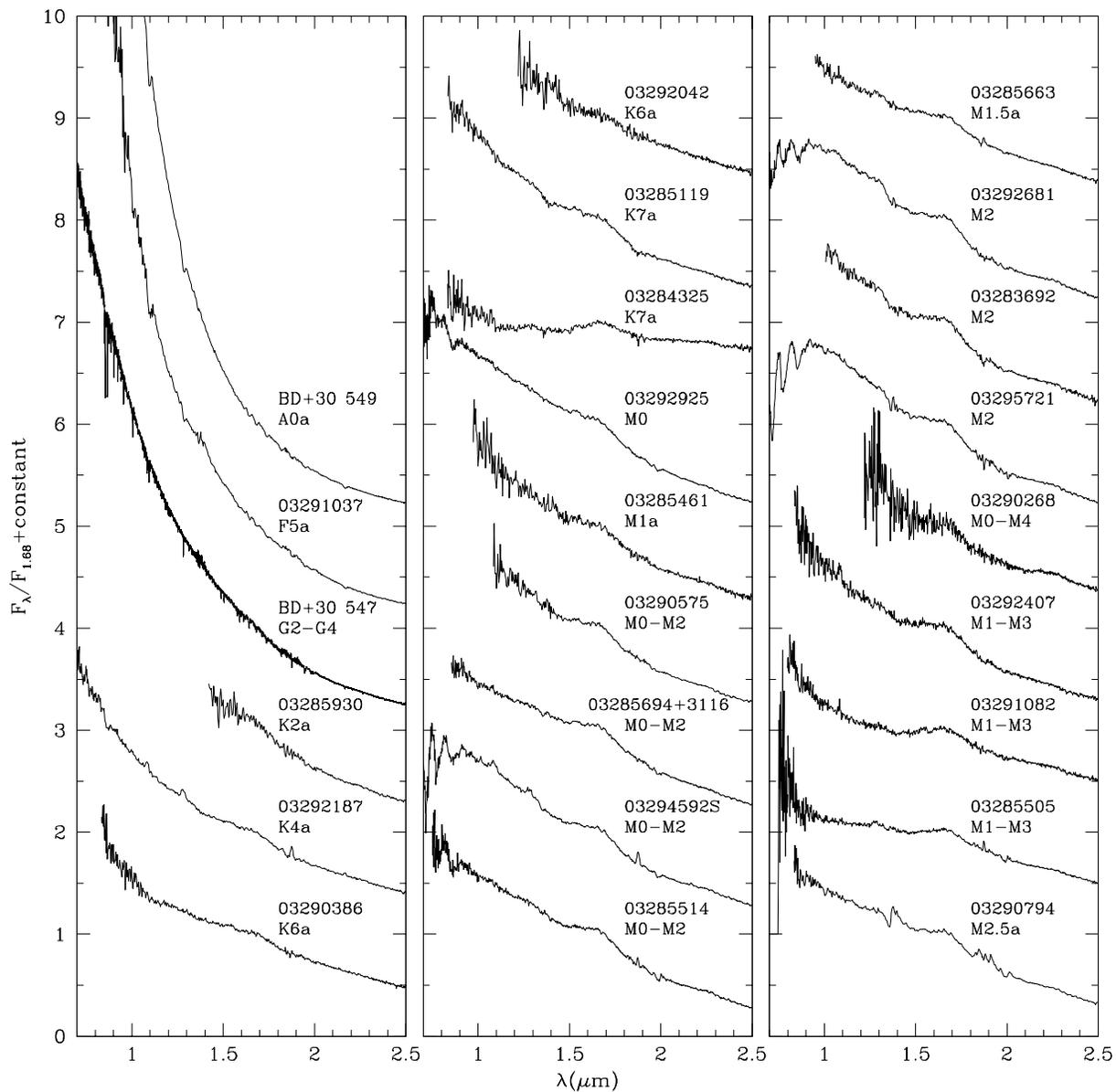}
\caption{
Near-IR spectra of members of NGC~1333.
It was not possible to measure accurate spectral types from some of these
data. Those spectra are labeled with the types that we have adopted from
previous studies (denoted by a suffix of ``a").
The remaining types have been measured from these spectra.
The spectra have been dereddened to match the slopes of standards
near 1~\micron. The spectrum of BD+$30\arcdeg$547 has a
resolution of $R=750$ while the other data have $R=150$.
Except for BD+$30\arcdeg$547, each spectrum is labeled with
the portion of the source name from Table~\ref{tab:mem1333} that
corresponds to right ascension. For one object, additional digits
of declination are listed to uniquely identify it among the members.
The data behind this figure have been provided as FITS files.
}
\label{fig:sp1333b}
\end{figure}

\begin{figure}
\epsscale{1}
\plotone{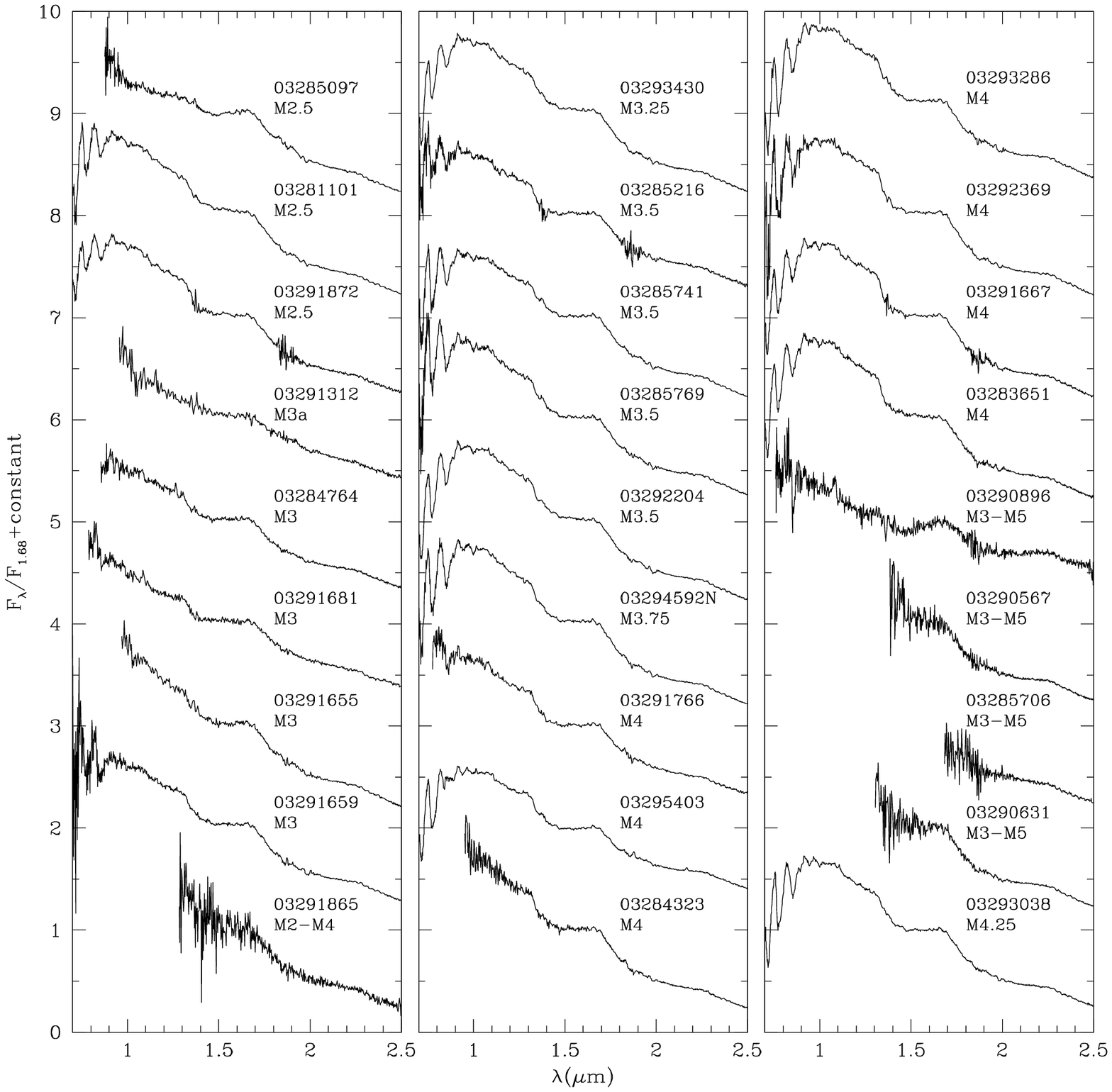}
\caption{
More near-IR spectra of members of NGC~1333 (see Figure~\ref{fig:sp1333b}).
The data behind this figure have been provided as FITS files.
}
\label{fig:sp1333c}
\end{figure}

\begin{figure}
\epsscale{1}
\plotone{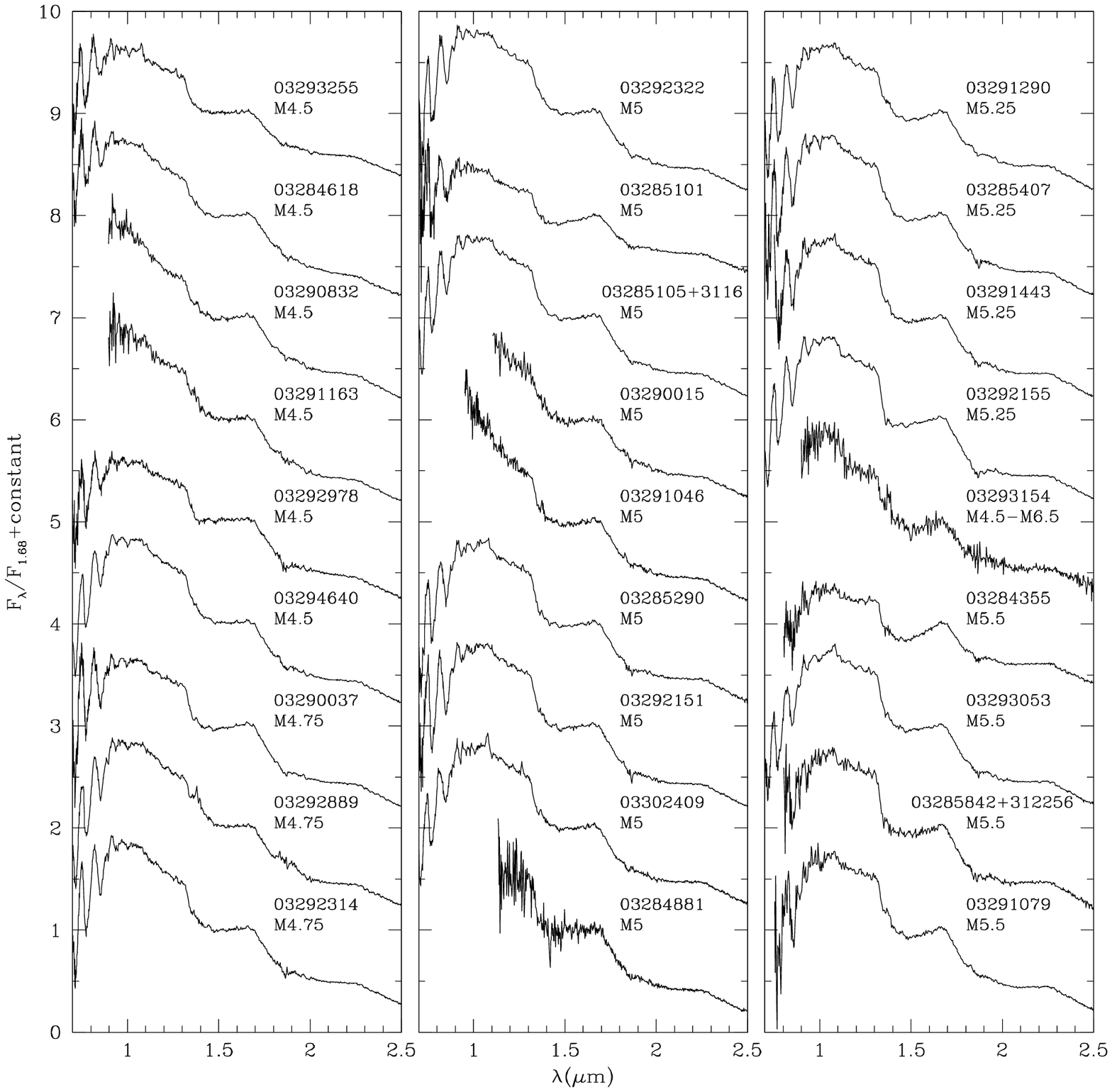}
\caption{
More near-IR spectra of members of NGC~1333 (see Figure~\ref{fig:sp1333b}).
The data behind this figure have been provided as FITS files.
}
\label{fig:sp1333d}
\end{figure}

\begin{figure}
\epsscale{1}
\plotone{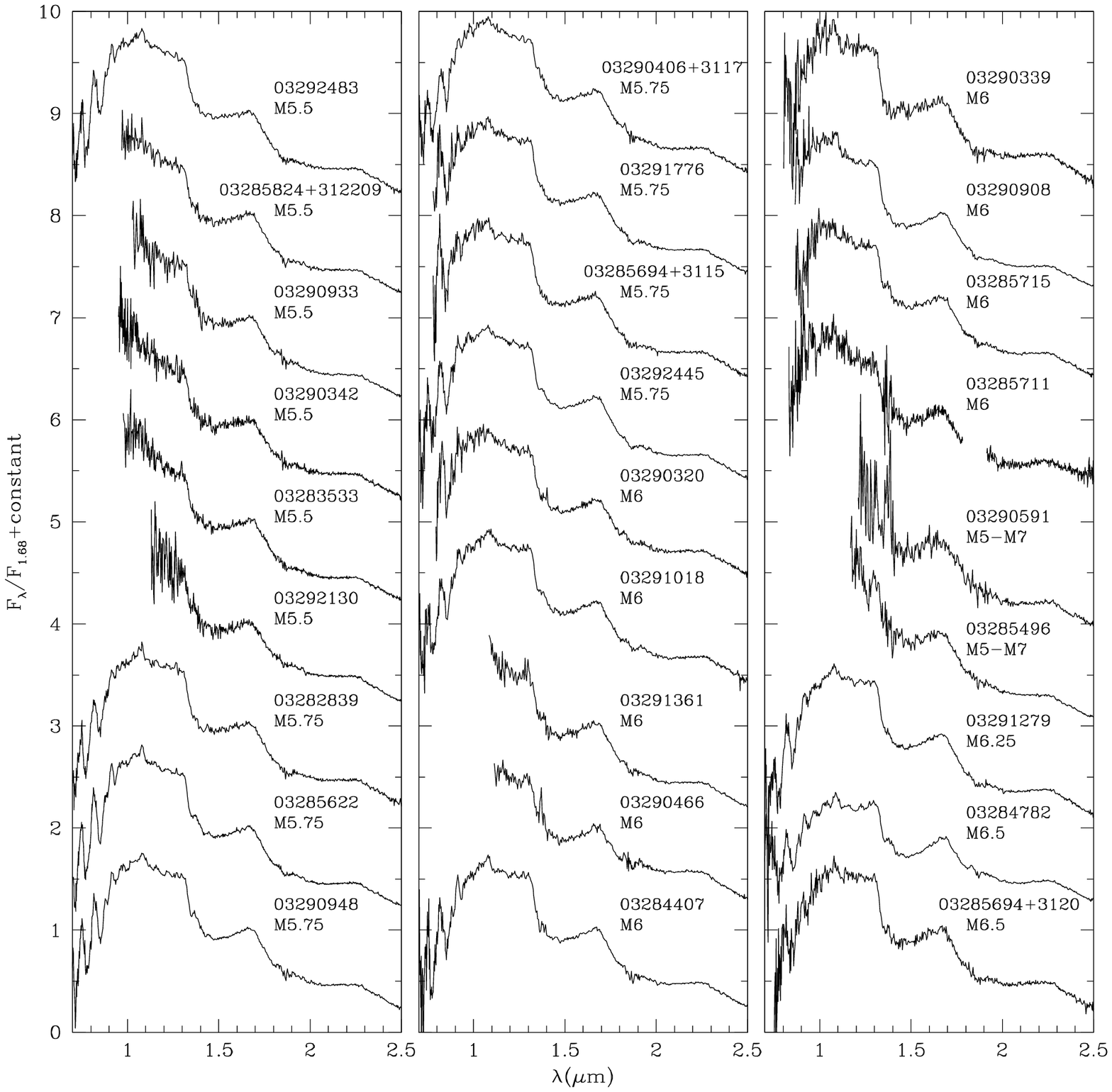}
\caption{
More near-IR spectra of members of NGC~1333 (see Figure~\ref{fig:sp1333b}).
The data behind this figure have been provided as FITS files.
}
\label{fig:sp1333e}
\end{figure}

\begin{figure}
\epsscale{1}
\plotone{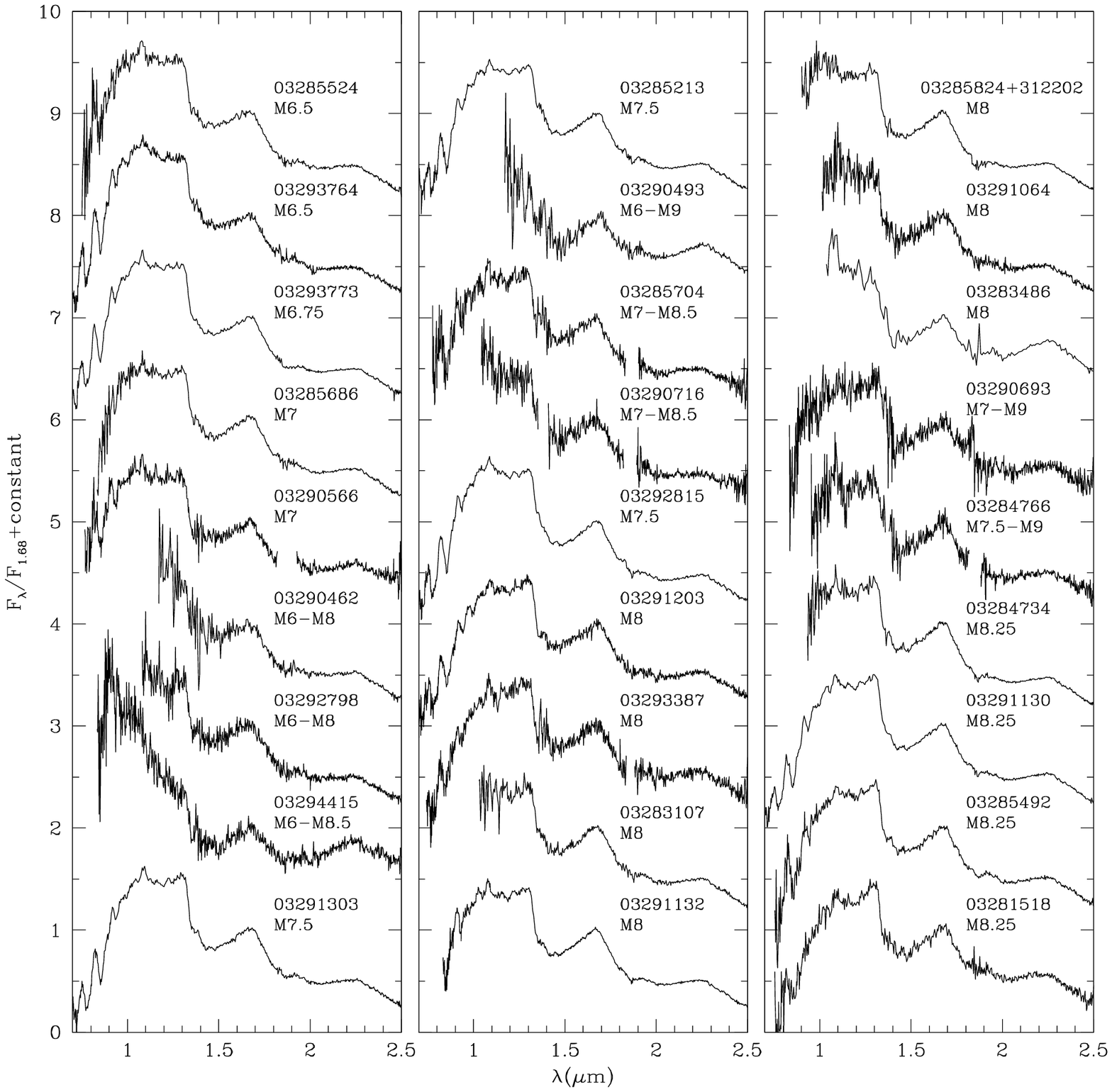}
\caption{
More near-IR spectra of members of NGC~1333 (see Figure~\ref{fig:sp1333b}).
The data behind this figure have been provided as FITS files.
}
\label{fig:sp1333f}
\end{figure}

\begin{figure}
\epsscale{1}
\plotone{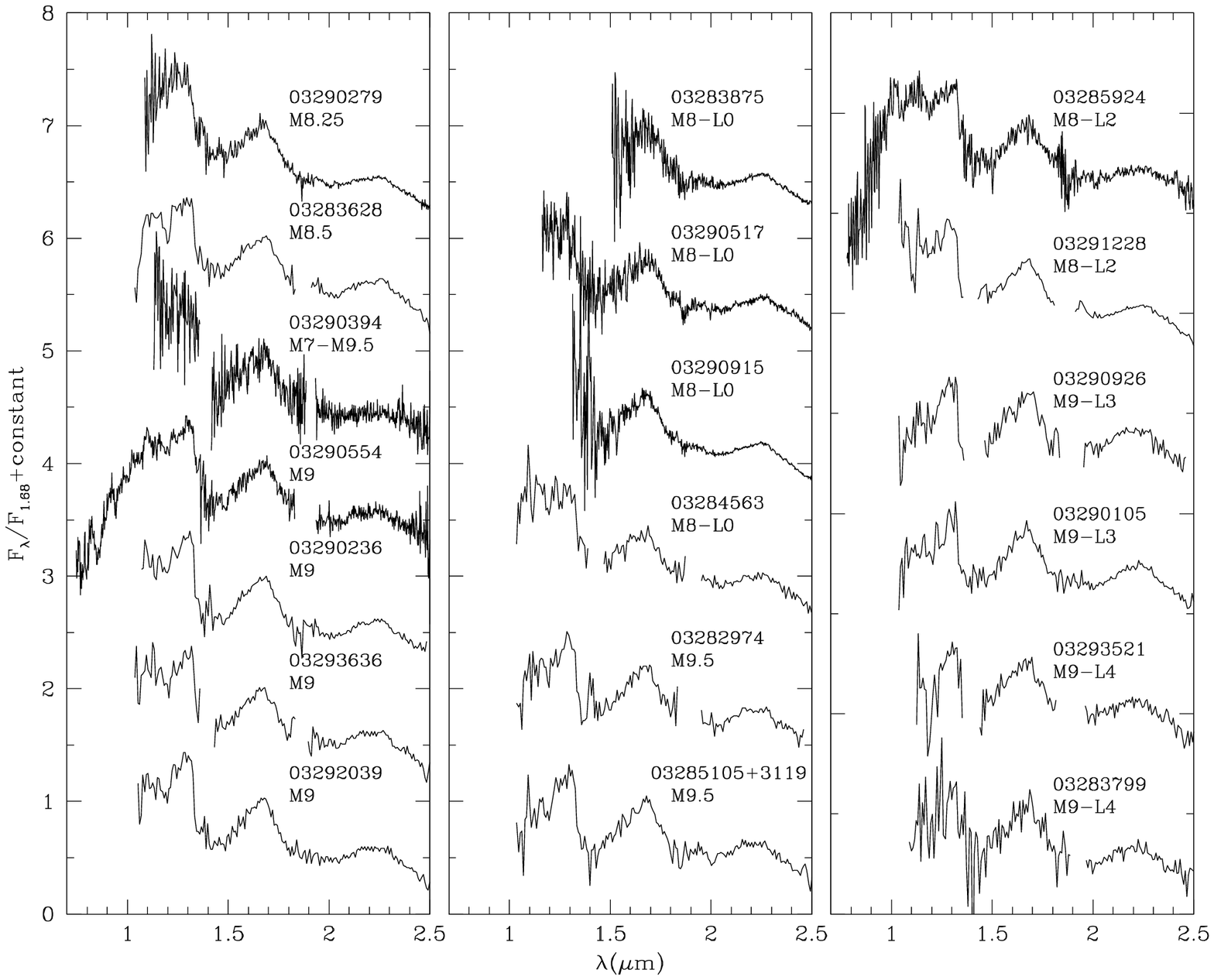}
\caption{
More near-IR spectra of members of NGC~1333 (see Figure~\ref{fig:sp1333b}).
The data behind this figure have been provided as FITS files.
}
\label{fig:sp1333g}
\end{figure}

The dereddened near-IR spectra of the new members of IC~348 and
NGC~1333 are presented in Figures~\ref{fig:sp348a}--\ref{fig:sp348f} and
Figures~\ref{fig:sp1333a}--\ref{fig:sp1333g}, respectively.
We also include all of our SpeX data for previously known members from this
work and previous studies \citep{luh05wfpc,mue07} with the exception of the
four featureless spectra in IC~348 that were presented by \citet{mue07}.
Within the lists of all known members of the clusters in
Tables~\ref{tab:mem348} and \ref{tab:mem1333}, we indicate the dates
and instruments for our new spectra and the previous SpeX data.
The candidates from our spectroscopic sample that we have classified as
non-members are found in Table~\ref{tab:non}.

\subsection{Spectral Classification}
\label{sec:class}

\begin{figure}
\epsscale{1}
\plotone{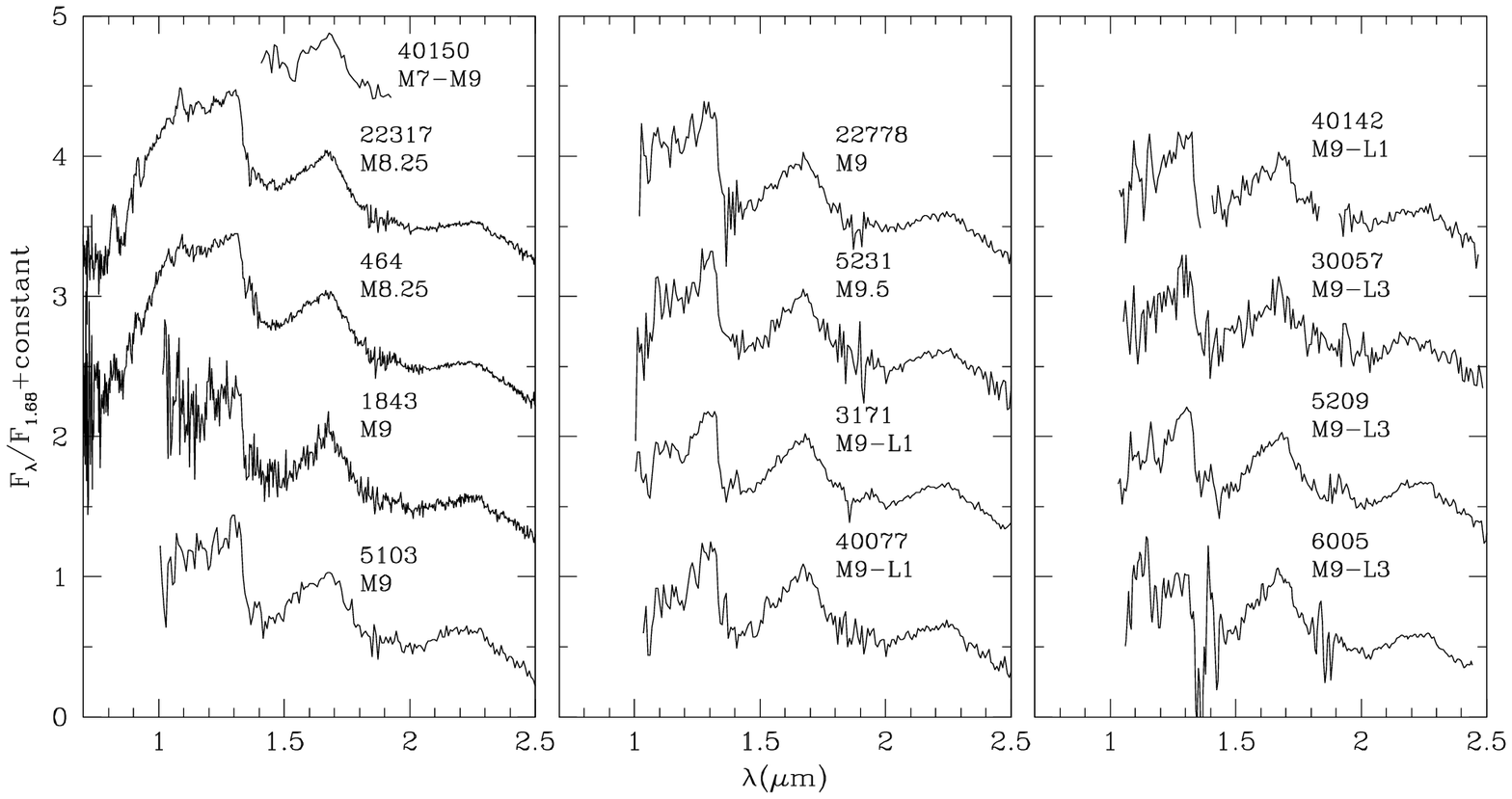}
\caption{
More near-IR spectra of members of IC~348 (see Figure~\ref{fig:sp348a}).
The data behind this figure have been provided as FITS files.
}
\label{fig:sp348f}
\end{figure}

We have used the spectra that we have collected to measure spectral types of
our targets and to help determine whether they are members of IC~348 and
NGC~1333. Seven of our targets of optical spectroscopy lack the M-type
spectral features (TiO, VO)
expected if they were members of IC~348, indicating that they are likely
early-type field stars or giants that are behind the cluster.
The remaining 14 objects with optical spectra do exhibit M-type
features, all of which are young based on their Na~I and K~I
absorption lines, which are sensitive to surface gravity.
We measured spectral types from these data through comparison to the
averages of dwarf and giant standards \citep{luh99}.
In previous studies \citep[e.g.,][]{mue07}, we have measured spectral
types from near-IR spectra of young late-type objects via comparison
to spectra of individual young late-type objects that we had classified at
optical wavelengths. Because of the relatively large number of young
objects that now have both optical types and near-IR spectra, we have
recently combined the spectra of several such objects for each subclass
(K. Luhman, in preparation).  We have used the resulting spectra as
standards for classifying the near-IR spectra in IC~348 and NGC~1333
that exhibit steam absorption ($\gtrsim$M0) and
evidence of youth \citep[e.g., triangular $H$-band continuum][]{luc01}.
For the remaining IR data, spectral types
were measured with spectra of standard dwarfs and giants from our previous
studies and from \citet{cus05} and \citet{ray09}.
In addition, we have revised our previous classifications of SpeX data for
members of IC~348 \citep{luh05wfpc,mue07} using the new standard spectra.
The resulting changes are $\leq0.25$~subclass for most objects.
We also measured spectral types from the near-IR spectra of candidate
members of IC~348 presented by \citet{alv13}.
For one of those objects, LRL~5209, the S/N appeared to be lower than expected
for its magnitude, so we performed our own reduction of the raw data.
We have included the new version of the spectrum with the data
that we have collected in Figure~\ref{fig:sp348f}. Most of our classifications
are similar to those from \citet{alv13}. For those cases, we adopt the
types from \citet{alv13}.
We do present our new spectral types for LRL 670, LRL 5209, and LRL 22443
since they differ noticeably from the previous measurements.
In addition, whereas \citet{alv13} classified the membership status of LRL 670,
LRL~5209, LRL 10378, LRL 22443, and LRL 54229 as uncertain, we find that the
spectra of these objects from \citet{alv13} do show sufficient evidence of
membership in the form of the gravity sensitive features. As a result, we
have included those objects in our census of members, as mentioned in
Section~\ref{sec:previous}.

For the members of IC~348 and NGC~1333 that have measured spectral types 
and that have SpeX data from this work and our previous studies, we
have estimated extinctions by comparing the observed spectral slopes at
1~\micron\ to the slopes our young standards and adopting
the extinction law of \citet{car89}.
If a spectral type could not be measured from a SpeX spectrum, but a
classification was available from another source (e.g., optical spectrum),
then we adopted that classification when estimating the extinction.
For a spectrum with low S/N at 1~\micron, we estimated the extinction from the
slope at longer wavelengths if a $K$-band excess was unlikely to be present
based on an absence of a mid-IR excess.
The resulting extinctions were used for dereddening the spectra in
Figures~\ref{fig:sp348a}--\ref{fig:sp1333g}.
Among those dereddened spectra, most of the objects with high extinctions
can be identified by their lower S/N at shorter wavelengths.

The spectral types that we have measured for members of IC~348 and NGC~1333
are included within the lists of all known members in Tables~\ref{tab:mem348}
and \ref{tab:mem1333}.
The errors for the optical and IR types are $\pm0.25$ and
0.5~subclass, respectively, unless indicated otherwise. The uncertainties
in the IR types tend to be large at $\gtrsim$M9 because of a degeneracy between
spectral type and reddening in near-IR spectra with low resolution and 
low-to-moderate S/N. For instance, a young M9 object with $A_V\sim3.5$ can
appear quite similar to an unreddened L3 object in data of this kind.
Our classifications for non-members are presented in Table~\ref{tab:non}.
Most of these non-members are giants and early-type stars, which are
difficult to distinguish from K and early-M cluster members in color-magnitude
diagrams. Meanwhile, cooler members have more distinctive colors, so
the yield of confirmed members can be higher at fainter magnitudes when
the appropriate bands of photometry are available. For instance, 
13 of the 14 candidate low-mass members of NGC~1333 observed with GNIRS
were confirmed as such.
Finally, we note that several of the candidate members of IC~348 that we
selected from our color-magnitude diagrams and confirmed with spectroscopy
were previously identified as photometric candidates by \citet{alv13},
consisting of LRL 6005, LRL 5231, LRL 1254, LRL 1824, and LRL 22778.

\subsection{Comparison to Previous Work}
\label{sec:compare}

\begin{figure}
\epsscale{1}
\plotone{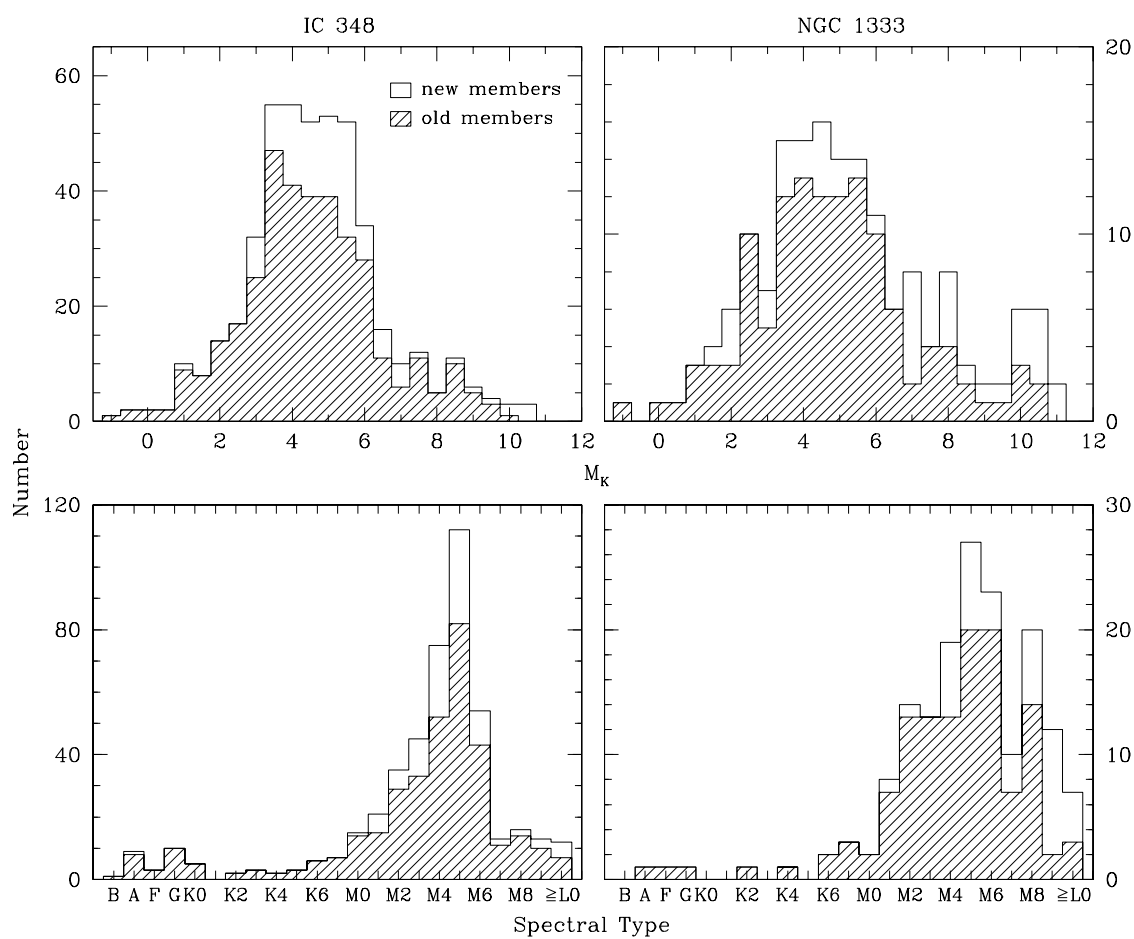}
\caption{
Distributions of spectral types and extinction-corrected $M_K$
for previously known members of IC~348 and NGC~1333 (shaded histograms) and
new members from this work (open histograms). Members that lack measured
spectral types are absent, which consist of protostars with featureless spectra.
}
\label{fig:histonew}
\end{figure}

We have compared our membership lists and spectral types for IC~348
and NGC~1333 to those from previous studies.
We have presented spectra for 100 and 42 members of these clusters,
respectively, that have not been previously classified, which we refer
to as new members. Two additional stars, LRL~60~B and LRL~187~B, also
lack previous classifications, but they are treated as previously
known members, as mentioned in Section~\ref{sec:previous}.
To illustrate the magnitudes and spectral types at which the census of
these clusters has expanded, we show in Figure~\ref{fig:histonew} the
distributions of extinction-corrected $M_K$ and spectral types for
previously known and new members. We have adopted the extinctions
estimated from our IR spectra in Section~\ref{sec:class} when available.
Otherwise, extinctions are estimated with photometry in the
manner described in Section~\ref{sec:hr}.
Members that lack measured spectral types
(and hence extinctions) are absent from Figure~\ref{fig:histonew}, which
consist of protostars with featureless spectra.
The new members of IC~348 are predominantly low-mass stars at M4--M6 in
the outskirts of the cluster, but they also include several objects that are
the faintest known members. In NGC~1333, the new members are
distributed more uniformly with magnitude and spectral type among the
low-mass stars and brown dwarfs.

\citet{reb15b} compiled a sample of 130 members of NGC~1333, 17 of which
are absent from our census of the cluster for the following reasons.
Seven of these missing objects are among the candidate members that we
have identified in Section~\ref{sec:select}, but they lack spectroscopic
confirmation of their membership. They consist of 2MASS J03291532+3129346
and NGC 1333 IRS J03284883+3117537, J03285358+3112147, J03285508+3114163,
J03291565+311911, J03285709+3121250, and J03291317+3119495.
Sources 12, 45, 113, and 174 from \citet{win10} have been detected in X-rays,
but are rejected as field stars by our color-magnitude diagrams and have
no other evidence of membership. Using the identifiers from \citet{gut08},
sources 38 and 76 are field stars based on our spectroscopy,
sources 8 and 38 are an outflow lobe and a galaxy, respectively
\citep{arn12}, source 95 has uncertain membership (Section~\ref{sec:comments}),
and source 26 is extended at both near- and mid-IR wavelengths, so it is
unclear whether it a star.

\citet{reb15b} identified five new candidate members of NGC~1333 based on
their mid-IR variability.
One of these candidates, SSTYSV J032903.46+311617.9, is in our spectroscopic
sample. Based on its featureless near-IR spectrum and red color in
$[3.6]-[4.5]$ (it is blended with a brighter star at longer wavelengths),
it is probably a protostar, so we have included it in our census of members.
\citet{reb15b} discussed at length a second of their candidates,
SSTYSV J032911.86+312155.7. They classified it as a possible protostar based
on its flux at 24~\micron\ relative to shorter wavelengths. However, we find
that no detection is apparent in the 24~\micron\ images.
Bright extended emission associated with the nearby star SVS~3 prevents
detections in the {\it Spitzer} bands longward of 4.5~\micron.
Nevertheless, given its close proximity to other members like SVS~3 and
the excess at 4.5~\micron\ relative to bands at shorter wavelengths,
it is a promising candidate member.
We include it in our list of candidates that have not been observed with
spectroscopy in Section~\ref{sec:completeness}.
Two other objects from \citet{reb15b}, SSTYSV J032918.65+312021.8 and
SSTYSV J032907.24+312409.7, are also in our sample of candidates based
on their variability, mid-IR excess emission, and their positions
in our color-magnitude diagrams.
The final candidate from \citet{reb15b}, SSTYSV J032836.43+312856.7,
does not exhibit mid-IR excess emission and is rejected by our color-magnitude
diagrams, so we do not consider it to be a candidate member.

As mentioned in Section~\ref{sec:obs}, we have performed spectroscopy on
a large number of the previously known members of NGC~1333 with the goal
of obtaining spectral classifications that are derived with the same
scheme applied to IC~348 in our previous studies and this work.
Spectral types are available from both our work and previous
studies for 90 objects. For a majority of the previous classifications,
there is not a systematic difference from our measurements. 
However, we do find that our spectral types are an average
of $\sim1$~subclass earlier than those from \citet{sch09,sch12a,sch12b},
which corresponds to $\sim50$\% higher
mass estimates when combined with evolutionary models \citep[e.g.,][]{bar98}.
There remain several (very faint) members from \citet{sch09,sch12a,sch12b}
for which we have not obtained spectra and measured spectral types.

\subsection{Comments on Individual Sources}
\label{sec:comments}

{\it LRL~62 and LRL~155}. The optical spectrum of LRL~62 from \citet{luh99}
exhibits a spectral type of M4.5 and evidence of youth in the form of weak K~I
and Na~I lines. However, its radial velocity of $\sim$5~km~s$^{-1}$ differs
significantly from the mean value of 15.4~km~s$^{-1}$
($\sigma=0.7$~km~s$^{-1}$) for a sample of known members \citep{cot15}.
We obtained a near-IR spectrum to verify its youth through additional
gravity-sensitive features. Our new classification based on that spectrum
is consistent with the optical result. Like LRL~62, 
LRL~155 also has a discrepant radial velocity ($\sim$23~km~s$^{-1}$).
Its near-IR spectrum indicates a spectral type of M1, but it is not sensitive
to signatures of low surface gravity for this type.
Both stars have higher extinctions ($A_J=0.7$ and 1) than
expected for foreground dwarfs ($A_J<0.1$) and are too bright for background
dwarfs. Therefore, we treat them as members of IC~348.
The stars may have been ejected from the cluster through dynamical interactions
with other members \citep{kro98,wei11}.

{\it IC 348 IRS J03442484+3213482}.
It was identified as a candidate protostar by \citet{eva09} and \citet{you15}
(source 405 in those studies). Its near-IR spectrum from SpeX is featureless.
Protostars can have featureless spectra if significant veiling is present.
However, such spectra are normally very red (see Fig.~\ref{fig:sp1333a}),
whereas the spectrum of this object is much bluer ($H-K_s\sim0.5$).
In addition, it is far from the area in the southern part
of the cluster where most of the known protostars are found.
As a result, we conclude that it is more likely to be a galaxy than
a protostar, and we classify it as a non-member for the purposes of this work.

{\it BD+$30\arcdeg$547}. \citet{pre97} described it as a likely foreground star,
but its proper motion is consistent with membership in NGC~1333
(E. Mamajek, private communication) and it exhibits X-ray emission.
Some previous studies have concluded that it has IR excess emission at
24~\micron\ \citep{eva09,you15,reb15b}, but that appears to be due to
contamination from a protostar at a separation of $3\farcs4$.

{\it 2MASS J03290575+3116396}. Our SpeX type (M0--M2) is much later than
previous types (late-G, A3).

{\it [SVS76] NGC 1333 7}. It has been classified as an early B giant
through optical spectroscopy \citep{tur80}. Meanwhile, several studies have
reported IR excess emission for this star based on images
from {\it Spitzer}. However, it is detected with only low S/N at 8~\micron\ and
is not detected at 24~\micron\ because of bright extended emission.
Among the bands at shorter wavelengths where better photometry is available,
only the [5.8] band exhibits a significant excess.
Given the excess in that band and its location near the center of the cluster,
the star is a promising candidate member, but we exclude it from
our census of confirmed members because of the uncertainty in its
spectral classification.

{\it NGC 1333 IRS J03290347+3116179}. It has a red, featureless near-IR
spectrum. It is too close to a brighter star (2MASS J03290375+3116039)
to be detected by {\it Spitzer} at $>5$~\micron, but its 
$[3.6]-[4.5]$ color is indicative of a protostar \citep{reb15b}, which
would be consistent with the appearance of its spectrum.
Its membership is further supported by its detection in X-rays and its close
proximity to known members of the cluster.

{\it 2MASS J03290895+3122562}. Its near-IR spectrum is red and featureless,
which is consistent with the protostellar nature implied by its
IRAC colors \citep{gut08,eva09}.

{\it 2MASS J03294415+3119478}. The strength of the steam bands for this
object imply a type of M7--M8, but the overall slope of the SpeX data
does not agree with that of any reddened standard. In addition,
it is fainter than most members near its type in color-magnitude diagrams.
With these characteristics, it is similar to 2MASS~J04381486+2611399, which
is a low-mass member of Taurus that is seen in scattered light from an edge-on
disk \citep{luh07edge}.

{\it 2MASS J03283695+3123121}.
\citet{pre97} suggested that it is a foreground star based on its proper
motion from \citet{her83}, but the motion that we have measured with
IRAC is consistent with membership in NGC~1333.

{\it NGC 1333 IRS J03284883+3117537}.
The S/N of our spectrum of this object is too low for classification. 
It is a candidate member based on its location on color-magnitude diagrams,
proper motion, and excesses in the IRAC bands. It also may 
be detected at low S/N at 24~\micron\ in images from MIPS, which would
further support the presence of excess emission. It is included in
our sample of remaining candidate members that lack classifications
(Section~\ref{sec:completeness}).

{\it 2MASS 03302246+3132403}. \citet{eva09} and \citet{you15} identified it
as a possible protostar based on its mid-IR excess emission.
However, it is detected in the optical bands of the Digitized Sky Survey,
whereas most protostars are too heavily reddened for detections in those data.
In addition, it is not near known protostars or high column densities of gas. 
Therefore, we conclude that it is probably a galaxy.

\section{Analysis of New Census}
\label{sec:analysis}

\subsection{Completeness}
\label{sec:completeness}

\begin{figure}
\epsscale{1}
\plotone{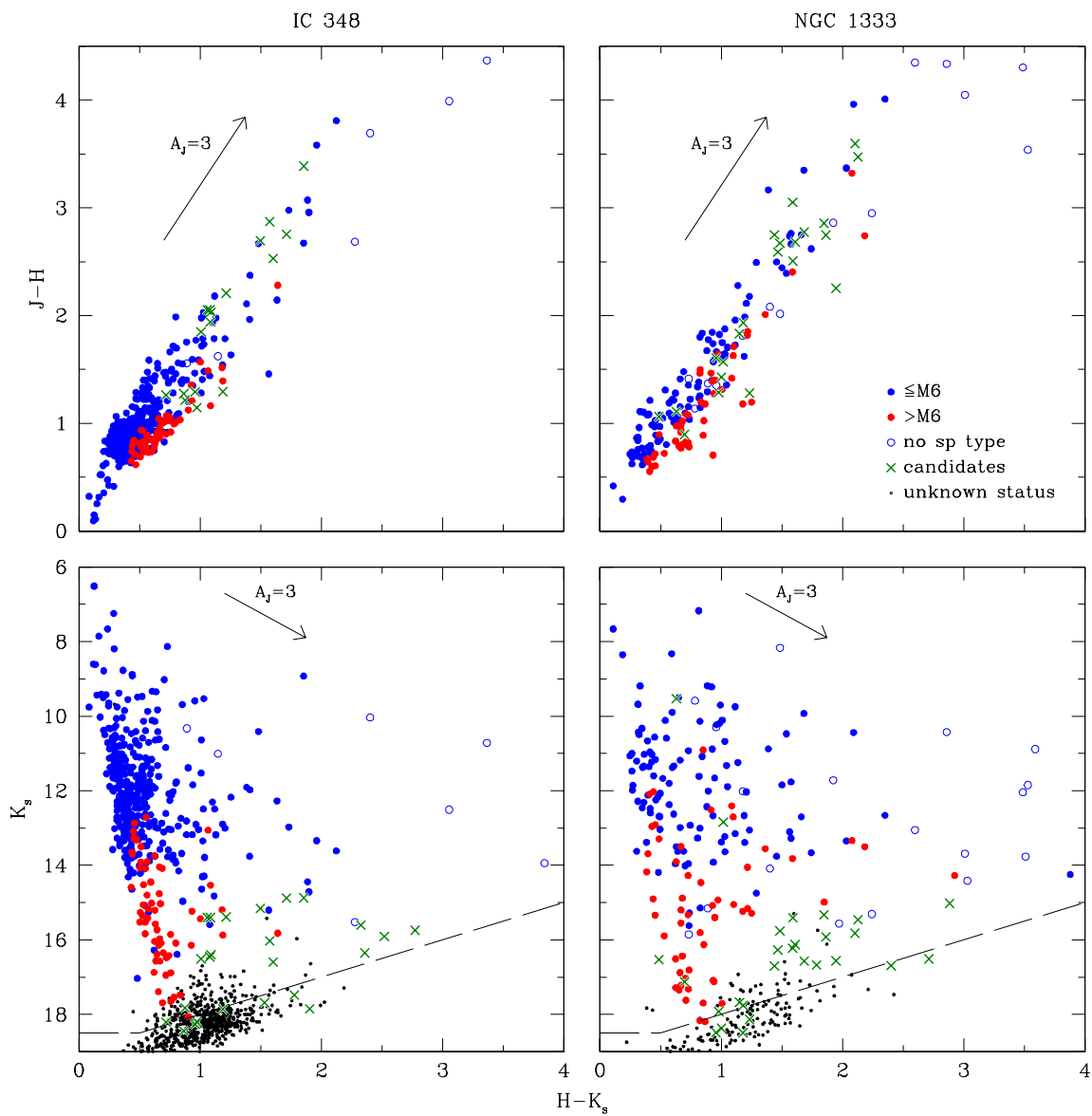}
\caption{
Near-IR color-color and color-magnitude diagrams for the known members of
IC~348 and NGC~1333 (filled and open circles), candidate members within the
$14\arcmin$ radius field in IC~348 and within the ACIS-I field in NGC~1333
(crosses, Tables~\ref{tab:cand348} and \ref{tab:cand1333}), 
and the remaining sources in those fields with unconstrained membership
(small points; shown only in the bottom diagram).
These data are from 2MASS, UKIDSS, WIRCam, and \citet{mue03}.
The completeness limits of the WIRCam images are indicated (long dashed lines).
}
\label{fig:hk}
\end{figure}

As discussed in Section~\ref{sec:fields}, we have focused our survey
for members of IC~348 and NGC~1333 within a radius of $14\arcmin$ from
BD+$31\arcdeg$643 in the former and within the ACIS-I images of the latter,
which cover a field with a size of $18\arcmin\times18\arcmin$ 
(see Fig.~\ref{fig:map}).
To characterize the completeness of our new census for each field,
we use a color-magnitude diagram in two bands that can
detect objects at both low masses and high extinctions.
Given the available data, the best options for these bands are $H$ and $K_s$.
In Figure~\ref{fig:hk}, we plot diagrams of $K_s$ versus $H-K_s$ for
all known members of IC~348 and NGC~1333, the remaining candidate members
identified in Section~\ref{sec:select} that lack spectra and are
within the survey fields, and all other objects within those fields that
are detected in $H$ and $K_s$ and that are not rejected as field stars
by any of the color-magnitude diagrams that we used in selecting candidates.
Thirteen and 20 members of IC~348 and NGC~1333, respectively, are absent from
those diagrams, which consist of companions that are unresolved from brighter
stars and protostars that are extended\footnote{Some of these protostars have
measurements in the 2MASS Point Source Catalog, but are found to be
dominated by extended emission in the higher resolution images from WIRCam.}
or are not detected in $H$ or $K_s$. In Figure~\ref{fig:hk}, there are few
remaining objects in the survey fields with undetermined membership status down
to rather faint magnitudes for low-to-moderate levels of extinction.
Specifically, our census appears to be nearly complete for extinction-corrected
magnitudes of $K_s<16.8$, 15.8, and 15.3 in the IC~348 field and for
$K_s<17.3$, 16.2, and 15.3 in the NGC~1333 field for $A_J<1.5$, 3, and 5,
respectively.

We can use our new census of IC~348 and NGC~1333 to examine the completeness
of previous surveys for members. \citet{luh03} estimated that their census
of a $16\arcmin\times14\arcmin$ field in IC~348 was nearly complete for
extinction-corrected magnitudes of $H<15.5$ ($\lesssim$M8) for $A_V<4$
($A_J<1.13$). Subsequent studies have not uncovered any additional members
in their survey field and in that range of magnitudes and extinctions.
\citet{alv13} noted that three of their new members were within the field
from \citet{luh03}. However, one of those objects, LRL~659, actually
falls slightly outside of that field. The other two members, LRL~2050 and
22528 (M8 and M9), are fainter than the completeness limit from \citet{luh03}.
Meanwhile, \citet{alv13} concluded that their survey for brown dwarfs 
($\gtrsim$M6.5) was complete down to $\sim0.013$~$M_\odot$ ($\lesssim$M9)
for $A_V\leq4$ for a field that encompasses the entire cluster.
However, we have found several new members at M6.5--M9 with $A_V<4$,
consisting of LRL~1254, LRL~1824, LRL~5103, LRL~10256, LRL~22185, LRL~22191,
LRL~22317, and LRL~22778.
\citet{sch09,sch12a,sch12b} searched for members of NGC~1333 within a
$30\arcmin\times30\arcmin$ field covering the entire cluster using images
that exhibited completeness limits of $J=20.8$ and $K=18$. We have
identified 32 additional members above those limits (13 at $\leq$M6,
19 at $>$M6). The incompleteness in their census
and the systematic offset between their spectral classification scheme
and that applied to other regions like IC~348 (Section~\ref{sec:class})
cast doubt on the validity of the statements by
\citet{sch09,sch12a,sch12b,sch13} regarding the mass function in NGC~1333.

\begin{figure}
\epsscale{1}
\plotone{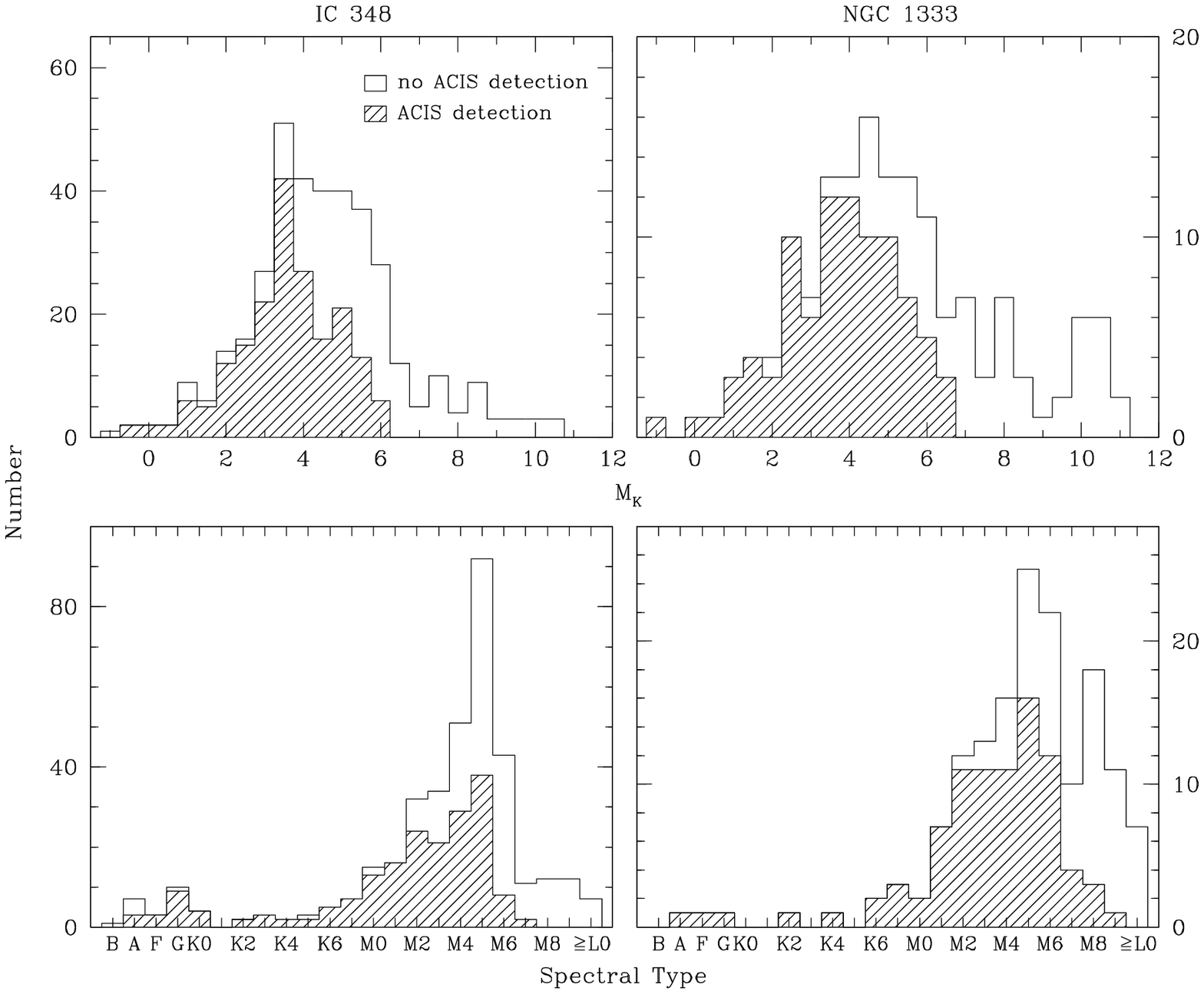}
\caption{
Distributions of spectral types and extinction-corrected $M_K$
for members of IC~348 and NGC~1333 that are detected by ACIS-I on {\it Chandra}
(shaded histograms) and that are within the ACIS-I fields but
are not detected
\citep[open histograms,][K. Getman, in preparation]{pre01,pre02,get02,win10,for11,ste12}.
}
\label{fig:histox}
\end{figure}

We also can characterize the fraction of members in our census that have
been detected in X-rays by ACIS-I on {\it Chandra}.
\citet{ste12} analyzed the four existing ACIS-I observations of IC~348,
arriving at list of 290 detected sources. They found X-ray counterparts
for 187 of the 316 members from their adopted cluster census that were
within the ACIS-I images. Using our updated census, 388 known members
were observed by ACIS-I, 197 of which were detected.
The members with X-ray detections can be identified in Table~\ref{tab:mem348}
via the presence of source designations from \citet{ste12}.
K. Getman (in preparation) has performed a similar analysis for the
existing ACIS-I data in NGC~1333.
Among the 186 known members within those images, 98 have counterparts
in their catalog of ACIS-I sources, as indicated in the column for evidence
of membership in Table~\ref{tab:mem1333}.
In Figure~\ref{fig:histox}, we plot the distributions of extinction-corrected
$M_K$ and spectral types for all known members of IC~348 and
NGC~1333 within the ACIS-I images and for the members detected in those data.
As expected, the fraction of members with X-ray detections decreases with
fainter magnitudes and later spectral types, quickly approaching zero
at $M_K>6$ and $>$M7.

In Tables~\ref{tab:cand348} and \ref{tab:cand1333}, we present our remaining
candidate members that lack spectra and that are within the $14\arcmin$ radius
field in IC~348 and within the ACIS-I field in NGC~1333. Although it has been
previously observed with spectroscopy, [SVS76] NGC 1333 7 is included with
these candidates since its membership is uncertain (Section~\ref{sec:comments}).
The probability of membership varies substantially among these candidates.
Those identified via both color-magnitude diagrams and proper motions are
promising while those selected by proper motions alone (i.e., they lack
the optical data needed for the color-magnitude diagrams) are much less likely 
to be members (Section~\ref{sec:pm}).
Based on their positions in the color-magnitude diagrams in
Figure~\ref{fig:hk}, most of the candidates should have spectral types
of $\gtrsim$M6 if they are members. To check whether they have the near-IR
colors expected for those types, we have included diagrams of $J-H$ versus
$H-K_s$ for each cluster in Figure~\ref{fig:hk}.
Some of the candidates at $H-K_s<1.5$ do resemble known late-type members
in both $J-H$ and $H-K_s$, but most of the candidates at $H-K_s>1.5$ have
colors indicative of earlier types, and thus are likely to be background stars.

\subsection{H-R Diagrams}
\label{sec:hr}

\begin{figure}
\epsscale{1}
\plotone{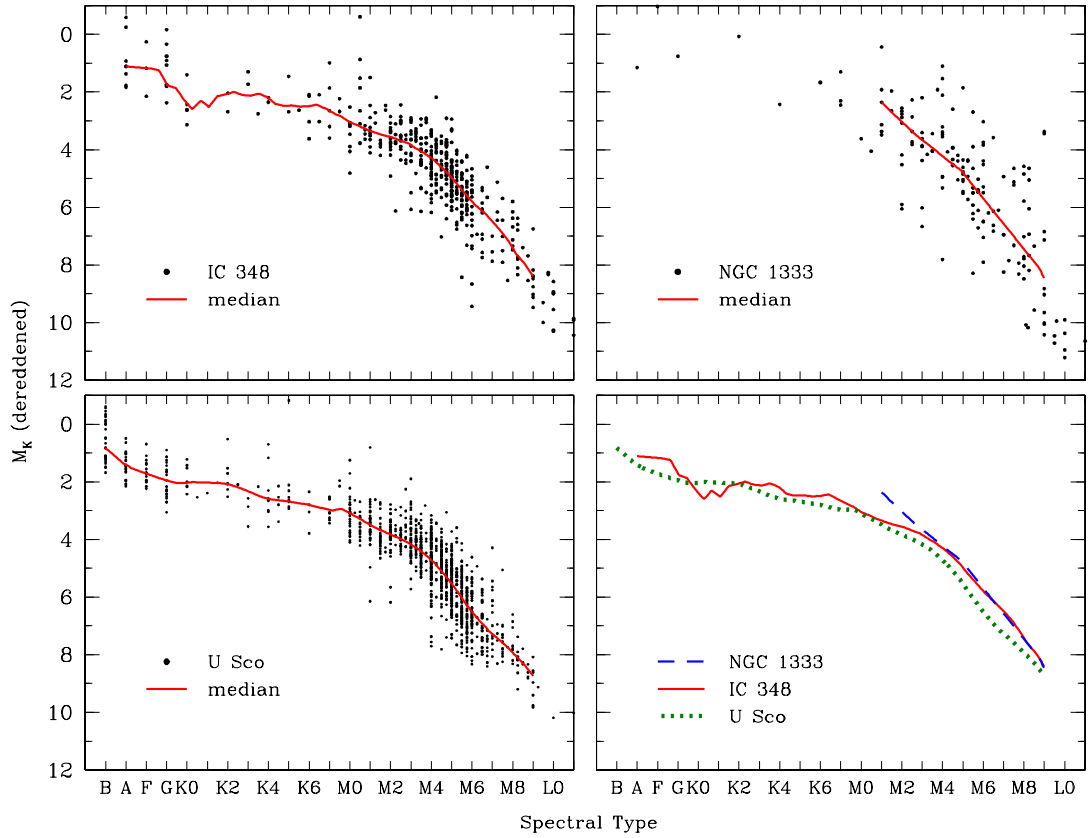}
\caption{
Top and left panels: Extinction-corrected $M_K$ versus
spectral type for the known members of IC~348, NGC~1333, and Upper Sco
(points) assuming distances of 300, 235, and 145~pc, respectively.
The median sequence is shown for each population (solid lines).
Bottom right panel:
The median sequences of IC~348 and NGC~1333 (solid and dashed lines)
suggest that the clusters have
similar ages. Those clusters are 0.4~mag brighter than Upper Sco (dotted line),
corresponding to ages that are younger by 0.25~dex based on evolutionary models.
}
\label{fig:hr}
\end{figure}

We have constructed H-R diagrams for the known members of IC~348 and NGC~1333
in terms of $M_K$ versus spectral type. We use absolute magnitude 
and spectral type instead of bolometric luminosity and effective temperature
to avoid uncertainties in bolometric corrections and conversions between
spectral type and temperature.
We choose $K_s$ for the band of the absolute magnitude because it is long
enough in wavelength that extinctions are relatively low for most objects
while short enough in wavelength that the fluxes are likely to be dominated
by stellar photospheres rather than circumstellar disks. In addition,
given the sensitivities of the available images, those at $K_s$ detect the
largest fraction of the clusters members. The analysis in this section
was also performed with the $J$ and $H$ bands, which produced identical
results to those from $K_s$.
We estimated extinctions for all known members of IC~348 and NGC~1333 that
have measured spectral types in a similar manner as done by \citet{fur11}
for members of Taurus. For the members that we observed with IR spectroscopy,
we have adopted the extinctions derived during the spectral classifications.
For each of the remaining objects, we calculated the extinction from the
excess in $J-H$ relative to the color expected for a young stellar
photosphere at the spectral type in question \citep{luh10tau}.
After correcting the $K_s$ measurements for extinction, we converted
them to absolute magnitudes using distances of 300~pc for IC~348
\citep{her08} and 235~pc for NGC~1333 \citep{hir08}.
The resulting values of $M_K$ are plotted as a function of spectral type in
Figure~\ref{fig:hr}. For comparison, we have included data for the members
of the Upper Sco association compiled by \citet{luh12usco}. We have 
adopted a distance of 145~pc for Upper Sco \citep{pm08} and have estimated
extinctions with the same methods that were applied to IC~348 and NGC~1333.

Several members of IC~348 and NGC~1333 are unusually faint for their
spectral types, appearing below the cluster sequences in Figure~\ref{fig:hr}.
These objects include LRL~276, LRL~435, LRL~621, LRL~622, LRL~725, and
LRL~4011 in IC~348 and sources 39, 92, 99, 110, and 122 from \citet{gut08}
in NGC~1333. The first five sources in IC~348 exhibited similar positions
in the H-R diagram from \citet{luh03}. As noted in Section~\ref{sec:midir},
stars that are occulted by circmumstellar disks often are observed
primarily in scattered light, which results in underestimates of their
luminosities. All of these stars do show evidence of disks in the form
of mid-IR excess emission, so it is plausible that they are occulted by
edge-on disks.

To compare the ages of IC~348, NGC~1333, and Upper Sco, we have computed the
median values of $M_K$ as a function of spectral type for each population.
This was done by applying local linear quantile regression with the
function {\tt lprq} in the {\it quantreg}
package \citep{koe16}
within R \citep{R} using a bandpass of one spectral type.
The resulting median sequences are plotted together in Figure~\ref{fig:hr}.
We do not include the median for $<$M1 in NGC~1333 because of the small
number of members at those types.
The sequences for IC~348 and NGC~1333 are not offset vertically
from each other, which would suggest that they have similar ages.
However, NGC~1333 exhibits clear evidence of a younger age in the form of
a greater abundance of protostars and
circumstellar disks \citep[][Section~\ref{sec:disks}]{mue07,gut08} and
higher extinction (Figure~\ref{fig:hk}).
In order for the H-R diagram to produce a younger age for NGC~1333, we would
need to adopt a larger distance for it \citep{her83} or a smaller distance
for IC~348 \citep{rip14}, e.g., if the clusters have similar distances.
Indeed, one would not expect the cluster distances to differ as much
as we have assumed (65~pc) given that their projected separation is only
$\sim$17~pc, although it is possible that they reside in separate clouds
along the line of site rather than a single cloud \citep{bal08}.
Meanwhile, the median sequences for IC~348 (at 300~pc) and NGC~1333 (at 235~pc)
are 0.4~mag brighter than the sequence for Upper Sco (at 145~pc),
which corresponds to an age difference of 0.25~dex based on evolutionary
models \citep[e.g.,][]{bar98,bar15}.
If Upper Sco has an age of 11~Myr \citep{pec12}\footnote{A younger
age of 4--5~Myr has also been proposed for Upper Sco based on its low-mass
stars \citep{deg89,pre02b,sle06,her15}.}, then IC~348 and NGC~1333 would have
ages of 6~Myr. The latter agrees with the value derived for
IC~348 by \citet{bel13} from color-magnitude diagrams and evolutionary models.
However, a distance of 250~pc was adopted for IC~348
in that study. Using that distance, the sequences in Figure~\ref{fig:hr} 
would indicate similar ages for IC~348 and Upper Sco, which would be difficult
to reconcile with the fact that IC~348 has higher abundances of disks and
protostars and, unlike Upper Sco, is still associated with a molecular cloud.
The {\it Gaia} mission \citep{per01} should soon help isolate the sources
of these discrepancies by providing accurate parallactic distances for IC~348,
NGC~1333, and Upper Sco, as well as other nearby clusters and associations.

\subsection{Initial Mass Functions}

Previous studies have estimated the initial mass functions (IMFs) in
IC~348 and NGC~1333 based on earlier samples of spectroscopically confirmed
members \citep{luh98,luh03,gre07,alv13,sch09,sch12a,sch12b}\footnote{The IMFs
in these clusters also have been constrained via IR luminosity functions
\citep{lad95,lad96,mue03}}.
Typically, the masses of individual objects were derived by combining estimates
of bolometric luminosities and effective temperatures with the values predicted
by evolutionary models.
As a result, the IMFs depended on the adopted bolometric corrections,
temperature scales, and models.
To avoid those dependencies, we examine the IMFs in IC~348 and NGC~1333 
in terms of observational parameters that should be roughly correlated with
stellar mass, spectral type and extinction-corrected $K_s$. Note that these
parameters still depend on the methods adopted for measuring spectral
types and extinctions.
As done in our previous studies of IMFs in star-forming regions, we
attempt to construct a sample of members in each cluster that is representative
and unbiased in terms of mass by considering all known members within
a field and an extinction threshold for which the current census has
a high level of completeness.
Guided by the analysis of completeness in Section~\ref{sec:completeness},
we select extinction thresholds that are high enough to encompass large
numbers of members while low enough that that the completeness extends to low
masses, arriving at $A_J<1.5$ for the $14\arcmin$ radius field in IC~348
and $A_J<3$ for the ACIS-I field in NGC~1333.
In Section~\ref{sec:completeness}, we found that the census of IC~348
and NGC~1333 within these fields and extinction limits should be
nearly complete for extinction-corrected magnitudes of $K_s<16.8$ and 16.2,
respectively. These samples contain 341 and 120 members, respectively,
which correspond to 71\% and 59\% of the known members.

\begin{figure}
\epsscale{1}
\plotone{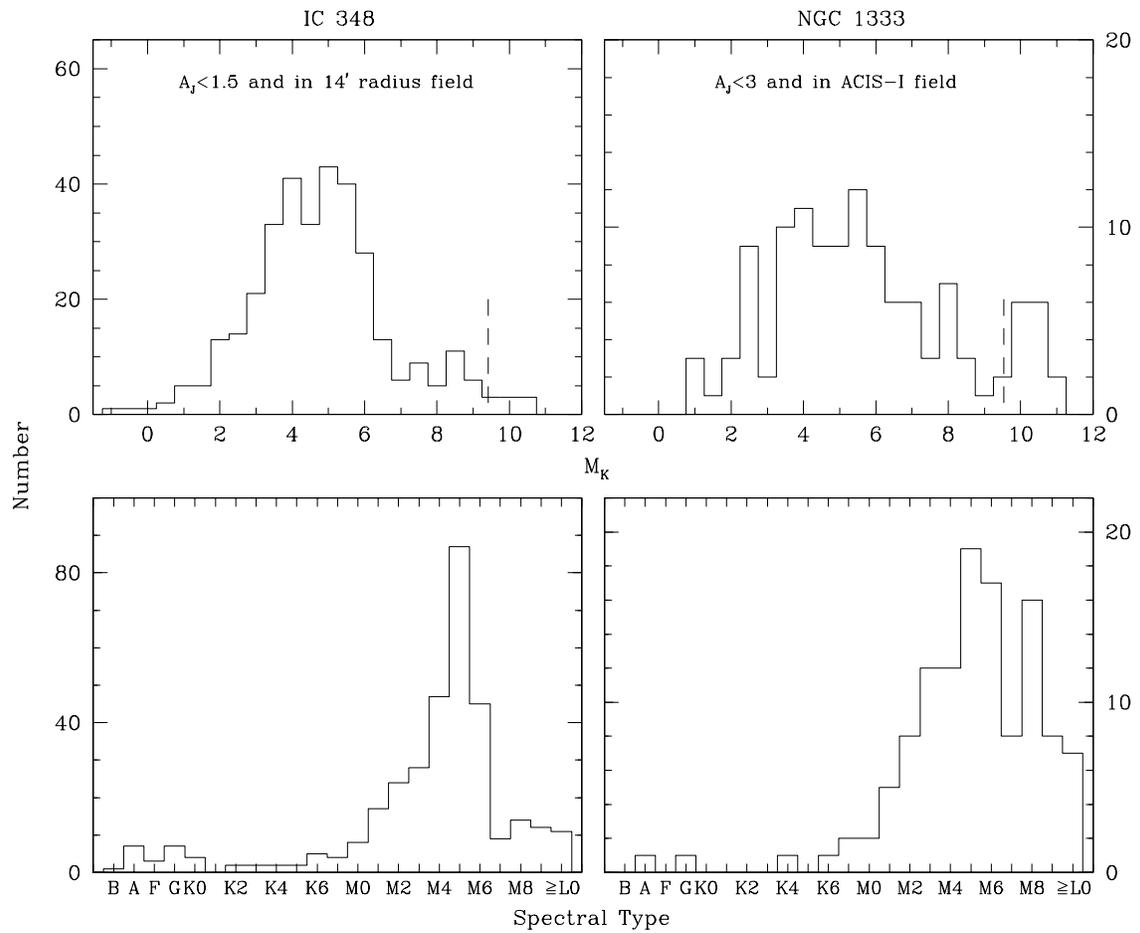}
\caption{
Distributions of spectral types and extinction-corrected $M_K$
for members of IC~348 within its $14\arcmin$ radius field that have $A_J<1.5$
and members of NGC~1333 within the ACIS-I field that have $A_J<3$. 
The dashed lines indicate the completeness limits of
these samples of members for the fields and ranges
of extinctions that they represent (see Fig.~\ref{fig:hk}).
}
\label{fig:histoav}
\end{figure}

The distributions of spectral types and extinction-corrected $M_K$
for our extinction-limited samples of members of IC~348 and NGC~1333 are
plotted in Figure~\ref{fig:histoav}.
Relative to IC~348, NGC~1333 exhibits a surplus of objects with late
spectral types and faint magnitudes.
For instance, N($\geq$M6.5)/N($<$M6.5)=54/287=0.188$^{+0.025}_{-0.02}$
and 42/78=0.538$\pm0.056$ and 
N($M_K\geq6.5$)/N($M_K<6.5$)=50/291=0.172$^{+0.025}_{-0.019}$
and 40/80=0.50$^{+0.056}_{-0.054}$ for IC~348 and NGC~1333, respectively.
There are multiple possible explanations for the differences in these ratios.
They could reflect a variation in the IMF between the two clusters, although
a significant variation would be surprising given that the clusters
have similar stellar densities and environments, and indeed have arisen
from the same cloud (or at least related clouds). 
Because most of the spectral types in NGC~1333 have been measured with
IR spectra, and the resulting classifications tend to have larger uncertainties
than the optical types that are more frequently available in IC~348, one
would expect the distribution of spectral types to be somewhat broader
in NGC~1333, which would result in a higher abundance of late spectral types.
However, this effect would not explain the difference
between the two clusters in their distributions of $M_K$.
Another possibility is that the average extinctions of members of star-forming
regions vary with stellar mass, in which case extinction-limited samples
would not be representative of the stellar populations in those clusters.
For instance, if members at lower masses tend to have less extinction
in the most embedded clusters like NGC~1333, then an extinction-limited sample
could capture most of the brown dwarfs but miss many of the stars.
In a cluster like IC~348 that has dispersed more of its natal cloud,
the dependence of extinction on stellar mass could be smaller, leading to
a larger, more representative ratio of stars to brown dwarfs in an
extinction-limited sample.
This scenario could account for the differences in the
extinction-limited samples for IC~348 and NGC~1333.
Finally, we note that both samples contain several known members that
are fainter than the completeness limits, and the degree of incompleteness
beyond those limits may differ between the two samples, which could
somewhat inflate the perceived abundance of low-mass objects in one cluster
relative to the other.
To determine whether these last two issues are
responsible for the surplus of low-mass objects in the extinction-limited
sample for NGC~1333 relative to the sample in IC~348, it will be necessary
to obtain additional data (e.g., photometry, spectroscopy, proper motions)
that can extend the completeness limits of the census to higher extinctions
and lower masses in both clusters.

Our work has provided new constraints on the minimum masses of the IMFs
in IC~348 and NGC~1333. In each cluster, members are present down to
and below the completeness limits, and thus the minimum of the IMF has
not been detected. The faintest known members have $M_K=10.4$ and 11.2,
which correspond to masses of $\sim$0.004--0.006 and 0.003--0.005~$M_\odot$,
respectively, for ages of 1--3~Myr according to the evolutionary models of
\citet{bur97} and \citet{cha00}.

\subsection{Disk Fractions}
\label{sec:disks}

\begin{figure}
\epsscale{1}
\plotone{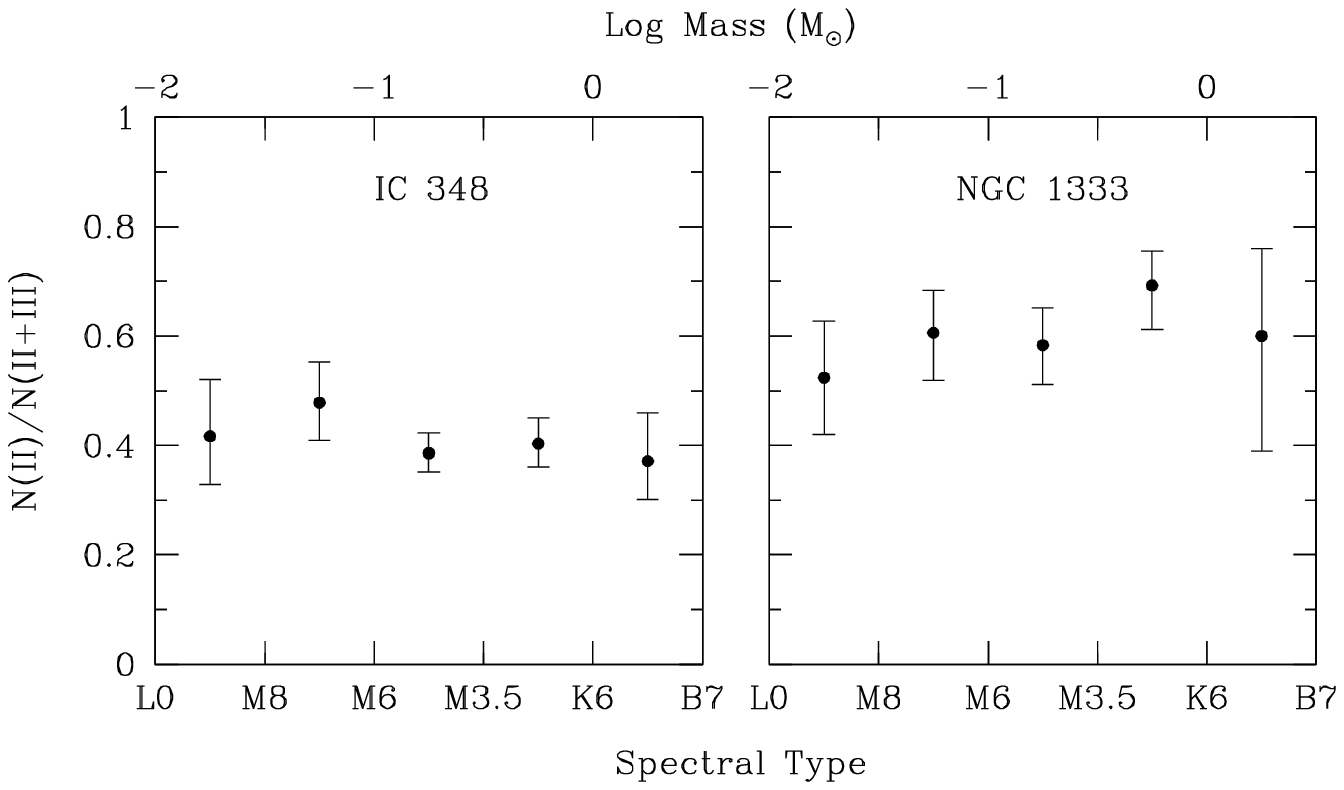}
\caption{
Fraction of sources with circumstellar disks (class~II)
as a function of spectral type within the $14\arcmin$ radius field in IC~348
and within the ACIS-I field in NGC~1333 based on mid-IR photometry from
{\it Spitzer}
\citep[][Table~\ref{tab:disks}]{luh05frac,lad06,mue07,gut08,cur09,eva09}.
The boundaries of the spectral type bins have been
chosen to correspond approximately to logarithmic intervals of mass.
}
\label{fig:disks}
\end{figure}

We can combine our census of IC~348 and NGC~1333 with the previous mid-IR
imaging of these clusters to measure the fractions of members that have
circumstellar disks.
A disk is present in the first three stages of a young stellar object, which
consist of classes~0 and I (protostar+disk+infalling envelope)
and class~II \citep[star+disk,][]{lw84,lad87,and93,gre94}.
A star that has fully cleared its primordial disk is in the class~III stage.
A disk fraction can be defined as either N(I+II)/N(II+III) or N(II)/N(II+III)
(class~0 objects are rare enough that their contribution is usually negligible).
Because the ages measured for young clusters (with H-R diagrams)
often apply to the class~II
and III sources, we choose the latter definition, as done in \citet{luh10tau}.
Since protostars normally have heavily veiled, featureless spectra
(Fig.~\ref{fig:sp1333a}), we can exclude them from our calculations of disk
fractions by considering only members that have measured spectral types.
Members of star-forming regions are often discovered based on mid-IR excess
emission from disks. As a result, samples of members can be biased in favor
of disks, making it difficult to measure disk fractions that are representative
of the stellar populations. However, IC~348 and NGC~1333 have been thoroughly
surveyed for members using a variety of methods, and we have shown that
the current membership samples for the $14\arcmin$ radius field in IC~348
and the ACIS-I field in NGC~1333 are nearly complete for a wide range
of masses and extinctions (Section~\ref{sec:completeness}).
We have assigned the presence or absence of mid-IR excess emission for each
member based on the results of previous disk surveys in these clusters
with {\it Spitzer}
\citep{luh05frac,lad06,mue07,gut08,cur09,eva09,arn12,reb15b,you15}.
A few of the faintest brown dwarfs lack sufficiently accurate photometry for
determining whether excess emission in present; they are excluded from our
calculations of disk fractions.
In Table~\ref{tab:disks} and Figure~\ref{fig:disks}, we list and plot
the fraction of members that have excess emission as a function of spectral
type for the $14\arcmin$ radius field in IC~348 and for the ACIS-I field in
NGC~1333.
The latter has a higher disk fraction, which agrees with previous
analysis of near-IR photometry \citep{lad95,lad96} and the {\it Spitzer} data
\citep{lad06,mue07,gut08}.
The higher disk fraction in NGC~1333 is consistent with its higher
abundance of protostars \citep{mue07,gut08} and the younger age implied by
its greater obscuration. In each cluster, the disk fraction is
roughly constant for the full range of spectral types.

\subsection{Spatial Distributions}

\begin{figure}
\epsscale{1}
\plotone{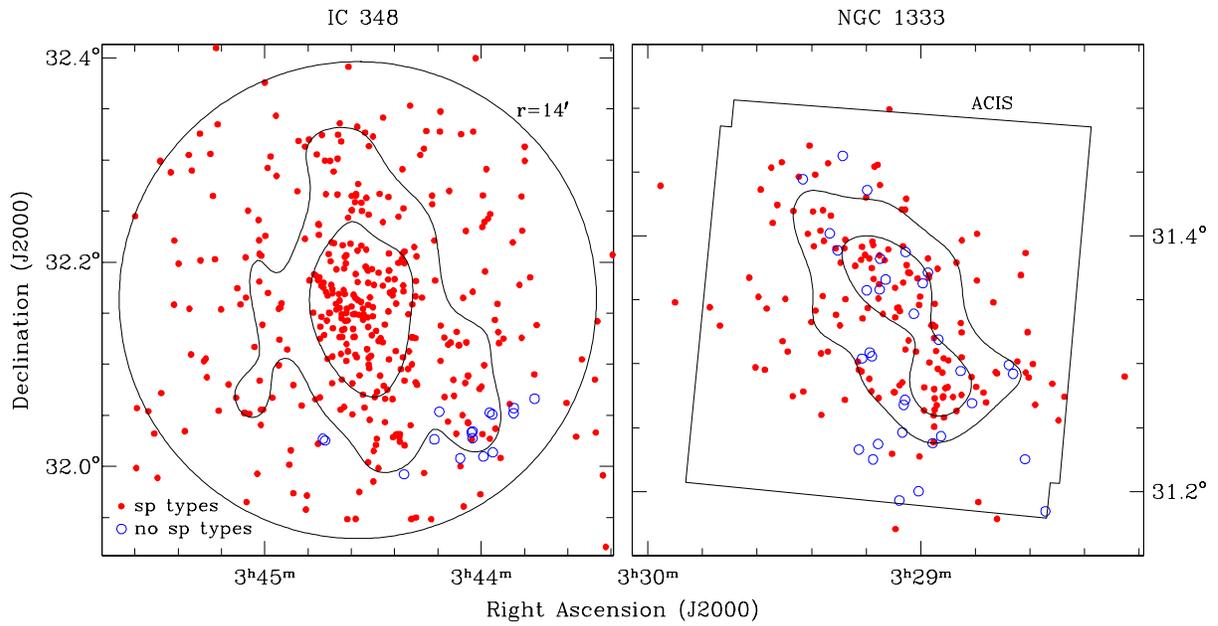}
\caption{
The positions of the known members of IC~348 and NGC~1333
that have measured spectral types (filled circles, $\approx$classes~II and III)
and those that do not (open circles, $\approx$classes~0 and I).
The surface density of members with spectral types is represented by the
contours.
We have marked the $14\arcmin$ radius field in IC~348 and the ACIS-I field
in NGC~1333 within which we have focused our survey for members.
}
\label{fig:map2}
\end{figure}

The spatial distributions of the stellar populations in IC~348 and NGC~1333 
has been previously studied through analysis of probable members detected
in near- and mid-IR imaging \citep{lad95,lad96,mue03,gut08}.
The current census of each cluster now offers confirmation of membership,
measurements of spectral types for most members, and a high level of
completeness for most locations, masses, and extinctions.
These features allow us to examine the spatial distributions of spectral types
and offsets in $M_K$ from the median cluster sequence at a given spectral
type ($\Delta M_K$), which
serve as proxies for stellar masses and ages, respectively. We can also
measure the spatial dependence of disk fractions using the mid-IR excess data
compiled in the previous section. We define $\Delta M_K$ as $M_K$(median
sequence at a star's spectral type)$-M_K$(star), i.e., higher
values of $\Delta M_K$ correspond to younger implied ages.
We consider only members within the $14\arcmin$ radius field in IC~348
and the ACIS-I field in NGC~1333 because of the well-defined completeness
in those areas. The class~0 and I objects exhibit distinct spatial distributions
compared to members in the more evolved classes, so we exclude them from our
analysis by considering only members that have measured spectral types.
We also omit members that
are later than M9 since their types tend to have large uncertainties.
For each of these samples of members, we have computed surface density 
as a function of position using the {\tt kde2d} function in the R package
{\it MASS} \citep{ven02},
which performs a two dimensional kernel density estimation with a bivariate 
normal kernel. We then identified the density contours that would divide the
sample for a given cluster into three subsets that have equal numbers.
We selected that number of sections to allow coarse measurements
of the variations of median spectral type, $\Delta M_K$, and disk fraction
with surface density while also providing good number statistics within
each section. The resulting contours are plotted in
Figure~\ref{fig:map2} with the locations of all known members of each cluster.
The average surface densities in these sections are 0.3, 1.0, and
3.0~arcmin$^{-2}$ in IC~348 and 0.2, 0.9, and 2.6~arcmin$^{-2}$ in NGC~1333.
As found in previous studies, IC~348 is more centrally concentrated than
NGC~1333, which exhibits a double cluster morphology \citep{lad96}.

\begin{figure}
\epsscale{1}
\plotone{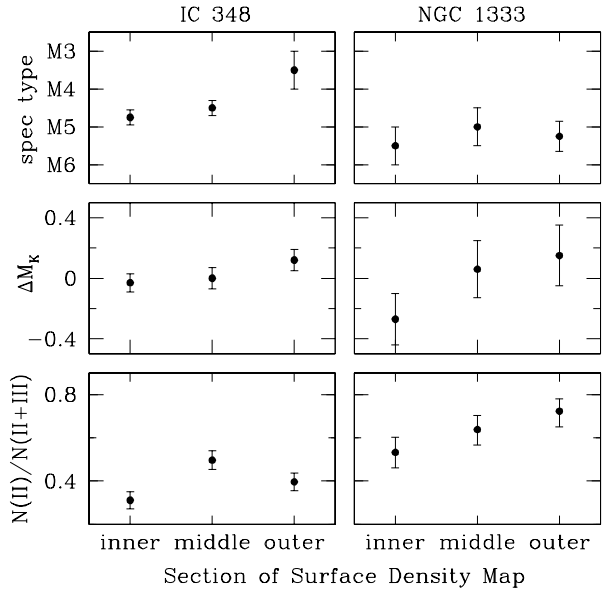}
\caption{
Median values of spectral type, $\Delta M_K$, and disk fraction 
for members of IC~348 and NGC~1333 within the three regions of each cluster
that are separated by the surface density contours in Figure~\ref{fig:map2}.
$\Delta M_K$ is defined as the difference in $M_K$ between the
median sequence of a cluster and an individual member in Figure~\ref{fig:hr}
where positive values of $\Delta M_K$ correspond to positions above
the median sequence (i.e., younger ages).
}
\label{fig:map3}
\end{figure}

For each of the three sections within IC~348 and NGC~1333 in
Figure~\ref{fig:map2}, we have computed the median spectral type,
the median $\Delta M_K$, and the disk fraction.
The errors in the medians were estimated with bootstrapping.
The resulting values are plotted for each section in Figure~\ref{fig:map3}.
None of these parameters exhibit significant variations among the
sections in either cluster. 
Through analysis of the near-IR luminosity function of IC~348, \citet{mue03}
found a higher abundance of solar-mass stars in the core relative to the
outskirts of the cluster.
That surplus is also detected when distributions of spectral types are
compared between the inner and outer portions of the cluster, 
although it is does not have a noticeable effect on the median types
because low-mass stars are the dominant component of the stellar population
throughout the cluster. 
Using X-ray and near-IR photometry, \citet{get14} have detected
age gradients in NGC~2024 and the Orion Nebula Cluster in which younger
stars are found in the cores of the clusters.
A trend of that kind is not present in the median values of $\Delta M_K$
for IC~348 and NGC~1333. For perspective, an age gradient like that
reported for Orion (1.2--1.9~Myr) should correspond to a difference of
0.3~mag in $M_K$ according to evolutionary models of low-mass stars.
The errors in median $\Delta M_K$ are larger for NGC~1333
than for IC~348 because the former has a broader sequence at a given
spectral type.

We note that a variety of more sophisticated methods are available for
characterizing the spatial distributions of members of star-forming clusters
\citep{gut09,kuh14} and searching for evidence of mass segregation
\citep{sag88,hil98,all09,mas11}.
The optimum approach for measuring the latter has been a subject of
debate in recent years \citep{asc09,olc11,par15}.

\subsection{Candidate Binary Systems}

Our census of IC~348 and NGC~1333 may contain resolved components of 
multiple systems. In Table~\ref{tab:pairs}, we have compiled
all pairs of objects from our census that have separations less than
$6\arcsec$. We have omitted binaries that have been resolved only
in high-resolution imaging and that lack spectral classifications of both
components \citep{duc99}. For some of these pairs, both components have
spectral types of late M, making them candidates for wide binary brown
dwarfs \citep{luh04bin,luh09fu}. The numbers of pairs are 23 and 8
for IC~348 and NGC~1333, respectively. To roughly estimate the fraction
of these pairs that comprise binary systems, we performed
a Monte Carlo simulation of the projected separations of unrelated cluster
members using surface density maps of the known members.
In $\sim$90\% of the realizations, the number of chance alignments with
separations of $<6\arcsec$ is between 5--15 for IC~348 and between 2--8
for NGC~1333. Thus, a significant fraction of the candidate binaries
could consist of unrelated cluster members.

\section{Conclusions}

We have sought to improve the completeness of the census of stars and
brown dwarfs in IC~348 and NGC~1333
and the accuracies of spectral types of known members of the latter.
The results of this study are summarized as follows:

\begin{enumerate}

\item
We have obtained optical and near-IR spectra of candidate members of
IC~348 and NGC~1333 that have been selected based on X-ray emission, mid-IR
excess emission, positions in color-magnitude diagrams, and proper motions.
We have classified 100 and 42 of the candidates as new members of IC~348
and NGC~1333, respectively.
The total numbers of known members are now 478 and 203.
Two stars in IC~348, LRL~62 and LRL~155, have radial velocities that
differ significantly from that of the bulk of the cluster,
but they exhibit other evidence of membership, which suggests that
they may have been ejected via dynamical interactions.

\item
We have searched for new members primarily within a radius of $14\arcmin$ from
the B5 star BD+$31\arcdeg$643 in IC~348 and within the area in NGC~1333 that
was observed by ACIS-I on {\it Chandra} ($18\arcmin\times18\arcmin$).
These fields are large enough to encompass most or all members of the clusters. 
The new census is nearly complete for extinction-corrected magnitudes of
$K_s<16.8$, 15.8, and 15.3 in the IC~348 field and for
$K_s<17.3$, 16.2, and 15.3 in the NGC~1333 field
for $A_J<1.5$, 3, and 5, respectively. For perspective, $K_s=15$ and 17
correspond to masses of $\sim$0.025 and 0.008~$M_\odot$, respectively, for
an age of 3~Myr according to evolutionary models \citep{bur97,cha00,bar15}.
IC~348 and NGC~1333 now have two of the most complete membership lists 
among star-forming clusters.

\item
The known members of IC~348 and NGC~1333 extend down to (and below)
the completeness limits of the current census. As a result, we have not
yet detected the low-mass cutoffs in the mass functions of these clusters.
The faintest known members have $M_K=10.4$ and 11.2, which imply masses of
$\sim$0.004--0.006 and 0.003--0.005~$M_{\odot}$, respectively, for ages of
1--3~Myr based on evolutionary models. 

\item
In addition to the candidate members, we have performed spectroscopy
on a large fraction (77\%) of the previously known members of NGC~1333.
These data provide greater uniformity in the spectral types among members
of this cluster and relative to stars in other young clusters like IC~348.

\item
To estimate the IMFs in IC~348 and NGC~1333, we have attempted to select
a sample of members in each cluster that is unbiased in terms of mass.
We have constructed extinction-limited samples for this purpose,
which should have a high level of completeness down to low masses.
The resulting sample for NGC~1333 has a higher abundance of low-mass objects
than the sample for IC~348.
For instance, N($\geq$M6.5)/N($<$M6.5)=0.188$^{+0.025}_{-0.02}$ in IC~348
and 0.538$\pm0.056$ in NGC~1333. Similar fractions are found when the clusters
are compared in terms of $M_K$.  A variation in the IMF between the clusters
would be surprising given their similar densities and environments.
Instead, it is possible that average extinctions are lower for
objects at lower masses, in which case extinction-limited samples may be
biased in favor of low-mass objects in heavily embedded clusters like NGC~1333.
To test that explanation, the completeness limits of the
census of IC~348 and NGC~1333 need to be extended to higher extinctions.

\item
We have constructed H-R diagrams for IC~348 and NGC~1333 in terms of spectral
type and $M_K$. For the adopted distances of 300 and 235~pc, the median
sequences of the clusters coincide, which suggests that they have similar ages. 
In contrast, NGC~1333 shows strong evidence of a younger age in the form of
higher abundances of disk-bearing stars and protostars and greater obscuration.
This discrepancy may indicate that IC~348 is closer or NGC~1333 is more
distant than we have assumed. The {\it Gaia} mission should soon test this
explanation by measuring parallactic distances for the clusters.

\item
Based on mid-IR photometry, the fraction of members that have circumstellar
disks is higher NGC~1333 than in IC~348 (N(II)/N(II+III)$\sim$0.6 and 0.4),
which agrees with results for earlier samples of members.
In each cluster, the disk fraction is roughly constant across the entire
range of stellar masses (0.01--3~$M_\odot$).

\item
For each cluster, we have examined the spatial distribution of stellar masses
and ages by computing the median spectral types and median offsets in $M_K$
from the median cluster sequence in three sections with differing stellar
densities. We also have measured the disk fraction in each of these areas.
None of these parameters exhibit significant variations with stellar density.

\end{enumerate}

\acknowledgements

This work was supported by grant AST-1208239 from the NSF.
We thank Konstantin Getman for providing his X-ray catalog of
NGC~1333 and Catarina Alves de Oliveira for providing her near-IR spectra of
candidate members of IC~348. We also thank Cameron Bell, Catarina Alves
de Oliveira, Eric Feigelson, Konstantin Getman, Charles Lada,
and Eric Mamajek for helpful discussions and comments on the manuscript.
The IRTF is operated by the University of Hawaii under contract
NNH14CK55B with NASA. The Gemini data were obtained through programs
GN-2008B-Q-21, GN-2014B-Q-55, GN-2015B-Q-43, and GN-2015B-FT-10.
Gemini Observatory is operated by AURA under a cooperative agreement with
the NSF on behalf of the Gemini partnership: the NSF (United States), the NRC
(Canada), CONICYT (Chile), the ARC (Australia),
Minist\'{e}rio da Ci\^{e}ncia, Tecnologia e Inova\c{c}\~{a}o (Brazil) and
Ministerio de Ciencia, Tecnolog\'{i}a e Innovaci\'{o}n Productiva (Argentina).
2MASS is a joint project of the University of Massachusetts and IPAC
at Caltech, funded by NASA and the NSF.  This work used data from 
the NASA/IPAC Infrared Science Archive, operated by JPL under contract
with NASA, and the SIMBAD database, operated at CDS, Strasbourg, France.
The Digitized Sky Survey was produced at the Space Telescope Science
Institute under U.S. Government grant NAG W-2166. The images of these
surveys are based on photographic data obtained using the Oschin Schmidt
Telescope on Palomar Mountain and the UK Schmidt Telescope. The plates
were processed into the present compressed digital form with the
permission of these institutions.
WIRCam is a joint project of CFHT, Taiwan, Korea, Canada, and France.
MegaCam is a joint project of CFHT and CEA/DAPNIA.
CFHT is operated by the NRC of Canada, the Institute
National des Sciences de l'Univers of the Centre National de la Recherche
Scientifique of France, and the University of Hawaii.
Subaru Telescope is operated by National Astronomical Observatory of Japan.
The W.M. Keck Observatory is operated as a scientific partnership among
Caltech, the University of California, and NASA. The Observatory was made
possible by the generous financial support of the W.M. Keck Foundation. 
The Center for Exoplanets and Habitable Worlds is supported by the
Pennsylvania State University, the Eberly College of Science, and the
Pennsylvania Space Grant Consortium.

\clearpage

\begin{deluxetable}{ll}
\tabletypesize{\scriptsize}
\tablewidth{0pt}
\tablecaption{Members of IC 348\label{tab:mem348}}
\tablehead{
\colhead{Column Label} &
\colhead{Description}}
\startdata
Name & Source name\tablenotemark{a} \\
Luhman & Name from \citet{luh98,luh03,luh05flam,luh05wfpc}, \citet{luh99}, \citet{mue07}, and this work  \\
Evans & Name from \citet{eva09} \\
Stelzer & Name from \citet{ste12} \\
AlvesdeOliveira & Name from \citet{alv13} \\
OtherNames & Other source names\\
SpType & Spectral type \\
r\_SpType & Spectral type reference\tablenotemark{b} \\
Spectrograph & Spectrograph for spectral classification \\
Date & Date of spectroscopy \\
Adopt & Adopted spectral type \\
Aj & Extinction in $J$ \\
IRexc & IR excess?\tablenotemark{c} \\
pmRA & Relative proper motion in right ascension\\
e\_pmRA & Error in pmRA\\
pmDec & Relative proper motion in declination\\
e\_pmDec & Error in pmDec\\
Jmag & $J$ magnitude \\
e\_Jmag & Error in Jmag \\
Hmag & $H$ magnitude \\
e\_Hmag & Error in Hmag \\
Ksmag & $K_s$ magnitude \\
e\_Ksmag & Error in Ksmag \\
JHKref & JHK reference\tablenotemark{d} \\
\enddata
\tablenotetext{a}{Coordinate-based identifications from the 2MASS Point
Source Catalog when available. Otherwise, identifications are based on the
coordinates measured from the WIRCam images in this work.}
\tablenotetext{b}{
(1) this work;
(2) \citet{mue07};
(3) \citet{alv13};
(4) \citet{luh03};
(5) \citet{str74b};
(6) \citet{luh98};
(7) \citet{luh99};
(8) \citet{har54};
(9) \citet{luh05flam};
(10) \citet{her98};
(11) \citet{luh05wfpc}.}
\tablenotetext{c}{Based on mid-IR photometry from the {\it Spitzer Space
Telescope} \citep{luh05frac,lad06,mue07,eva09,cur09,you15}.}
\tablenotetext{d}{2 = 2MASS Point Source Catalog; m = \citet{mue03};
u = UKIDSS Data Release 10; w = WIRCam data from this work.}
\tablecomments{
The table is available in a machine-readable format in the online journal.}
\end{deluxetable}

\begin{deluxetable}{ll}
\tabletypesize{\scriptsize}
\tablewidth{0pt}
\tablecaption{Members of NGC 1333\label{tab:mem1333}}
\tablehead{
\colhead{Column Label} &
\colhead{Description}}
\startdata
Name & Source name\tablenotemark{a} \\
Strom & Name from \citet{str74a} \\
Aspin & Name from \citet{asp94} \\
Lada & Name from \citet{lad96} \\
Getman & Name from \citet{get02} \\
Wilking & Name from \citet{wil04} \\
Gutermuth & Name from \citet{gut08} \\
Oasa & Name from \citet{ots08} \\
Evans & Name from \citet{eva09} \\
Scholz & Name from \citet{sch09} \\
Winston & Name from \citet{win10} \\
Rebull & Name from \citet{reb15b} \\
OtherNames & Other source names\\
SpType & Spectral type \\
r\_SpType & Spectral type reference\tablenotemark{b} \\
Spectrograph & Spectrograph for spectral classification \\
Date & Date of spectroscopy \\
Adopt & Adopted spectral type \\
Aj & Extinction in $J$ \\
IRexc & IR excess?\tablenotemark{c} \\
Evidence & Additional membership evidence\tablenotemark{d} \\
pmRA & Relative proper motion in right ascension\\
e\_pmRA & Error in pmRA\\
pmDec & Relative proper motion in declination\\
e\_pmDec & Error in pmDec\\
Jmag & $J$ magnitude \\
e\_Jmag & Error in Jmag \\
Hmag & $H$ magnitude \\
e\_Hmag & Error in Hmag \\
Ksmag & $K_s$ magnitude \\
e\_Ksmag & Error in Ksmag \\
JHKref & JHK reference\tablenotemark{e} \\
\enddata
\tablenotetext{a}{Coordinate-based identifications from the 2MASS Point
Source Catalog when available. Otherwise, identifications are based on the
coordinates measured from the WIRcam images in this work.}
\tablenotetext{b}{
(1) \citet{win09};
(2) this work;
(3) \citet{sch12a};
(4) \citet{sch09};
(5) \citet{coh80};
(6) \citet{asp03};
(7) \citet{wil04};
(8) \citet{gre07};
(9) \citet{sch12b};
(10) \citet{mer10};
(11) \citet{con10};
(12) \citet{str74a};
(13) \citet{str02};
(14) \citet{rac68};
(15) \citet{cie12}.}
\tablenotetext{c}{Based on mid-IR photometry and spectroscopy from the
{\it Spitzer Space Telescope} \citep{gut08,eva09,arn12,you15,reb15b}.}
\tablenotetext{d}{
Membership in NGC~1333 is indicated by
strong emission lines \citep[e,][this work]{win09,win10},
Li absorption \citep[Li,][]{win09},
X-ray emission \citep[X,][K. Getman, in preparation]{get02,win10},
the shape of the gravity-sensitive steam bands 
\citep[H$_2$O,][this work]{sch09,sch12a,sch12b},
or proper motions in Figure~\ref{fig:pm} (pm and pm?).}
\tablenotetext{e}{2 = 2MASS Point Source Catalog; u = UKIDSS Data Release 10;
w = WIRCam data from this work.}
\tablecomments{
The table is available in a machine-readable format in the online journal.}
\end{deluxetable}

\begin{deluxetable}{lllll}
\tabletypesize{\scriptsize}
\tablewidth{0pt}
\tablecaption{Field Stars in Spectroscopic Samples for IC~348 and NGC~1333\label{tab:non}}
\tablehead{
\colhead{Name} &
\colhead{Spectral Type} &
\colhead{Ref\tablenotemark{a}} &
\colhead{Telescope/Instrument} &
\colhead{Date}}
\startdata
2MASS J03283231+3127079\tablenotemark{b} & G--K & 1 & IRTF/SpeX & 2015 Jan 6 \\
2MASS J03283954+3121571\tablenotemark{b} & $<$M0 & 1 & IRTF/SpeX & 2015 Jan 6 \\
2MASS J03284197+3112171 & $<$M3,$<$M0 & 2,1 & IRTF/SpeX & 2015 Dec 14 \\
2MASS J03284316+3126061\tablenotemark{b} & $<$M3,F--K & 2,1 & IRTF/SpeX & 2015 Jan 7 \\
2MASS J03284622+3112034 & $<$M3,$<$K0 & 2,1 & IRTF/SpeX & 2015 Dec 14 \\
2MASS J03284624+3130120\tablenotemark{b} & M3.5,$<$M3,M3.5 & 3,2,1 & IRTF/SpeX & 2013 Aug 27 \\
2MASS J03284687+3120277\tablenotemark{b} & M5V & 1 & IRTF/SpeX & 2015 Jan 7 \\
2MASS J03285521+3125223 & $<$M3,G--K & 2,1 & IRTF/SpeX & 2015 Jan 6 \\
2MASS J03285750+3113162\tablenotemark{b} & giant & 1 & IRTF/SpeX & 2015 Jan 8 \\
2MASS J03290216+3116114\tablenotemark{b} & M4.2,M,M3.5,M3.5 & 4,5,3,1 & IRTF/SpeX & 2013 Aug 27 \\
2MASS J03290862+3122297 & $<$M0 & 1 & IRTF/SpeX & 2015 Dec 14 \\
2MASS J03291793+3114535\tablenotemark{b} & $<$M3,M1--M3: & 2,1 & IRTF/SpeX & 2015 Jan 7 \\
2MASS J03291987+3118478\tablenotemark{b} & $<$M3,F--K & 2,1 & IRTF/SpeX & 2015 Jan 8 \\
2MASS J03292760+3121100 & $<$M3,$<$M0 & 2,1 & IRTF/SpeX & 2015 Jan 6 \\
2MASS J03292805+3118391\tablenotemark{b} & $<$M3,A--F & 2,1 & IRTF/SpeX & 2015 Jan 7 \\
2MASS J03293084+3123529\tablenotemark{b} & A--F & 1 & IRTF/SpeX & 2015 Jan 6 \\
2MASS J03293219+3117074\tablenotemark{b} & $<$M3,A--F & 2,1 & IRTF/SpeX & 2015 Jan 6 \\
2MASS J03293240+3113011\tablenotemark{b} & F & 1 & IRTF/SpeX & 2015 Jan 6 \\
2MASS J03293441+3119106 & G--K & 1 & IRTF/SpeX & 2015 Jan 6 \\
2MASS J03293476+3129081\tablenotemark{b} & $<$M3,A--F & 2,1 & IRTF/SpeX & 2015 Jan 6 \\
2MASS J03293654+3129465 & giant & 1 & IRTF/SpeX & 2015 Jan 6 \\
2MASS J03293740+3117415\tablenotemark{b} & $<$M3,F--G & 2,1 & IRTF/SpeX & 2015 Jan 6 \\
2MASS J03293974+3114525 & G--K & 1 & IRTF/SpeX & 2015 Jan 6 \\
2MASS J03293976+3121144\tablenotemark{b} & G--K & 1 & IRTF/SpeX & 2015 Jan 6 \\
2MASS J03294283+3120147 & B--A & 1 & IRTF/SpeX & 2015 Jan 6 \\
2MASS J03295048+3118305\tablenotemark{b} & M3.0,M2.5 & 3,1 & IRTF/SpeX & 2013 Aug 27 \\
2MASS J03430722+3207169 & early or giant & 1 & Magellan/IMACS & 2005 Jan 4 \\
2MASS J03430800+3201275 & giant & 1 & Magellan/IMACS & 2005 Jan 4 \\
2MASS J03431143+3200276 & early or giant & 1 & Magellan/IMACS & 2005 Jan 4 \\
2MASS J03431541+3211037 & giant & 1 & Magellan/IMACS & 2005 Jan 4 \\
2MASS J03431928+3208537 & giant & 1 & Magellan/IMACS & 2005 Jan 4 \\
2MASS J03433322+3213047 & giant & 1 & Magellan/IMACS & 2005 Jan 4 \\
2MASS J03434503+3208479\tablenotemark{b} & $<$M0 & 1 & IRTF/SpeX & 2011 Oct 4 \\
2MASS J03434546+3201042 & $<$M0 & 1 & IRTF/SpeX & 2015 Dec 14 \\
2MASS J03434712+3213211 & $<$M0 & 1 & IRTF/SpeX & 2015 Dec 14 \\
2MASS J03435067+3213065\tablenotemark{b} & $<$M0 & 1 & IRTF/SpeX & 2011 Oct 5 \\
2MASS J03435160+3215565\tablenotemark{b} & $<$M0 & 1 & IRTF/SpeX & 2011 Oct 4 \\
2MASS J03435860+3218392 & $<$M0 & 1 & IRTF/SpeX & 2011 Oct 4 \\
2MASS J03440462+3220269\tablenotemark{b} & giant & 1 & Magellan/IMACS & 2005 Jan 4 \\
2MASS J03440616+3220420 & $<$M0 & 1 & IRTF/SpeX & 2011 Oct 4 \\
2MASS J03440973+3217130 & $<$M0 & 1 & IRTF/SpeX & 2011 Oct 4 \\
2MASS J03441829+3218588\tablenotemark{b} & $<$M0 & 1 & IRTF/SpeX & 2011 Oct 5 \\
IC 348 IRS J03441847+3206421\tablenotemark{b} & L & 1 & Gemini/GNIRS & 2015 Jan 7 \\
IC 348 IRS J03442070+3222489 & $<$M0 & 1 & IRTF/SpeX & 2011 Oct 5 \\
2MASS J03442086+3220439 & $<$M0 & 1 & IRTF/SpeX & 2011 Oct 4 \\
IC 348 IRS J03442188+3223370 & $<$M0 & 1 & IRTF/SpeX & 2011 Oct 5 \\
2MASS J03442250+3157054 & M0-M2? & 1 & IRTF/SpeX & 2015 Dec 14 \\
IC 348 IRS J03442484+3213482 & $<$M0 & 1 & Gemini/GNIRS & 2015 Nov 3 \\
2MASS J03442761+3156369 & $<$M0 & 1 & IRTF/SpeX & 2016 Jan 4 \\
2MASS J03443036+3221528 & $<$M0 & 1 & IRTF/SpeX & 2011 Oct 4 \\
IC 348 IRS J03443089+3200154 & $<$M0 & 1 & Gemini/GNIRS & 2015 Oct 25 \\
2MASS J03443157+3157288 & $<$K0 & 1 & IRTF/SpeX & 2016 Jan 4 \\
2MASS J03443537+3156081 & $<$M0 & 1 & IRTF/SpeX & 2011 Oct 5 \\
2MASS J03443675+3220239 & $<$M0 & 1 & IRTF/SpeX & 2011 Oct 5 \\
2MASS J03444410+3158091 & $<$M0 & 1 & IRTF/SpeX & 2015 Dec 14 \\
2MASS J03444525+3156379 & A & 1 & IRTF/SpeX & 2011 Oct 4 \\
2MASS J03444724+3156498\tablenotemark{b} & $<$M0 & 1 & IRTF/SpeX & 2011 Dec 3 \\
IC 348 IRS J03445102+3223094 & $<$M0 & 1 & IRTF/SpeX & 2015 Dec 14 \\
2MASS J03445462+3221405\tablenotemark{b} & A & 1 & IRTF/SpeX & 2011 Oct 4 \\
IC 348 IRS J03445516+3212136 & $<$M0 & 1 & IRTF/SpeX & 2015 Dec 14 \\
2MASS J03450578+3159339 & $<$M0 & 1 & IRTF/SpeX & 2011 Oct 5 \\
2MASS J03450980+3219303 & $<$M0 & 1 & IRTF/SpeX & 2015 Dec 14 \\
2MASS J03451154+3217358 & $<$M0 & 1 & IRTF/SpeX & 2011 Oct 4 \\
2MASS J03451566+3212090 & $<$M0 & 1 & IRTF/SpeX & 2015 Dec 14 \\
2MASS J03451737+3213591\tablenotemark{b} & $<$M0 & 1 & IRTF/SpeX & 2011 Oct 4 \\
2MASS J03451851+3206153\tablenotemark{b} & $<$M0 & 1 & IRTF/SpeX & 2011 Oct 5 \\
2MASS J03452158+3203289\tablenotemark{b} & $<$M0 & 1 & IRTF/SpeX & 2011 Oct 5 \\
2MASS J03452222+3203379 & $<$M0 & 1 & IRTF/SpeX & 2015 Dec 14 \\
2MASS J03452475+3209488 & $<$M0 & 1 & IRTF/SpeX & 2015 Dec 14 \\
2MASS J03452900+3203483 & M3-M5 & 1 & IRTF/SpeX & 2016 Jan 4 \\
2MASS J03452934+3200296\tablenotemark{b} & $<$M0 & 1 & IRTF/SpeX & 2011 Oct 5 \\
2MASS J03453078+3214320 & M3.5V & 1 & IRTF/SpeX & 2011 Oct 5 \\
2MASS J03453279+3208070 & $<$M0 & 1 & IRTF/SpeX & 2011 Oct 4 \\
2MASS J03453669+3213041 & $<$M0 & 1 & IRTF/SpeX & 2011 Oct 5 \\
2MASS J03453789+3208249 & $<$M0 & 1 & IRTF/SpeX & 2011 Oct 4 \\
2MASS J03453925+3210093 & $<$G0 & 1 & IRTF/SpeX & 2011 Oct 4 \\
2MASS J03453998+3208334 & $<$M0 & 1 & IRTF/SpeX & 2015 Dec 14 \\
2MASS J03454159+3214225 & $<$M0 & 1 & IRTF/SpeX & 2011 Oct 4 \\
\enddata
\tablenotetext{a}{
(1) this work;
(2) \citet{sch12a};
(3) \citet{win09};
(4) \citet{wil04};
(5) \citet{sch09}.}
\tablenotetext{b}{Proper motion differs from cluster median by
$>6$~mas~yr$^{-1}$ (outside of outer circle in Fig.~\ref{fig:pm}).}
\end{deluxetable}

\begin{deluxetable}{llllll}
\tabletypesize{\scriptsize}
\tablewidth{0pt}
\tablecaption{Candidate Members of IC~348\label{tab:cand348}}
\tablehead{
\colhead{Name\tablenotemark{a}} & \colhead{LRL} & \colhead{Basis of}
& \colhead{$J$\tablenotemark{c}} & \colhead{$H$\tablenotemark{c}} &
\colhead{$K_s$\tablenotemark{c}}\\
\colhead{} & \colhead{} & \colhead{Selection\tablenotemark{b}}
& \colhead{(mag)} & \colhead{(mag)} & \colhead{(mag)}}
\startdata
IC 348 IRS J03434428+3203424 & 54299 & pm & \nodata & 18.52$\pm$0.04 & 15.74$\pm$0.03 \\
2MASS J03435901+3158282 & 1869 & pm? & 19.34$\pm$0.03 & 16.59$\pm$0.03 & 14.88$\pm$0.03 \\
IC 348 IRS J03440704+3159242 & 54486 & pm? & \nodata & 19.21$\pm$0.05 & 17.68$\pm$0.03 \\
IC 348 IRS J03441057+3157004 & 54502 & pm? & 20.47$\pm$0.05 & 17.60$\pm$0.03 & 16.03$\pm$0.03 \\
IC 348 IRS J03441387+3157170 & 54532 & pm & 20.73$\pm$0.06 & 18.20$\pm$0.03 & 16.60$\pm$0.03 \\
IC 348 IRS J03442154+3157381 & 52567 & pm & 18.81$\pm$0.03 & 16.60$\pm$0.03 & 15.39$\pm$0.03 \\
2MASS J03442217+3159371 & 10418 & pm? & \nodata & 18.42$\pm$0.03 & 15.90$\pm$0.03 \\
IC 348 IRS J03442355+3157339 & 1865 & CMD,pm? & 18.42$\pm$0.03 & 16.48$\pm$0.03 & 15.40$\pm$0.03 \\
IC 348 IRS J03442539+3207245 & 30053 & CMD & 20.58$\pm$0.04 & 19.31$\pm$0.04 & 18.44$\pm$0.03 \\
IC 348 IRS J03442750+3200477 & 22286 & pm? & 19.37$\pm$0.03 & 17.52$\pm$0.03 & 16.51$\pm$0.03 \\
IC 348 IRS J03442790+3157489 & 52620 & pm & 19.52$\pm$0.03 & 17.48$\pm$0.03 & 16.39$\pm$0.03 \\
2MASS J03442811+3158306 & 1886 & pm? & 20.11$\pm$0.03 & 16.73$\pm$0.03 & 14.87$\pm$0.03 \\
IC 348 IRS J03443171+3204326 & 30024 & CMD & 20.18$\pm$0.03 & 18.92$\pm$0.03 & 18.20$\pm$0.03 \\
IC 348 IRS J03443276+3158294 & 52641 & pm & \nodata & 19.25$\pm$0.04 & 17.48$\pm$0.03 \\
IC 348 IRS J03443516+3211052 & 596 & CMD & 19.91$\pm$0.03 & 18.70$\pm$0.03 & 17.82$\pm$0.03 \\
IC 348 IRS J03443537+3158233 & 52657 & pm? & \nodata & 19.75$\pm$0.10 & 17.85$\pm$0.03 \\
IC 348 IRS J03443631+3205066 & 30030 & CMD & 20.31$\pm$0.03 & 19.16$\pm$0.03 & 18.19$\pm$0.03 \\
2MASS J03444037+3157292 & 1929 & pm & 19.35$\pm$0.03 & 16.65$\pm$0.03 & 15.15$\pm$0.03 \\
IC 348 IRS J03444668+3201010 & 40172 & pm & \nodata & 17.93$\pm$0.03 & 15.61$\pm$0.03 \\
IC 348 IRS J03450146+3213418 & 30124 & CMD & 20.52$\pm$0.05 & 19.23$\pm$0.04 & 18.27$\pm$0.03 \\
IC 348 IRS J03450384+3200235 & 40182 & pm & \nodata & 18.70$\pm$0.04 & 16.35$\pm$0.03 \\
IC 348 IRS J03450403+3158129 & 22639 & pm & 20.33$\pm$0.07 & 19.04$\pm$0.05 & 17.85$\pm$0.04 \\
IC 348 IRS J03451226+3205502 & 22766 & CMD,pm? & 18.51$\pm$0.03 & 16.46$\pm$0.03 & 15.40$\pm$0.03 \\
IC 348 IRS J03451871+3205310 & 22898 & pm? & 19.59$\pm$0.03 & 17.53$\pm$0.03 & 16.45$\pm$0.03 \\
\enddata
\tablenotetext{a}{Coordinate-based identifications from the 2MASS Point
Source Catalog when available. Otherwise, identifications are based on the
coordinates measured from the WIRCam images in this work.}
\tablenotetext{b}{Sources were selected as candidate members based on
the color-magnitude diagrams in Figure~\ref{fig:cmd348} (CMD) or
proper motions in Figure~\ref{fig:pm} (pm or pm?).}
\tablenotetext{c}{WIRCam data from this work.}
\end{deluxetable}

\begin{deluxetable}{ll}
\tabletypesize{\scriptsize}
\tablewidth{0pt}
\tablecaption{Candidate Members of NGC~1333\label{tab:cand1333}}
\tablehead{
\colhead{Column Label} &
\colhead{Description}}
\startdata
Name & Source name\tablenotemark{a} \\
Aspin & Name from \citet{asp94} \\
Lada & Name from \citet{lad96} \\
Wilking & Name from \citet{wil04} \\
Gutermuth & Name from \citet{gut08} \\
Oasa & Name from \citet{ots08} \\
Evans & Name from \citet{eva09} \\
Winston & Name from \citet{win10} \\
Rebull & Name from \citet{reb15b} \\
Selection & Basis of selection\tablenotemark{b} \\
Jmag & $J$ magnitude \\
e\_Jmag & Error in Jmag \\
Hmag & $H$ magnitude \\
e\_Hmag & Error in Hmag \\
Ksmag & $K_s$ magnitude \\
e\_Ksmag & Error in Ksmag \\
JHKref & JHK references\tablenotemark{c} \\
\enddata
\tablenotetext{a}{Coordinate-based identifications from the 2MASS Point
Source Catalog when available. Otherwise, identifications are based on the
coordinates measured from the WIRcam images in this work.}
\tablenotetext{b}{Sources were selected as candidate members based on
X-ray emission \citep[X,][K. Getman, in preparation]{get02,win10},
mid-IR excess emission \citep[IR,][]{gut08,eva09}, the color-magnitude diagrams
in Figure~\ref{fig:cmd1333} (CMD), or proper motions in Figure~\ref{fig:pm} (pm
or pm?).}
\tablenotetext{c}{2 = 2MASS Point Source Catalog; u = UKIDSS Data Release 10;
w = WIRCam data from this work.}
\tablecomments{
The table is available in a machine-readable format in the online journal.}
\end{deluxetable}

\begin{deluxetable}{lll}
\tabletypesize{\scriptsize}
\tablewidth{0pt}
\tablecaption{Disk Fractions in IC~348 and NGC~1333\tablenotemark{a}
\label{tab:disks}}
\tablehead{
\colhead{Spectral Type} & 
\colhead{IC~348} &
\colhead{NGC~1333}}
\startdata
$<$K6 & 13/35=$0.37^{+0.09}_{-0.07}$ & 3/5=$0.60^{+0.16}_{-0.21}$\\
K6--M3.5 & 48/119=$0.40^{+0.05}_{-0.04}$ & 27/39=$0.69^{+0.06}_{-0.08}$\\
M3.75--M5.75 & 71/184=$0.39^{+0.04}_{-0.03}$ & 28/48=$0.58\pm0.07$\\
M6--M8 & 22/46=$0.48\pm0.07$ & 20/33=$0.61^{+0.08}_{-0.09}$\\
$>$M8 & 10/24=$0.42^{+0.11}_{-0.09}$ & 11/21=$0.52\pm0.10$\\
\enddata
\tablenotetext{a}{Fraction of sources with measured spectral types
and within the $14\arcmin$ radius field in IC~348 and within the ACIS-I field
in NGC~1333 that exhibit mid-IR excess emission.}
\end{deluxetable}

\begin{deluxetable}{lllll}
\tabletypesize{\scriptsize}
\tablewidth{0pt}
\tablecaption{Pairs of Members of IC~348 and NGC~1333 Separated by $<6\arcsec$
\label{tab:pairs}}
\tablehead{
\colhead{Name} &
\colhead{Spectral} &
\colhead{Name} &
\colhead{Spectral} &
\colhead{Separation}\\
\colhead{} &
\colhead{Type} &
\colhead{} &
\colhead{Type} &
\colhead{(arcsec)}}
\startdata
LRL 12 A & G0 & LRL 12 B & A3 &  1.29 \\
LRL 16 & G6 & LRL 33 & M2.5 &  5.41 \\
LRL 24 A & K6.5 & LRL 24 B & M0 &  4.32 \\
LRL 42 A & M4.25 & LRL 42 B & M2.5 &  2.63 \\
LRL 60 A & M2 & LRL 60 B & M2 &  1.21 \\
LRL 78 A & M0.5 & LRL 78 B & M6.5 &  1.64 \\
LRL 99 A & M3.75 & LRL 99 B & M5.25 &  3.13 \\
LRL 138 & M4 & LRL 3093 & M5.5 &  2.80 \\
LRL 160 & M4.75 & LRL 55400 & \nodata &  5.65 \\
LRL 165 & M5.25 & LRL 366 & M5 &  5.65 \\
LRL 166 A & M4.25 & LRL 166 B & M5.75 &  0.79 \\
LRL 187 & M4.25 & LRL 9187 & M4.25 &  1.29 \\
LRL 192 & M4.5 & LRL 1684 & M5.75 &  4.74 \\
LRL 210 & M3.5 & LRL 761 & M7 &  4.51 \\
LRL 259 A & M5 & LRL 259 B & M5 &  2.08 \\
LRL 1937 & M0 & LRL 1928 & M5.5 &  4.35 \\
LRL 233 & M4.75 & LRL 3171 & L0 &  2.85 \\
LRL 265 & M3.25 & LRL 148 & M4 &  5.68 \\
LRL 102 & M0 & LRL 22356 & M3.75 &  1.90 \\
LRL 104 & M4 & LRL 22317 & M8.25 &  2.00 \\
LRL 10148 & M5 & LRL 23177 & M5.75 &  1.90 \\
LRL 264 & M3 & LRL 4035 & M5.5 &  2.56 \\
LRL 54459 & \nodata & LRL 54460 & \nodata &  5.54 \\
2MASS J03284325+3117330 & K7 & 2MASS J03284355+3117364 & M5.5 &  5.06 \\
2MASS J03285505+3116287 & M2 & 2MASS J03285514+3116247 & M1 &  4.24 \\
2MASS J03285720+3114189 & G3 & NGC 1333 IRS J03285737+3114162 & \nodata &  3.44 \\
2MASS J03285769+3119481 & M3.5 & 2MASS J03285741+3119505 & M3.5 &  4.32 \\
2MASS J03290279+3122172 & M8.25 & NGC 1333 IRS J03290236+3122159 & M9 &  5.67 \\
2MASS J03290493+3120385 & M7.5 & NGC 1333 IRS J03290517+3120370 & M9 &  3.40 \\
2MASS J03290575+3116396\tablenotemark{a} & M1 & NGC 1333 IRS J03290591+3116403 & M6 &  2.14 \\
2MASS J03294592+3104406S & M1 & 2MASS J03294592+3104406N & M3.75 &  2.77 \\
\enddata
\tablenotetext{a}{This star has an additional candidate companion,
[OTS2008] 48, at a separation of $2\arcsec$.}
\end{deluxetable}

\clearpage

\clearpage

\clearpage

\begin{thebibliography}{}

\bibitem[Allison et al.(2009)]{all09}
Allison, R. J., Goodwin, S. P., Parker, R. J., et al. 2009, \mnras, 395, 1449

\bibitem[Alves de Oliveira et al.(2013)]{alv13}
Alves de Oliveira, C., Moraux, E., Bouvier, J., et al. 2013, \aap, 549, A123

\bibitem[Andr\'e et al.(1993)]{and93}
Andr\'e, P., Ward-Thompson, D., \& Barsony, M. 1993, \apj, 406, 122

\bibitem[Arnold et al.(2012)]{arn12}
Arnold, L. A., Watson, D. M., Kim, K. H., et al. 2012, \apjs, 201, 12

\bibitem[Ascenso et al.(2009)]{asc09}
Ascenso, J., Alves, J., \& Lago, M. T. V. T. 2009, \aap, 495, 147

\bibitem[Aspin(2003)]{asp03}
Aspin, C. 2003, \aj, 125, 1480

\bibitem[Aspin et al.(1994)]{asp94}
Aspin, C., Sandell, G., \& Russell, A. P. G. 1994, \aaps, 106, 165 

\bibitem[Bally et al.(2008)]{bal08}
Bally, J., Walawender, J., Johnstone, D., Kirk, H., \& Goodman, A.
2008, in ASP Monograph Series 4, Handbook of Star Forming Regions, Vol. 1,
The Northern Sky, ed. B. Reipurth (San Francisco, CA: ASP), 308

\bibitem[Baraffe et al.(1998)]{bar98}
Baraffe, I., Chabrier, G., Allard, F., \& Hauschildt, P. H. 1998, \aap, 337, 403

\bibitem[Baraffe et al.(2015)]{bar15}
Baraffe, I., Hormeier, D., Allard, F., \& Chabrier, G. 2015, \aap, 577, 42


\bibitem[Bell et al.(2013)]{bel13}
Bell, C. P. M., Naylor, T., Mayne, N. J., Jeffries, R. D., \& Littlefair, S. P.
2013, \mnras, 434, 806

\bibitem[Burgess et al.(2009)]{bur09}
Burgess, A. S. M., Moraux, E., Bouvier, J., et al. 2009, \aap, 508, 823

\bibitem[Burrows et al.(1997)]{bur97}
Burrows, A., Marley, M., Hubbard, W. B., et al. 1997, \apj, 491, 856

\bibitem[Cardelli et al.(1989)]{car89}
Cardelli, J. A., Clayton, G. C., \& Mathis, J. S. 1989, \apj, 345, 245

\bibitem[Chabrier et al.(2000)]{cha00}
Chabrier, G., Baraffe, I., Allard, F., \& Hauschildt, P. 2000, \apj, 542, 464

\bibitem[Cieza et al.(2007)]{cie07}
Cieza, L., Padgett, D. L., Stapelfeldt, K. R., et al. 2007, \apj, 667, 308

\bibitem[Cieza et al.(2012)]{cie12}
Cieza, L. A., Schreiber, M. R., Romero, G. A., et al. 2012, \apj, 750, 157

\bibitem[Cody \& Hillenbrand(2014)]{cod14}
Cody, A. M., \& Hillenbrand, L. A. 2014, \apj, 796, 129

\bibitem[Cohen(1980)]{coh80}
Cohen, M. 1980, \aj, 85, 29

\bibitem[Connelley \& Greene(2010)]{con10}
Connelley, M. S. \& Greene, T. P. 2010, \aj, 140, 1214

\bibitem[Cottaar et al.(2015)]{cot15}
Cottaar, M., Covey, K. R., Foster, J. B., et al. 2015, \apj, 807, 27



\bibitem[Currie \& Kenyon (2009)]{cur09}
Currie, T., \& Kenyon, S. J. 2009, \aj, 138, 703

\bibitem[Cushing et al.(2005)]{cus05} 
Cushing, M. C., Rayner, J. T., \& Vacca, W. D. 2005, \apj, 623, 1115

\bibitem[Cushing et al.(2004)]{cus04} 
Cushing, M. C., Vacca, W. D., \& Rayner, J. T. 2004, \pasp, 116, 362

\bibitem[de Geus et al.(1989)]{deg89}
de Geus, E. J., de Zeeuw, P. T., \& Lub, J. 1989, \aap, 216, 44


\bibitem[Duch\^ene et al.(1999)]{duc99}
Duch\^ene, G., Bouvier, J., \& Simon, T. 1999, \aap, 343, 831

\bibitem[Elias et al.(2006)]{eli06}
Elias, J. H., Joyce, R. R., Liang, M., et al. 2006, SPIE, 6269, 62694C


\bibitem[Esplin \& Luhman(2016)]{esp16}
Esplin, T. L., \& Luhman, K. L. 2016, \aj, 151, 9

\bibitem[Evans et al.(2003)]{eva03}
Evans, N. J., II, Allen, L. E., Blake, G. A., et al. 2003, \pasp, 115, 965

\bibitem[Evans et al.(2009)]{eva09}
Evans, N. J., Dunham, M. M., J{\o}rgensen, J. K., et al. 2009, \apjs, 181, 321

\bibitem[Fazio et al.(2004)]{faz04}
Fazio, G. G., Hora, J. L., Allen, L. E., et al. 2004, \apjs, 154, 10

\bibitem[Feigelson et al.(1987)]{fei87}
Feigelson, E. D., Jackson, J. M., Mathieu, R. D., et al. 1987, \aj, 94, 1251

\bibitem[Foster et al.(2015)]{fos15}
Foster, J. B., Cottaar, M., Covey, K. R., et al. 2015, \apj, 799, 136

\bibitem[Flaherty et al.(2012)]{fla12}
Flaherty, K. M., Muzerolle, J., Rieke, G. H., et al. 2012, \apj, 748, 71

\bibitem[Flaherty et al.(2013)]{fla13}
Flaherty, K. M., Muzerolle, J., Rieke, G. H., et al. 2013, \aj, 145, 66

\bibitem[Forbrich et al.(2011)]{for11}
Forbrich, J., Osten, R., \& Wolk, S. J. 2011, \apj, 736, 25

\bibitem[Fredrick(1956)]{fre56}
Fredrick, L. W. 1956, \aj, 61, 437

\bibitem[Furlan et al.(2011)]{fur11}
Furlan, E., Luhman, K. L., Espaillat, C., et al. 2011, \apjs, 195, 3

\bibitem[Getman et al.(2014)]{get14}
Getman, K. V., Feigelson, E. D., \& Kuhn, M. A. 2014, \apj, 787, 109

\bibitem[Getman et al.(2002)]{get02}
Getman, K. V., Feigelson, E. D., Townsley, L., et al. 2002, \apj, 575, 354

\bibitem[Greene et al.(1994)]{gre94}
Greene, T. P., Wilking, B. A., Andr\'{e}, P., Young, E. T., \& Lada, C. J.
1994, \apj, 434, 614

\bibitem[Greissl et al.(2007)]{gre07}
Greissl, J., Meyer, M. R., Wilking, B. A., et al. 2007, \aj, 133, 1321

\bibitem[Gutermuth et al.(2009)]{gut09}
Gutermuth, R. A., Megeath, S. T., Myers, P. C., et al. 2009, \apjs, 184, 18

\bibitem[Gutermuth et al.(2008)]{gut08}
Gutermuth, R. A., Myers, P. C., Megeath, S. T., et al. 2008, \apj, 674, 336

\bibitem[Harris et al.(1954)]{har54}
Harris, D. L., Morgan, W. W., \& Roman, N. G. 1954, \apj, 119, 622

\bibitem[Hatchell \& Dunham(2009)]{hat09}
Hatchell, J., \& Dunham, M. M. 2009, \aap, 502, 139

\bibitem[Herbig(1954)]{her54}
Herbig, G. H. 1954, \pasp, 66, 19

\bibitem[Herbig(1998)]{her98}
Herbig, G. H. 1998, \apj, 497, 736

\bibitem[Herbig(1983)]{her83}
Herbig, G. H., \& Jones, B. F. 1983, \aj, 88, 1040

\bibitem[Herbst(2008)]{her08}
Herbst, W. 2008, in Handbook of Star Forming Regions, Vol. 1, The Northern Sky,
ASP Monograph Series 4, ed. B. Reipurth (San Francisco, CA: ASP), 372

\bibitem[Herczeg \& Hillenbrand(2015)]{her15}
Herczeg, G. J., \& Hillenbrand, L. A. 2015, \apj, 808, 23

\bibitem[Hillenbrand \& Hartmann(1998)]{hil98}
Hillenbrand, L. A., \& Hartmann, L. W. 1998, \apj, 492, 540

\bibitem[Hirota et al.(2008)]{hir08}
Hirota, T., Bushimata, T., Choi, Y. K., et al. 2008, \pasj, 60, 37

\bibitem[Hodapp et al.(2003)]{hod03}
Hodapp, K. W., Jensen, J. B., Irwin, E. M., et al. 2003, \pasp, 115, 1388

\bibitem[J{\o}rgensen et al.(2006)]{jor06}
J{\o}rgensen, J. K., Harvey, P. M., Evans, N. J., II, et al. 2006, \apj, 645,
1246

\bibitem[J{\o}rgensen et al.(2007)]{jor07}
J{\o}rgensen, J. K., Johnstone, D., Kirk, H., \& Myers, P. C. 2007, \apj, 656,
293






\bibitem[Koenker(2016)]{koe16}
Koenker, R. 2016, quantreg: Quantile Regression. R package version 5.21,
\url{http://CRAN.R-project.org/package=quantreg}

\bibitem[Kuhn et al.(2014)]{kuh14}
Kuhn, M. A., Feigelson, E. D., Getman, K. V., et al. 2014, \apj, 787, 107

\bibitem[Kroupa(1998)]{kro98}
Kroupa, P. 1998, \mnras, 298, 231

\bibitem[Lada(1987)]{lad87}
Lada, C. J. 1987, in IAU Symp. 115, Star Forming Regions, ed. M. Peimbert \&
J. Jugaku (Dordrecht: Reidel), 1

\bibitem[Lada et al.(1996)]{lad96}
Lada, C. J., Alves, J., \& Lada, E. A. 1996, \aj, 111, 1964

\bibitem[Lada \& Lada(1995)]{lad95}
Lada, E. A., \& Lada, C. J. 1995, \aj, 109, 1682

\bibitem[Lada et al.(2006)]{lad06}
Lada, C. J., Muench, A. A., Luhman, K. L., et al. 2006, \aj, 131, 1574

\bibitem[Lada \& Wilking(1984)]{lw84}
Lada, C. J., \& Wilking, B. A. 1984, \apj, 287, 610

\bibitem[Lawrence et al.(2007)]{law07}
Lawrence, A., Warren, S. J., Almaini, O., et al. 2007, \mnras, 379, 1599

\bibitem[Lucas et al.(2001)]{luc01}
Lucas, P. W., Roche, P. F., Allard, F., \& Hauschildt, P. H. 2001, \mnras,
326, 695

\bibitem[Luhman(1999)]{luh99} 
Luhman, K. L. 1999, \apj, 525, 466


\bibitem[Luhman(2004)]{luh04bin} 
Luhman, K. L. 2004, \apj, 614, 398





\bibitem[Luhman(2012)]{luh12} 
Luhman, K. L. 2012, \araa, 50, 65

\bibitem[Luhman et al.(2007)]{luh07edge} 
Luhman, K. L., Adame, L., D'Alessio, P., et al. 2007, \apj, 666, 1219

\bibitem[Luhman et al.(2010)]{luh10tau} 
Luhman, K. L., Allen, P. R., Espaillat, C., Hartmann, L., \& Calvet, N. 2010,
\apjs, 186, 111 

\bibitem[Luhman et al.(2003a)]{luh03tau} 
Luhman, K. L., Brice\~{n}o, C., Stauffer, J. R., et al. 2003a, \apj, 590, 348

\bibitem[Luhman et al.(2005c)]{luh05frac}
Luhman, K. L., Lada, C. J., Hartmann, L., et al. 2005c, \apj, 631, L69

\bibitem[Luhman et al.(2005a)]{luh05flam}
Luhman, K. L., Lada, E. A., Muench, A. A., \& Elston, R. J. 2005a, \apj, 618,
810

\bibitem[Luhman \& Mamamjek(2012)]{luh12usco}
Luhman, K. L., \& Mamajek, E. E. 2012, \apj, 758, 31

\bibitem[Luhman et al.(2009)]{luh09fu}
Luhman, K. L., Mamajek, E. E., Allen, P. R., Muench, A. A., \& 
Finkbeiner, D. P.  2009, \apj, 691, 1265

\bibitem[Luhman et al.(2005b)]{luh05wfpc}
Luhman, K. L., McLeod, K. K., \& Goldenson, N. 2005b, \apj, 623, 1141


\bibitem[Luhman et al.(1998)]{luh98}
Luhman, K. L., Rieke, G. H., Lada, C. J., \& Lada, E. A. 1998, \apj, 508, 347

\bibitem[Luhman et al.(2003b)]{luh03} 
Luhman, K. L., Stauffer, J. R., Muench, A. A., et al. 2003b, \apj, 593, 1093




\bibitem[Mainzer \& McLean(2003)]{mai03}
Mainzer, A. K.,, \& McLean, I. S. 2003, \apj, 597, 555


\bibitem[Maschberger \& Clarke(2011)]{mas11}
Maschberger T., \& Clarke C. J., 2011, \mnras, 416, 541

\bibitem[Matthews \& Soiffer(1994)]{mat94}
Matthews, K., \& Soifer, B. T. 1994, Exp. Astron., 3, 77

\bibitem[Merin et al.(2010)]{mer10}
Merin, B., Brown, J. M., Oliveira, I., et al. 2010, \apj, 718, 1200

\bibitem[Monet et al.(2003)]{mon03}
Monet, D. G., Levine, S. E., Canzian, B., et al. 2003, \aj, 125, 984

\bibitem[Muench et al.(2003)]{mue03}
Muench, A. A., Lada, E. A., Lada, C. J., et al. 2003, \aj, 125, 2029

\bibitem[Muench et al.(2007)]{mue07}
Muench, A. A., Lada, C. J., Luhman, K. L., Muzerolle, J., \& Young, E. 2007,
\aj, 134, 411

\bibitem[Najita et al.(2000)]{naj00}
Najita, J., Tiede, G. P., \& Carr, J. S. 2000, \apj, 541, 977

\bibitem[Oasa et al.(2008)]{ots08}
Oasa, Y., Tamura, M., Sunada, K., \& Sugitani, K. 2008, \aj, 136, 1372

\bibitem[Olczak et al.(2011)]{olc11}
Olczak, O., Spurzem, R., \& Henning, Th. 2011, \aap, 532, A119

\bibitem[Palau et al.(2014)]{pal14}
Palau, A., Zapata, L. A., Rodriguez, L. F., et al. 2014, \mnras, 444, 833

\bibitem[Parker \& Goodwin(2015)]{par15}
Parker, R. J., \& Goodwin, S. P. 2015, \mnras, 449, 3381

\bibitem[Pecaut et al.(2012)]{pec12}
Pecaut, M. J., Mamajek, E. E., Bubar, E. J. 2012, \apj, 746, 154

\bibitem[Perryman et al.(2001)]{per01}
Perryman, M. A. C., de Boer, K. S., Gilmore, G., et al. 2001, \aap, 369, 339

\bibitem[Preibisch(1997)]{pre97}
Preibisch, T. 1997, \aap, 324, 690

\bibitem[Preibisch(2003)]{pre03}
Preibisch, T. 2003, \aap, 401, 543

\bibitem[Preibisch et al.(2002)]{pre02b}
Preibisch, T., Brown, A. G. A., Bridges, T. Guenther, E., \& Zinnecker, H.
2002, \aj, 124, 404

\bibitem[Preibisch \& Mamajek(2008)]{pm08}
Preibisch, T., \& Mamajek, E. 2008,
in Handbook of Star Forming Regions, Vol. 2, The Southern Sky,
ASP Monograph Series 5, ed. B. Reipurth (San Francisco, CA: ASP), 235

\bibitem[Preibisch \& Zinnecker(2001)]{pre01}
Preibisch, T., \& Zinnecker, H. 2001, \aj, 122, 866

\bibitem[Preibisch \& Zinnecker(2002)]{pre02}
Preibisch, T., \& Zinnecker, H. 2002, \aj, 123, 1613

\bibitem[Preibisch \& Zinnecker(2004)]{pre04}
Preibisch, T., \& Zinnecker, H. 2004, \aap, 422, 1001

\bibitem[Preibisch et al.(1996)]{pre96}
Preibisch, T., Zinnecker, H., \& Herbig, G. H. 1996, \aap, 310, 456

\bibitem[R Core Team(2013)]{R}
R Core Team, 2013, R Foundation for Statistical Computing, Vienna, Austria,
\url{http://www.R-project.org}

\bibitem[Racine(1968)]{rac68}
Racine, R. 1968, \aj, 73, 233

\bibitem[Rayner et al.(2009)]{ray09}
Rayner, J. T., Cushing, M. C., \& Vacca, W. D. 2009, \apjs, 185, 289

\bibitem[Rayner et al.(2003)]{ray03}
Rayner, J. T., Toomey, D. W., Onaka, P. M., et al. 2003, \pasp, 115, 362


\bibitem[Rebull et al.(2007)]{reb07}
Rebull, L. M., Stapelfeldt, N. J., Evans, N. J., II, et al. 2007, \apjs, 171,
447

\bibitem[Rebull et al.(2015)]{reb15b}
Rebull, L. M., Stauffer, J. R., Cody, A. M., et al. 2015, \aj, 150, 175


\bibitem[Rieke et al.(2004)]{rie04}
Rieke, G. H., Young, E. T.; Engelbracht, C. W., et al. 2004, \apjs, 154, 25

\bibitem[Ripepi et al.(2014)]{rip14}
Ripepi, V., Molinaro, R., Marconi, M., et al. 2014, \mnras, 437, 906

\bibitem[Sager et al.(1988)]{sag88}
Sagar R., Miakutin V. I., Piskunov A. E., Dluzhnevskaia O. B., 1988, \mnras,
234, 831

\bibitem[Schlafly et al.(2014)]{sch14}
Schlafly, E. F., Green, G., Finkbeiner, D. P., et al. 2014, \apj, 786, 29

\bibitem[Scholz et al.(1999)]{sch99}
Scholz, R.-D., Brunzendor, J., Ivanov, G., et al. 1999, \aaps, 137, 305

\bibitem[Scholz et al.(2013)]{sch13}
Scholz, A., Geers, V., Clark, P., Jayawardhana, R., \& Muzic, K. 2013, \apj,
775, 138

\bibitem[Scholz et al.(2009)]{sch09}
Scholz, A., Geers, V., Jayawardhana, R., et al. 2009, \apj, 702, 805

\bibitem[Scholz et al.(2012b)]{sch12b}
Scholz, A., Jayawardhana, R., Muzic, K., et al. 2012b, \apj, 756, 24

\bibitem[Scholz et al.(2012a)]{sch12a}
Scholz, A., Muzic, K., Geers, V., et al. 2012a, \apj, 744, 6

\bibitem[Skrutskie et al.(2006)]{skr06}
Skrutskie, M., Cutri, R. M., Stiening, R., et al. 2006, \aj, 131, 1163

\bibitem[Slesnick et al.(2006)]{sle06}
Slesnick, C. L., Carpenter, J. M., \& Hillenbrand, L. A. 2006, \aj, 131, 3016

\bibitem[Stelzer et al.(2012)]{ste12}
Stelzer, B., Preibisch, T., Alexander, F., et al. 2012, \aap, 537, 135

\bibitem[Strai{\v z}ys et al.(2002)]{str02}
Strai{\v z}ys, Corbally, C. J., Kazlauskas, A., \& {\v C}ernis, K. 2002,
Baltic Astronomy, 11, 261

\bibitem[Strom et al.(1974a)]{str74a}
Strom, S. E., Grasdalen, G. L., \& Strom, K. M. 1974a, \apj, 191, 111

\bibitem[Strom et al.(1974b)]{str74b}
Strom, S. E., Strom, K. M., \& Carrasco, L. 1974b, \pasp, 86, 798

\bibitem[Strom et al.(1976)]{str76}
Strom, S. E., Vrba, F. J., \& Strom, K. M. 1976, \aj, 81, 314

\bibitem[Turnshek et al.(1980)]{tur80}
Turnshek, D. A., Turnshek, D. E., \& Craine, E. R. 1980, \aj, 85, 1638

\bibitem[Vacca et al.(2003)]{vac03} 
Vacca, W. D., Cushing, M. C., \& Rayner J. T., 2003, \pasp, 115, 389

\bibitem[Venables \& Ripley(2002)]{ven02}
Venables, W. N., \& Ripley, B. D. 2002, Modern Applied Statistics
with S. Fourth Edition. Springer, New York, 
\url{http://CRAN.R-project.org/package=MASS}

\bibitem[Walawender et al.(2008)]{wal08}
Walawender, J., Bally, J., Franceso, J. D., J{\o}rgensen, J., \& Getman, K.
2008, in ASP Monograph Series 4, Handbook of Star Forming Regions, Vol. 1,
The Northern Sky, ed. B. Reipurth (San Francisco, CA: ASP), 346

\bibitem[Walawender et al.(2006)]{wal06}
Walawender, J., Bally, J., Kirk, H., et al. 2006, \aj, 132, 467

\bibitem[Walter et al.(1988)]{wal88}
Walter, F. M., Brown, A., Mathieu, R. D., Myers, P. C., \& Vrba, F. J. 1988,
\aj, 96, 297

\bibitem[Weidner et al.(2011)]{wei11}
Weidner, C., Bonnell, I. A., \& Moeckel, N. 2011, \mnras, 410, 1861

\bibitem[Werner et al.(2004)]{wer04}
Werner, M. W., Roellig, T. L., Low, F. J., et al. 2004, \apjs, 154, 1

\bibitem[Wilking et al.(2004)]{wil04}
Wilking, B. A., Meyer, M. R., Green, T. P., Mikhail, A., \& Carlson, G. 2004,
\aj, 127, 1131

\bibitem[Winston et al.(2009)]{win09}
Winston, E., Megeath, S. T., Wolk, S. J., et al. 2009, \aj, 137, 4777

\bibitem[Winston et al.(2010)]{win10}
Winston, E., Megeath, S. T., Wolk, S. J., et al. 2010, \aj, 140, 266

\bibitem[Young et al.(2015)]{you15}
Young, K. E., Young, C. H., Lai, S.-P., Dunham, M. M., \& Evans, N. J.
2015, \aj, 150, 40

\end{thebibliography}
\end{document}